\documentclass[11pt]{article}
\usepackage{graphicx} 
\usepackage{feynmf}
\usepackage{amssymb}
\usepackage{amsmath}
\usepackage{xcolor}
\usepackage{soul}
\usepackage{dsfont}
\usepackage{multirow}
\usepackage{array}
\usepackage{cancel}
\usepackage{soul}
\newcommand{\cals}{\text{$\cal S$}}

\newcommand{\calt}{\text{$\cal T$}}
\newcommand{\calc}{\text{$\cal C$}}

\newcommand{\calf}{\text{${\cal F}$}}
\newcommand{\call}{\text{${\cal L}$}}

\newcommand{\calftp}{\text{$\tilde{{\cal F}^{\prime}}$}}
\newcommand{\calft}{\text{$\tilde{{\cal F}}$}}
\newcommand{\calh}{\text{${\cal H}$}}

\newcommand{\cali}{\mbox{${\cal I}$}}
\newcommand{\calj}{\mbox{${\cal J}$}}

\newcommand{\dilog}{\text{Li}_2}
\newcommand{\trilog}{\text{Li}_3}

\newcommand{\bbar}{\overline{b}}
\newcommand{\detg}{\det{(G)}}

\newcommand{\sign}{\mbox{sign}}

\newcommand{\baru}{\bar{u}}
\newcommand{\barxi}{\bar{\xi}}

\newcommand{\hr}{\hat{r}}
\newcommand{\br}{\bar{r}}
\newcommand{\hk}{\hat{k}}
\newcommand{\tD}{\widetilde{D}}

\newcommand{\detgj}[1]{\det{(G^{\{#1\}})}}

\renewcommand\Re{\operatorname{Re}}
\renewcommand\Im{\operatorname{Im}}
\newcommand{\tuz}{\widetilde{u}_0}
\newcommand{\tuzr}{\widetilde{u}_{0 \, R}}
\newcommand{\tuzi}{\widetilde{u}_{0 \, I}}

\newcommand{\bR}{\bar{R}}

\newcommand{\bQ}{\bar{Q}}
\newcommand{\bQp}{\bar{Q}^{\prime}}
\newcommand{\bRt}{\bar{R}^{(2)}}
\newcommand{\bRtp}{\bar{R}^{\prime \, (2)}}
\newcommand{\cG}{\check{G}}
\newcommand{\cV}{\check{V}}
\newcommand{\cC}{\check{C}}

\newcommand{\bB}{\bar{B}}
\newcommand{\tG}{\tilde{G}}
\newcommand{\tV}{\tilde{V}}
\newcommand{\tC}{\tilde{C}}
\newcommand{\hG}{\hat{G}}
\newcommand{\hV}{\hat{V}}
\newcommand{\hC}{\hat{C}}
\newcommand{\bbsj}[2]{\overline{b}_{#1}^{[#2]}}
\newcommand{\bbsjsq}[2]{\overline{b}_{#1}^{[#2] \, 2}}
\newcommand{\detgsj}[1]{\det{(G^{[#1]})}}

\newcommand{\wtI}{\widetilde{I}}
\numberwithin{equation}{section}

\begin{document}

\setlength{\unitlength}{1mm}
\begin{fmffile}{samplepics}

\begin{titlepage}

\vspace{1.cm}

\long\def\symbolfootnote[#1]#2{\begingroup%
\def\thefootnote{\fnsymbol{footnote}}\footnote[#1]{#2}\endgroup} 

\begin{center}

{\Large \bf Two-loop scalar functions with $I$ internal lines, $I \le 5$}\\[2cm]

{\large  J.~Ph.~Guillet$^{a}$, E.~Pilon$^{a}$, 
Y.~Shimizu$^{b}$ and M. S. Zidi$^{c}$ } \\[.5cm]

\normalsize
{$^{a}$ Univ. Grenoble Alpes, Univ. Savoie Mont Blanc, CNRS, LAPTH, 74000 Annecy, France}\\
{$^{b}$ KEK, Oho 1-1, Tsukuba, Ibaraki 305-0801, Japan\symbolfootnote[2]{Y. Shimizu passed away during the completion of this series of articles.}}\\
{$^{c}$ LPTh, Universit\'e de Jijel, B.P. 98 Ouled-Aissa, 18000 Jijel, Alg\'erie}\\
      
\end{center}

\vspace{1cm}

\begin{abstract} 
\noindent
This article displays a proof of concept of the mixed analytical/numerical method, presented in previous publications, to compute two-loop 
functions with up to five massive propagators in a scalar theory having three- and four-leg vertices as the Higgs sector of the Standard Model. 
Several amplitudes are considered with two, three and four external legs. 
Some of them diverge in the UV region and we demonstrate that the method works in that case.
It is shown that all these classes of amplitudes can be generated by four master topologies.
Results of a numerical evaluation for some kinematics are presented, they are compared to a public code and agree well within the error bars quoted by the different programs.
\end{abstract}

\vspace{1cm}

\begin{flushright}
LAPTH-009/21\\
LPTh-Jijel-01/21
\end{flushright}

\vspace{2cm}

\end{titlepage}

\section{Introduction}\label{intro}

The control of quantum corrections in quantum field theory requires high performance tools and techniques to calculate the physical observables as precisely as possible. This is necessary to reduce the unphysical scale dependency from the theoretical predictions and make comparison with the precise data collected at TeV colliders. These data are decisive in inspecting the Standard Model and exploring the physics beyond the Standard Model. To achieve precise predictions, it is required to evaluate scattering amplitudes, which are the quantities connecting theory to experiments, beyond leading and next-to-leading orders. However, the increasing number of produced particles in the final state (processes with many jets), and the number of required loops at the wished order (order beyond NLO) lead to a dramatic increase in the number of contributing Feynman diagrams. These diagrams can have very complicated structure due to multi-loop topologies with massive internal propagators. Therefore, the only way out of this serious challenge is the automatic computation of the scattering amplitudes.
So, it is of great importance to be able to compute scattering amplitudes beyond leading and next-to-leading order accuracy in an efficient, fast and highly automatic way. \\

\noindent
The automation of the next-to-next-to leading order calculation (NNLO) has received a growing attention in the last few years. The major challenge of automated computation of two-loop multi-leg processes, which is one of the fundamental ingredients of NNLO calculation, is the stable and fast numerical calculation of the scalar Feynman integrals, since the full analytic evaluation is not available, so far, in the general multi-leg massive case. There has been an enormous progress in this task but it still extremely challenging, and the available results are derived for very specific processes, see for example refs. \cite{Anastasiou2015,Bonciani2020}. There are many approaches to evaluate two-loop Feynman diagrams, each one has its own advantages and disadvantages. It seems that the numerical approach for multi-loop amplitudes calculation is more practical and useful, since the analytical way, especially the differential equation method, which is most promising in this domain \cite{Kotikov:1991,Gehrmann:2000,Henn:2013}, becomes very complicated and remains always an issue with the growing number of external legs and scales in the general massive case because of the emergence of elliptical integrals. In alternative methods, based on Mellin–Barnes techniques ~\cite{Czakon:2005rk,Gluza:2007rt,Smirnov:2009up,Freitas:2010nx,Gluza:2016fwh} part of the integration is performed analytically and the remaining part is integrated numerically. As far as we know, the number of integrals left depends on the topologies under consideration which might be rather costly. The sector decomposition \cite{Borowka:2015mxa,Borowka:2012yc,Soper:1999xk,Bogner:2007cr,Smirnov:2008py}, which is a fully numerical multidimensional integration based method, gives a reliable results but it has a computing cost which can be high. This method is implemented in the public codes {\tt SecDec} \cite{secdec1,secdec2,secdec3} and {\tt Fiesta} \cite{fiesta}. There are other methods, developed in the last few years, to deal with the numerical multi-loop Feynman amplitudes, we cite for example: the numerical extrapolation approaches \cite{extrapolation1,extrapolation2}, the numerical solution of the differential equation \cite{numdiff1,numdiff2}, the numerical unitarity inspired methods \cite{numunitarity}, etc.  
It would therefore be useful to perform part of the Feynman parameter integrations analytically in a systematic way to reduce the number of numerical quadratures as proposed in ref. \cite{letter} which is the main subject of this paper. \\

\noindent
This paper is a proof of concept of the applicability of the semi-numerical method of the two-loop multi-leg calculation, in the general massive case, presented in \cite{letter}. In this approach, every two-loop scalar Feynman diagram, with $N$ external legs in $n$ dimensions, is expressed as a sum of some two dimensional integrals as follows
\[
^{(2)}I_{N}^{n}
\sim 
\sum \int_{0}^{1} d \rho \int_{0}^{1} d \xi \, W(\rho,\xi) \;
^{(1)}\widetilde{I}_{N^\prime}^{n^\prime}(\rho,\xi)
\]
where $W(\rho,\xi)$ are some weighting functions and the 
$^{(1)}\widetilde{I}_{N^{\prime}}^{n^\prime}(\rho,\xi)$ are``generalised
one-loop type'' $N^{\prime}$-point Feynman-type integrals. The 
``effective number of external legs'' $N^{\prime}$ and the ``effective dimension'' 
$n^{\prime}$ depend on the two-loop topology in particular the number of 
internal lines $I$, and on the dimension $n$. The generalised one-loop function $^{(1)}\widetilde{I}_{N^\prime}^{n^\prime}(\rho,\xi)$ is calculated analytically by the help of the ''Stokes-type`` identity introduced in ref. \cite{paper1}, see also \cite{paper2,paper3}. Once this function is calculated analytically, the two-loop function is obtained by numerical integration only over the sole two remaining variables $\rho$ and $\xi$, which, compared to the direct numerical methods, may represent a significant gain on numerical and computational efficiency. 
In this article, we present several two-loop functions with a number of internal lines less or equal to five ($I\leq 5$) in a scalar theory having three- and four-leg vertices. Such a scalar theory could be, for example, the Higgs sector in the Standard Model. The goal of this paper is not to cover all the possible topologies  and all possible kinematic configurations but rather to discuss specific ones as a proof of concept for the method presented in ref.~\cite{letter}. We construct two codes, one written in {\tt Fortran} and the other written in {\tt Mathematica}, to perform the numerical integration over the two remaining variables. We cross checked our results, obtained by these codes, with {\tt SecDec}. We showed that the results obtained by the three different codes are in a excellent agreement within the error bars. \\

\noindent
 The outline of this article is the following. The section~\ref{sec2} concerns preliminaries.
  We discuss the superficial degree of UV divergences of the two-, three- and four-point two-loop scalar Feynman integrals, in theories involving three- and four-leg vertices. 
  Then, we remind the reader of how the re-parameterisation of the Feynman parameters works. This is discussed in more details in~\cite{letter} leading to an integrand which is a second order polynomial in some new parameters. The coefficients of this polynomial may factorise under certain conditions which are presented here.
  The different topologies are classified in four main categories to which corresponds a formula in terms of weight functions and ``generalised one-loop functions''. 
  In section~\ref{secIeq3}, the first master topology is presented, this is a two-loop two-point function with three internal lines which is globally UV divergent. It is the only one of its category.
  Section~\ref{secIeq4} is dedicated to topologies with four internal lines. There is essentialy one master topology, a two-loop two-point function also globally UV divergent. Another topology, a two-loop three-point function, is a child of the first. Some other topologies are just mentioned because they are trivial extensions of the one presented.
  In section~\ref{secIeq5}, we discussed the topologies  involving five internal lines. There are two master topologies both two-loop two-point functions, one contains a UV divergent sub-graph, the other is UV finite. Four child topologies, two-loop three- and four-point functions, are also evaluated because the kinematics related cannot be guessed from the mother topologies. Here also, other child topologies, corresponding to trivial extensions of the mother topologies, are just quoted.
 In section~\ref{numres}, we give the numerical results for the nine topologies, presented in the previous sections, for five different kinematic configurations. 
 To facilitate the reading of this paper, we have postponed details of the calculations to many appendices. 
 The appendix~\ref{det_cal_t22021} contains the details of the computation of the master topology of section~\ref{secIeq3}.
 In appendices~\ref{intsecpar} and \ref{intspe}, we give the results, and how to derive them,  for many integrals appearing in globally UV divergent topologies. 
 The appendix~\ref{ol2Oe} shows how to calculate the one-loop two-point functions at order $\mathcal{O}(\varepsilon^0)$ and $\mathcal{O}(\varepsilon^1)$ which appear in the $\varepsilon$ expansion of the UV divergent topologies of section~\ref{secIeq4}. 
 In appendix~\ref{ol3ptsq}, we present the evaluation of the one-loop three-point function whose integration domain is the square and not the usual simplex as in the case of the ordinary one-loop functions. 
 The appendices~\ref{new_integral} and \ref{cut_log} are dedicated to the computation of the one-loop three-point function at order $\mathcal{O}(\varepsilon)$ appearing in some topologies of section~\ref{secIeq5}. 
 In appendix~\ref{fcheck}, we provide further checks of formulae found for the different scalar two-loop amplitudes.

\section{Preliminaries}\label{sec2}

Let us discuss the different topologies presented in this article. The scalar topologies are labelled as $\calt_{2n\_i\_j\_k}$ where $n$ is the number of external legs, $i$ is the number of vertices with three-legs, $j$ is the number of vertices with four-legs. The knowledge of these numbers does not fix uniquely the diagram, thus another integer $k$ is introduced labelling the different two-loop diagrams having the same number of external legs, three-leg vertices and four-leg vertices. We also provide in the table~\ref{tab_corr}, the correspondence between the labelling used in this article and the Nickel index following ref.~\cite{Bogner:2017xhp}.

\begin{table}[h!]
\begin{minipage}[t]{.4\linewidth}
  \hspace{1.5cm}\begin{tabular}{|l|l|}
\hline
$\calt_{2n\_i\_j\_k}$ Notation & Nickel Index \\
\hline
\multicolumn{2}{|c|}{Class of topologies 1 ($I=3$)}\\
\hline
$\calt_{22\_0\_2\_1}$ & $e111|e|$\\
\hline
\multicolumn{2}{|c|}{Class of topologies 2 ($I=4$)}\\
\hline
$\calt_{22\_2\_1\_1}$ & $e112|2|e|$\\
\hline
$\calt_{23\_1\_2\_1}$ & $e112|e2|e|$\\
\hline
$\calt_{23\_1\_2\_2}$ & $ee12|e22||$\\
\hline
$\calt_{24\_0\_3\_1}$& $ee12|e22|e|$\\
\hline
\multicolumn{2}{|c|}{Class of topologies 3 ($I=5$)}\\
\hline
$\calt_{22\_4\_0\_1}$&$e12|e3|33||$ \\
\hline
$\calt_{23\_3\_1\_1}$&$e112|3|e3|e|$ \\
\hline
$\calt_{24\_2\_2\_1}$&$e112|e3|e3|e|$ \\
\hline
$\calt_{23\_3\_1\_3}$&$ee12|e3|33||$\\
\hline
$\calt_{24\_2\_2\_3}$ &$ee12|ee3|33||$\\
\hline
$\calt_{24\_2\_2\_4}$&$e112|3|e3|ee|$\\
\hline
 \end{tabular}
\end{minipage}%
\hfill
\begin{minipage}[t]{.4\linewidth}
\hspace{-1.5cm}\begin{tabular}{|l|l|}
\hline
$\calt_{2n\_i\_j\_k}$ Notation & Nickel Index \\
\hline
$\calt_{24\_2\_2\_5}$&$e112|3|ee3|e|$\\
\hline
$\calt_{25\_1\_3\_1}$&$e112|3|ee3|ee|$\\
\hline
$\calt_{25\_1\_3\_2}$&$e12|ee3|e33|e|$\\
\hline
$\calt_{26\_0\_4\_1}$&$ee12|ee3|e33|e|$ \\
\hline
\multicolumn{2}{|c|}{Class of topologies 4 ($I=5$)}\\
\hline
$\calt_{22\_4\_0\_2}$&$e12|23|3|e|$ \\
\hline
$\calt_{23\_3\_1\_2}$&$e12|e23|3|e|$ \\
\hline
$\calt_{24\_2\_2\_2}$&$e12|e23|e3|e|$ \\
\hline
$\calt_{23\_3\_1\_4}$& $ee12|23|3|e|$\\
\hline
$\calt_{24\_2\_2\_6}$& $ee12|23|3|ee|$\\
\hline
$\calt_{24\_2\_2\_7}$& $e123|e2|3|ee|$\\
\hline
$\calt_{25\_1\_3\_3}$& $e123|ee2|3|ee|$\\
\hline
$\calt_{25\_1\_3\_4}$&$e12|e23|e3|ee|$\\
\hline
$\calt_{26\_0\_4\_2}$& $ee12|e23|e3|ee|$\\
\hline
 \end{tabular}
\end{minipage}
\caption{\small The correspondence between the labelling of topologies in this article and the Nickel index. The topologies are gathered by classes.}
\label{tab_corr}
\end{table}

\subsection{Superficial degree of UV divergence}\label{gen_topo}
Let us discuss in a general way the superficial degree of divergence. For that, let us note $g_3$, resp. $g_4$ the coupling constant for the three-leg (resp.~four-leg) vertices, the dimension (in terms of energy) of these two coupling constants is: $[g_3]=1$ and $[g_4]=0$.
The superficial degree of divergence of a Feynman diagram $G$ in this scalar theory, in a four dimensional space-time, is
\begin{equation}
  w(G) = 4 - N - \sum_{i=3,4} n_i \, [g_i] = 4 -N - n_3
  \label{equvscal1}
\end{equation}
where $N$ is the number of external bosons of the diagram $G$, $n_i$ is the number of vertices having the coupling constant $g_i$.

\subsubsection{Two-point functions}

For a two-point function, the superficial degree of divergence becomes
\begin{equation}
  w(G) = 2 - n_3
  \label{equvscal2}
\end{equation}
If $n_3 > 2$ then $w(G) < 0$ and the Feynman integral associated to the diagram $G$ is convergent in the UV region\footnote{It may contain some divergent subgraphs.}. So a UV divergent two-loop two-point function, in this theory, has 2, 1 or zero three-leg vertex. But once the number of three-leg vertices is determined, since the number of loop  and the number of external legs are fixed, the number of four-leg vertices is also known. Indeed, let us note $I$ the number of internal lines of the diagram $G$ and $L$ the number of internal loops, there are two equations which give some constraints on the different number of vertices
\begin{equation}
  \left\{ \begin{aligned}
  N + 2 \, I &= 3 \, n_3 + 4 \, n_4 \\
  L &= I - (n_3 + n_4 -1)
  \end{aligned} \right.
 \label{equscal3}
\end{equation}
In our case, taking $n_3=0$, eqs.~(\ref{equscal3}) reduces to:
\begin{equation}
  \left\{ \begin{aligned}
  I &= 2 \, n_4 -1 \\
  1 &= I - n_4
  \end{aligned} \right.
  \label{equscal4}
\end{equation}
which implies that $n_4 = 2$ and $I=3$. For $n_3=1$, the system of equations (\ref{equscal3}) has no solution in $\mathbb{N}$. For $n_3 = 2$, the system of equations (\ref{equscal3}) becomes
\begin{equation}
  \left\{ \begin{aligned}
  I &= 2 + 2\, n_4 \\
  3 &= I - n_4
  \end{aligned} \right.
  \label{equvscal5}
\end{equation}
leading to the following solution: $n_4=1$ and $I = 4$.
Therefore, the two-loop two-point functions which diverge globally in the UV domain have either zero three-leg vertex and two four-leg vertices with three internal lines or two three-leg vertices and one four-leg vertex with four internal lines.

\subsubsection{Three-point functions}

For a three-point function, the superficial degree of divergence becomes:
\begin{equation}
  w(G) = 1 - n_3
  \label{equvscal6}
\end{equation}
If $n_3 > 1$ then $w(G) < 0$ and the Feynman integral associated to the diagram $G$ is globally convergent in the UV region. Taking $n_3=1$, eqs. (\ref{equscal3}) reduce to
\begin{equation}
  \left\{ \begin{aligned}
  I &= 2 \, n_4 \\
  2 &= I - n_4
  \end{aligned} \right.
  \label{equscal7}
\end{equation}
which implies that $n_4 = 2$ and $I=4$. Note that for $n_3=0$, the system of equations (\ref{equscal3}) has no solution in $\mathbb{N}$. Therefore, the two-loop three-point functions which globally diverge in the UV domain have one three-leg vertex and two four-leg vertices with four internal propagators.

\subsubsection{Four-point functions}

The same study can be followed for a two-loop four-point function. In this case, the superficial degree of divergence of a two-loop four-point Feynman diagram $G$ is given by
\begin{equation}
  w(G) = - n_3
  \label{equscal8}
\end{equation}
The only possibility to have a global UV divergence is to set $n_3=0$. Let us determine $I$ and $n_4$ with the help of eqs.~(\ref{equscal3}), we have to solve
\begin{equation}
  \left\{ \begin{aligned}
  2 + I &= 2 \, n_4 \\
  1 &= I - n_4
  \end{aligned} \right.
  \label{equscal9}
\end{equation}
which implies that $I=4$ and $n_4=3$.\\

\subsection{Skeleton for the re-parametrisation of the Feynman parameters}\label{secskel}

The goal of this section is to give details on the way the Feynman parameters are re-parameterised for a two-loop scalar function. All the common formulae will be collected here, the specific ones will be discussed for each topologies. \\

\noindent
We mainly follow the discussion of ref.~\cite{letter}. Namely, the $n$-momentum $q_i$ carried by each propagator $i$ of a two-loop scalar function is parameterised in the following way $q_i = \hk_i + \hr_i$, where the $n$-momenta $\hr_i$ depend on the external $n$-momenta $p_i$ whereas the $\hk_i$ are some linear combinations of the loop $n$-momenta $k_1$ and $k_2$ which are specific for each topology\footnote{For the sake of simplicity, we use the dimensional regularisation scheme.}. 
Once this linear combination is fixed, the energy-momentum conservation at each vertex determines the $\hr_i$ except two, let us note them $\br_1$ and $\br_2$\footnote{This choice is also specific to each topology.}. 
Each $n$-vector $\hr_i$ can be decomposed as
\begin{equation}
  \hr_i = \alpha_i \, \br_1 + \beta_i \, \br_2 + Q_i(\{p_k\})
  \label{eqdecomp0}
\end{equation}
where $\alpha_i, \, \beta_i = 0, \, 1$ and $Q_i(\{p_k\})$ is a linear combination of the external $n$-momenta $p_k$. 
We further introduce a $2 \times 2$ matrix $A$, two other $n$-momenta $r_1$ and $r_2$ and a quantity $\calc$ such that the combination $\sum_{j=1}^{I} \, \tau_j \, (q_j^2 - m_j^2)$ reads
\begin{align}
  \sum_{j=1}^{I} \, \tau_j \, (q_j^2 - m_j^2) &= [k_1 \quad k_2] \cdot A \cdot \left[
  \begin{array}{c}
    k_1 \\
    k_2
  \end{array}
\right] + 2 \, [r_1 \quad r_2] \cdot \left[
\begin{array}{c}
  k_1 \\
  k_2
\end{array}
\right] + \calc
  \label{eqdefAr1r2}
\end{align}
The symbol $\tau_j$, resp.~$m_j$, is the Feynman parameter, resp.~the mass, associated to the internal line having a $n$-momenta $q_j$.
According to its definition, cf.~eq.~(\ref{eqdefAr1r2}), the matrix $A$ is equal to
  \begin{align}
    A &= \left[
    \begin{array}{cc}
      \sum_{i \in S_1} \tau_i + \sum_{i \in S_3} \tau_i & \sum_{i \in S_3} \tau_i \\
      \sum_{i \in S_3} \tau_i & \sum_{i \in S_2} \tau_i + \sum_{i \in S_3} \tau_i
    \end{array}
  \right]
    \label{eqdefA0}
  \end{align}
where the three sets $S_1$, $S_2$ and $S_3$ are defined as: $S_1$ is the set of propagator labels whose momentum involves only $k_1$ not $k_2$, $S_2$ is the set of propagator labels whose momentum involves only $k_2$ not $k_1$ and $S_3$ is the set of propagator labels whose momentum involves both $k_1$ and $k_2$\footnote{We choose the orientation of the $n$-momentum flow of the internal lines such that the linear combination of the loop momenta is always $k_1 + k_2$.}. Let us choose $N-1$ independent linear combinations of the external $n$-momenta\footnote{We ignore the fact that for $N \ge n+2$ there are further constraints which apply to the external $n$-momenta in addition to the global energy-momentum conservation, cf.~for example ref.~\cite{Binoth:2005ff}.} $t_i$ $i=1 \cdots N-1$. This choice is of course specific of the topology too. The $n$-momenta $r_1$ and $r_2$ can be expressed in terms of $\br_1$, $\br_2$ and the $n$-momenta $t_i$ as
\begin{align}
  \left[
  \begin{array}{c}
    r_1 \\
    r_2
  \end{array}
\right] &= A \cdot \left[
\begin{array}{c}
  \br_1 \\
  \br_2
\end{array}
\right] + \bB \cdot T
  \label{eqdefBbar}
\end{align}
where
\begin{align}
  T &= \left[
  \begin{array}{c}
    t_1 \\
    \vdots \\
    t_{N-1}
  \end{array}
\right]
  \label{eqdefT}
\end{align}
and $\bB$ is a $2 \times N-1$ matrix depending on the Feynman parameters $\tau_i$. 
Furthermore, the term $\calc$ which is equal to
\begin{equation}
  \calc = \sum_{j=1}^{I} \tau_j \, (\hr_j^2 - m_j^2)
  \label{eqdefcalc0}
\end{equation}
can be written as
\begin{align}
  \calc &= [\br_1 \quad \br_2] \cdot A \cdot \left[
  \begin{array}{c}
    \br_1 \\
    \br_2
  \end{array}
\right] + 2 \, [\br_1 \quad \br_2] \cdot \bB \cdot T + T^T \cdot \Gamma \cdot T - \sum_{j=1}^{I} \, \tau_j \, m_j^2
  \label{eqdefGamma}
\end{align}
where $\Gamma$ is a $N-1 \times N-1$ matrix depending only on the Feynman parameters $\tau_i$. 
Once the integration over the loop $n$-momenta $k_1$ and $k_2$ is performed, the two-loop amplitude is a $I$-dimensional integral over the Feynman parameters $\tau_k$, constrained by $\sum_{k=1}^{I} \tau_k = 1$, of an integrand proportional to $\det(A)^{I-3/2 \, n} \, [ \calf(\{\tau_k\}) - i \, \lambda]^{n-I}$ cf.~\cite{letter}.
This function $\calf(\{\tau_k\})$ is given by
\begin{align}
  \calf(\{\tau_k\}) &= [r_1 \quad r_2] \cdot \text{Cof}[A] \cdot \left[
  \begin{array}{c}
    r_1 \\
    r_2
  \end{array}
\right]
  - \det(A) \, \calc
  \label{eqintegrand0}
\end{align}
where $\text{Cof}[A]$ is the matrix of the cofactors of $A$. One can trade the $n$-momenta $r_1$ and $r_2$ in eq.~(\ref{eqintegrand0}) against $\br_1$ and $\br_2$ using eq.~(\ref{eqdefBbar}). 
But this new dependence in the $n$-momenta $\br_1$ and $\br_2$ cancels against the one coming from the term $\calc$ (cf.~eq.~(\ref{eqdefGamma})) in the polynomial $\calf(\{\tau_k\})$ and the latter becomes
\begin{align}
  \calf(\{\tau_k\}) &= \left( \bB \cdot T \right)^T \cdot \text{Cof}[A] \cdot \left( \bB \cdot T \right) - \det(A) \, \left[ T^T \cdot \Gamma \cdot T - \sum_{j=1}^{I} \tau_j \, m_j^2 \right]
  \label{eqdefpolycalf}
\end{align}
Thus the knowledge of the matrices $\bB$ and $\Gamma$ which are specific to each topology enables to determine $\calf(\{\tau_k\})$.
Note that although the polynomial $\calf(\{\tau_k\})$ is unique for a given Feynman diagram, the matrices $\bB$ and $\Gamma$ are not because parts of $\bB$ can be reabsorbed into $\Gamma$, cf.~the discussion in the Appendix A of ref.~\cite{letter}. An example will be given in the next section. 
Finally, the amplitude related to a scalar two-loop diagram with $N$ external legs is given by
\begin{align}
  ^{(2)}I_{N}^{n}\left( \kappa;\calt \right) &= (-1)^{I+1} \, (4 \, \pi)^{-n} \, \Gamma(I - n) \int_{(R^+)^I} \prod_{j=1}^{I} d \tau_i  \, \delta(1 - \sum_{j=1}^{I} \tau_i) \, \left( \det(A) \right)^{I - \frac{3}{2} \, n} \notag \\
  &\quad {} \times \left( \calf(\{\tau_k\}) - i \, \lambda \right)^{n - I }
  \label{eqdefT22n}
\end{align}
The symbol $\kappa$ stands for the kinematics, it is a set containing the independent invariants and internal masses squared while the symbol $\calt$ indicates the topology under consideration.\\

\noindent
The next step is to introduce three auxiliary parameters $\rho_i$ for $i=1,2,3$ defined by
\begin{align}
  \rho_i &= \sum_{k \in S_i} \tau_k 
  \label{eqdefrhoi}
\end{align}
Because of its definition in eq.~(\ref{eqdefA0}), the matrix $A$ is given in terms of the parameters $\rho_i$ by
\begin{align}
  A &= \left(
  \begin{array}{c c}
    \rho_1 + \rho_3 & \rho_3 \\
    \rho_3 & \rho_2 + \rho_3
  \end{array}
\right)
  \label{eqelemmatA}
\end{align}
and thus
\begin{align}
  \det(A) &= \rho_1 \, \rho_2 + \rho_2 \, \rho_3 + \rho_3 \, \rho_1
  \label{eqdefdetA}
\end{align}
Note that, due to their definitions, the $\rho_i$ parameters sum to $1$.
Then the Feynman parameters $\tau_k$ whose label belong to the set $S_j$ are re-parameterised such that $\tau_k = \rho_j \, u_k$. Since the $\sum_{k \in S_j} \tau_k = \rho_j$ the parameters $u_k$ sum to $1$. Eq.~(\ref{eqdefT22n}) becomes then 
\begin{align}
^{(2)}I_{N}^{n} \left( \kappa ; \calt \right)
&= 
(-1)^{I+1} \, (4 \, \pi)^{-n} \, \Gamma(I - n) \, \int_{(I\!\!R^{+})^{3}} 
\left[ \prod_{k=1}^{3} d \rho_{k} \, \rho_k^{|S_k|-1} \right] 
\delta \left( 1 - \sum_{l=1}^{3} \rho_{\, l} \right) \, 
\det(A)^{I - \, \frac{3 \, n}{2}} \notag \\
&\quad {} \times \int_{(I\!\!R^{+})^{I}} 
\prod_{k=1}^{3} \left[ 
\prod_{j \in S_k} d u_{j} \, \delta \left( 1 - \sum_{l \in S_k} u_{j} \right) \right] \,
\left[ \,
 \overline{\cal F}(\{u_{k}\},\{\rho_l\}) - i \, \lambda 
\right]^{n-I}
\label{eqG1a}
\end{align}
with
$\overline{\cal F}(\{u_{k}\},\{\rho_l\})
= 
{\cal F}(\{\tau_{i}(\{u_{k}\},\{\rho_l\})\})
$.
In eq.~(\ref{eqG1a}), $|S_k|$ denotes the cardinal of the set $S_k$.
Lastly, to get rid of the constraint on the sum of the parameter $\rho_j$, the following change of variable is performed $\rho_{j_1} = \rho \, \xi$, $\rho_{j_2} = \rho \, (1-\xi)$ where $j_1$ and $j_2$ are two distinct elements of the set $\{1,2,3\}$, this yields
\begin{align}
^{(2)}I_{N}^{n} \left( \kappa ; \calt \right)
&= 
\int_0^1 d \rho \, \int_0^1 d \xi \, W(\rho,\xi) \; 
^{(1)}\widetilde{I}_{N^{\prime}}^{n^{\prime}}\left( E;\tG,\tV,\tC,\rho,\xi \right)
\label{eqG1ap}
\end{align}
where $^{(1)}\widetilde{I}_{N^{\prime}}^{n^{\prime}}$ is the ``generalised'' one-loop $N^\prime$-point function given by 
\begin{align}
^{(1)}\widetilde{I}_{N^{\prime}}^{n^{\prime}}\left( E;\tG,\tV,\tC,\rho,\xi \right)
& =
\int_{(I\!\!R^{+})^{I}} 
\prod_{k=1}^{3} \left[ 
\prod_{j \in S_k} d u_{j} \, \delta \left( 1 - \sum_{l \in S_k} u_{j} \right) \right] \,
\left[
  \calft(\{u_{k}\},\rho,\xi) - i \, \lambda 
\right]^{\frac{n^{\prime}}{2}-N^{\prime}}
\label{eqG1b}
\end{align}
with $N^{\prime} = I - 2$ and $ n^{\prime} = 2 \, (n - 2)$. 
The weight function $W(\rho,\xi)$, in eq.~(\ref{eqG1ap}), has the following expression 
\begin{align}
  W(\rho,\xi) &= \rho^{1-\frac{n}{2}} \, \left( \rho \, \xi \right)^{|S_{j_1}|-1} \, \left( \rho \, (1 - \xi) \right)^{|S_{j_2}|-1} \, \left( 1 - \rho \right)^{|S_{j_3}|-1} \, \left( 1 - \rho + \rho \, \xi \, (1-\xi) \right)^{I - \frac{3 \, n}{2}}
  \label{eqweightfunc1}
\end{align}
with $j_3 = \{1,2,3\} \setminus \{j_1,j_2\}$.
In eq.~(\ref{eqG1b}), the quantity $\calft(\{u_{k}\},\rho,\xi)$ which is a second order polynomial in the $u_i$, is given by\footnote{Whatever the choice of $j_1$ and $j_2$, it is always possible to factorise $\rho$ from $\overline{\cal F}(\{u_{k}\},\{\rho_l(\rho,\xi)\})$.}
\begin{align}
  \calft(\{u_k\},\rho,\xi) &\equiv \frac{\overline{\cal F}(\{u_{k}\},\{\rho_l(\rho,\xi)\})}{\rho} \notag \\
  &= U^T \cdot \tG \cdot U - 2 \, \tV^T \cdot U - \tC
  \label{eqreparamcalf}
\end{align}
where $U$ is a $I-3$ vector gathering the leftover parameters $u_i$ after one got rid of the constraints $\sum_{k \in S_j} u_k = 1$.
In eqs.~(\ref{eqG1ap}) and (\ref{eqG1b}), the argument $E$ denotes the integration domain over the $u_i$ once these latter constraints have been taken into account, this domain is not always a simplex as in the case of the genuine one-loop functions. In the cases treated in this article it can be the 1-simplex $\Sigma_{(1)} = \{0 \leq u \leq 1\}$, the 2-simplex $\Sigma_{(2)} = \{ 0 \leq u_1,u_2 \leq 1, u_1+u_2 \leq 1\}$ or the square $K_{(2)} = \{ 0 \leq u_1,u_2 \leq 1\}$. 
In eq.~(\ref{eqreparamcalf}), $\tG$ is a $(I-3) \times (I-3)$ matrix, $\tV$ is a $I-3$ vector and $\tC$ is a scalar. 
They all depend on the parameters $\rho$ and $\xi$ and on the kinematics.
Note that, since the dependence on the variables $\rho$ and $\xi$ is only implicit, in the cases where these variables are fixed, we drop them from the list of the arguments of the generalised one-loop functions for the sake of simplicity.
Note also that some quantities introduced in this section depend implicitly on the topology. Strictly speaking, they have to be labelled differently for each topology. To make the notations lighter, we will ignore this except for the matrix $\tG$, the vector $\tV$ and the scalar $\tC$ to which we will associate a different letter for each topology.\\

\noindent
The ``generalised one-loop functions'' in space-time of dimension $n^{\prime} = 4 - 4 \, \varepsilon$ will be expanded around $\varepsilon=0$ as
\begin{align}
  ^{(1)}\widetilde{I}_{N^{\prime}}^{n^{\prime}}(E;G,V,C,\rho,\xi) &= \sum_{k=0}^{\infty} \, \frac{(-2 \, \varepsilon)^{k}}{k!} \, \wtI_{N^{\prime}}^{\,(k)}(E;G,V,C,\rho,\xi)
  \label{eqdeftI1}
\end{align}
where the quantities $\wtI_{N^{\prime}}^{\,(k)}(E;G,V,C,\rho,\xi)$ have the general expression
\begin{align}
  \wtI_{N^{\prime}}^{\,(k)}(E;G,V,C,\rho,\xi) &= \int_{(I\!\!R^{+})^{I}} 
\prod_{k=1}^{3} \left[
\prod_{j \in S_k} d u_{j} \, \delta \left( 1 - \sum_{l \in S_k} u_{j} \right) \right] \, \left[ \calft(\{u_k\},\rho,\xi) - i \, \lambda \right]^{2-N^{\prime}} \notag \\
&\qquad \qquad {} \times \ln^k\left( \calft(\{u_k\},\rho,\xi) - i \, \lambda \right)
  \label{eqformgeneItildek}
\end{align}
\\

\noindent
Since we limit ourselves to topologies with a number of internal lines $I \leq 5$, $N^{\prime} \leq 3$.
For $N^{\prime} = 2$, the coefficients of the expansion of the right-hand side of eq.~(\ref{eqdeftI1}) is given by
\begin{align}
  \wtI_{2}^{\,(k)}(E;G,V,C,\rho,\xi) &= \int_{E} d w \, \ln^k\left( G \, w^2 - 2 \, V \, w - C - i \, \lambda \right)
  \label{eqdeftI2}
\end{align}
while for $N^{\prime}=3$, they are given by
\begin{align}
  \wtI_{3}^{\,(k)}(E;G,V,C,\rho,\xi) &= \int_{E} d w_1 \, d w_2 \, \frac{\ln^k\left( W^{T} \cdot G \cdot W - 2 \, V^{T} \cdot W - C - i \, \lambda \right)}{W^{T} \cdot G \cdot W - 2 \, V^{T} \cdot W - C - i \, \lambda}
  \label{eqdeftI3}
\end{align}
with
\[
  W = \left[
  \begin{array}{c}
    w_1 \\
    w_2
  \end{array}
\right]
\]
\\

\noindent
In eq.~(\ref{eqG1b}), the way to choose the parameters $u_i$ is guided by the fact that in some case $\det(A)$ can be factorised from the vector $\tV$ and the scalar $\tC$. In ref.~\cite{letter}, we gave a sufficient condition for this property to hold. If we achieved to build a $\bB$ matrix homogeneous of degree $1$ in the parameter $u_i$\footnote{To be more precise, it is all the non zero entries of the matrix $\bB$ which have to be homogeneous of degree $1$ in the $u_i$ parameters.} then, the first term of eq.~(\ref{eqdefpolycalf}) which is not proportional to $\det(A)$ will contribute only to $U^T \cdot \tG \cdot U$, and thus the factorisation of $\det(A)$ will hold for the vector $\tV$ and the scalar $\tC$.\\

\noindent
Let us try to draw the condition to build a $\bB$ matrix homogeneous in the $u_i$ parameters. 
For that, let us discuss two-loop planar diagrams in a scalar theory having three- and four-leg vertices. These vertices are either external vertices where one or two external legs can be branched, they are $N$ of them and are represented by blobs; or internal vertices which connect only internal lines.
The topologies of these diagrams fall into three categories depicted on figs.~\ref{fig01} and \ref{fig02}.\\

\begin{figure}[h]
\centering
\parbox[c][63mm][t]{63mm}{\begin{fmfgraph*}(80,80)
  \fmfstraight
  \fmfsurroundn{e}{20}
  \fmfset{arrow_len}{3mm}
  \fmf{phantom,tension=16.0}{e8,vv1}
  \fmf{phantom,tension=16.0}{e20,vv3}
  \fmf{phantom,tension=7.0}{vv1,vv2}
  \fmf{phantom,tension=7.0}{vv3,vv4}
  \fmfdot{vv2,vv4}
  \fmf{phantom,right}{vv1,vv3}
  \fmf{phantom,right}{vv3,vv1}
  \fmf{dots,right}{vv2,vv4}
  \fmf{dots,right}{vv4,vv2}
  \fmfposition
  \fmffreeze
  \fmfipath{p[]}
  \fmfiset{p1}{vpath (__vv2,__vv4)}
  \fmfiset{p2}{vpath (__vv4,__vv2)}
  \fmfipath{pp[]}
  \fmfiset{pp1}{vpath (__vv1,__vv3)}
  \fmfiset{pp2}{vpath (__vv3,__vv1)}
  \fmfipair{a[]}
  \fmfiequ{a2}{(.50w,.75h)}
  \fmfiequ{a3}{(.604w,.47h)}
  \fmfi{phantom,tension=1.0}{point length(pp1)/6 of pp1 -- point length(p1)/6 of p1}
  \fmfi{phantom,tension=1.0}{point 4length(pp1)/6 of pp1 -- point 4length(p1)/6 of p1}
  \fmfiv{decor.shape=circle,decor.filled=full,decor.size=2thick}{point 4length(p1)/6 of p1}
  \fmfi{fermion,label={\small $q_{i}$},label.side=left}{subpath (4length(p1)/6,5length(p1)/6) of p1}
  \fmfi{fermion}{subpath (6length(p1)/6,5length(p1)/6) of p1}
  \fmfiv{label={\small $q_{i+1}$},label.angle=-180}{point 6length(p1)/6 of p1}
  \fmfi{phantom,tension=2.0}{point 3length(pp2)/6 of pp2 -- point 3length(p2)/6 of p2}
  \fmfi{phantom,tension=2.0}{point 4length(pp2)/6 of pp2 -- point 4length(p2)/6 of p2}
  \fmfiv{decor.shape=circle,decor.filled=full,decor.size=2thick}{point 4length(p2)/6 of p2}
  \fmfi{fermion,label={\small $q_{N+2}$},label.side=right}{subpath (5length(p2)/6,4length(p2)/6) of p2}
  \fmfiset{p4}{point 5length(pp2)/6 of pp2 -- point 5length(p2)/6 of p2}
  \fmfi{fermion,label={\small $q_{1}$},label.side=left}{subpath (5length(p2)/6,6length(p2)/6) of p2}
  \fmfiset{p3}{point 5length(p1)/6 of p1 .. a3 .. a2 .. point length(p4) of p4}
  \fmfi{fermion,rubout,label={\small $q_{N+3}$}}{subpath (0,3length(p3)/4) of p3}
  \fmfi{plain,rubout}{subpath (3length(p3)/4,length(p3)) of p3}
  \fmfv{label={\small Class of topologies $\calt_{1}$},label.dist=0.20w,label.angle=-115}{vv4}
\end{fmfgraph*}}
\qquad
\parbox[c][63mm][t]{63mm}{\begin{fmfgraph*}(80,80)
  \fmfstraight
  \fmfsurroundn{e}{20}
  \fmfset{arrow_len}{3mm}
  \fmf{phantom,tension=16.0}{e8,vv1}
  \fmf{phantom,tension=16.0}{e20,vv3}
  \fmf{phantom,tension=7.0}{vv1,vv2}
  \fmf{phantom,tension=7.0}{vv3,vv4}
  \fmfdot{vv2,vv4}
  \fmf{phantom,right}{vv1,vv3}
  \fmf{phantom,right}{vv3,vv1}
  \fmf{dots,right}{vv2,vv4}
  \fmf{dots,right}{vv4,vv2}
  \fmfposition
  \fmffreeze
  \fmfipath{p[]}
  \fmfiset{p1}{vpath (__vv2,__vv4)}
  \fmfiset{p2}{vpath (__vv4,__vv2)}
  \fmfipath{pp[]}
  \fmfiset{pp1}{vpath (__vv1,__vv3)}
  \fmfiset{pp2}{vpath (__vv3,__vv1)}
  \fmfipair{a[]}
  \fmfiequ{a3}{(.494w,.69h)}
  \fmfi{phantom,tension=1.0}{point length(pp1)/6 of pp1 -- point length(p1)/6 of p1}
  \fmfi{fermion,label={\small $q_{i}$},label.side=left}{subpath (4length(p1)/6,5length(p1)/6) of p1}
  \fmfiv{decor.shape=circle,decor.filled=full,decor.size=2thick}{point 4length(p1)/6 of p1}
  \fmfi{fermion,label={\small $q_{i+1}$},label.side=right}{subpath (6length(p1)/6,5length(p1)/6) of p1}
  \fmfi{phantom,tension=2.0}{point 4length(pp2)/6 of pp2 -- point 4length(p2)/6 of p2}
  \fmfi{fermion,label={\small $q_{N+1}$},label.side=right}{subpath (5length(p2)/6,4length(p2)/6) of p2}
  \fmfiv{decor.shape=circle,decor.filled=full,decor.size=2thick}{point 4length(p2)/6 of p2}
  \fmfiset{p4}{point 5length(pp2)/6 of pp2 -- point 5length(p2)/6 of p2}
  \fmfi{phantom,tension=1.0}{subpath (0,length(p4)) of p4}
  \fmfiv{decor.shape=circle,decor.filled=full,decor.size=2thick}{point length(p4) of p4}
  \fmfi{fermion,label={\small $q_{1}$},label.side=left}{subpath (5length(p2)/6,6length(p2)/6) of p2}
  \fmfiset{p5}{point 5length(pp1)/6 of pp1 -- point 5length(p1)/6 of p1}
  \fmfi{phantom,tension=1.0}{subpath (0,length(p5)) of p5}
  \fmfiset{p3}{point length(p5) of p5 .. a3 .. point length(p4) of p4}
  \fmfi{fermion,rubout,label={\small $q_{N+2}$},label.side=right}{subpath (0,length(p3)) of p3}
  \fmfv{label={\small Class of topologies $\calt_{2}$},label.dist=0.20w,label.angle=-115}{vv4}
\end{fmfgraph*}}
\caption{\small Diagrams picturing two classes of planar topologies with $N$ external vertices and one or two internal vertices.}
\label{fig01} 
\end{figure}
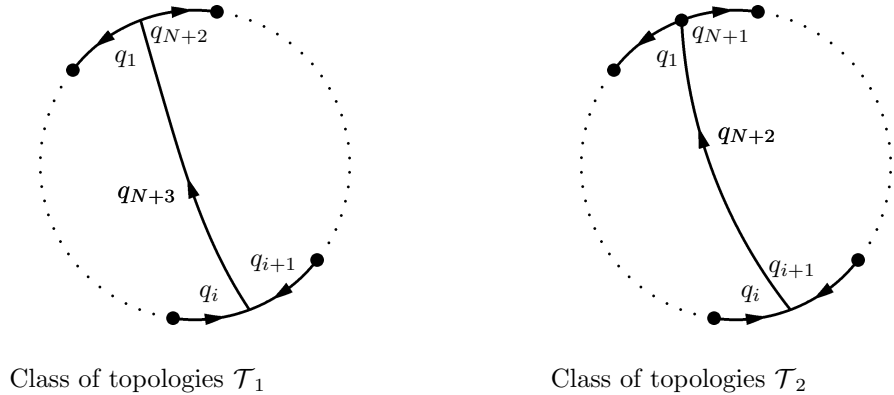

\begin{figure}[h]
\centering
\parbox[c][63mm][t]{63mm}{\begin{fmfgraph*}(80,80)
  \fmfstraight
  \fmfsurroundn{e}{20}
  \fmfset{arrow_len}{3mm}
  \fmf{phantom,tension=16.0}{e8,vv1}
  \fmf{phantom,tension=16.0}{e20,vv3}
  \fmf{phantom,tension=7.0}{vv1,vv2}
  \fmf{phantom,tension=7.0}{vv3,vv4}
  \fmfdot{vv2,vv4}
  \fmf{phantom,right}{vv1,vv3}
  \fmf{phantom,right}{vv3,vv1}
  \fmf{dots,right}{vv2,vv4}
  \fmf{dots,right}{vv4,vv2}
  \fmfposition
  \fmffreeze
  \fmfipath{p[]}
  \fmfiset{p1}{vpath (__vv2,__vv4)}
  \fmfiset{p2}{vpath (__vv4,__vv2)}
  \fmfipath{pp[]}
  \fmfiset{pp1}{vpath (__vv1,__vv3)}
  \fmfiset{pp2}{vpath (__vv3,__vv1)}
  \fmfipair{a[]}
  \fmfiequ{a3}{(.494w,.69h)}
  \fmfi{phantom,tension=1.0}{point length(pp1)/6 of pp1 -- point length(p1)/6 of p1}
  \fmfi{fermion,label={\small $q_{i}$},label.side=left}{subpath (4length(p1)/6,5length(p1)/6) of p1}
  \fmfiv{decor.shape=circle,decor.filled=full,decor.size=2thick}{point 4length(p1)/6 of p1}
  \fmfi{fermion,label={\small $q_{i+1}$},label.side=right}{subpath (6length(p1)/6,5length(p1)/6) of p1}
  \fmfi{phantom,tension=2.0}{point 4length(pp2)/6 of pp2 -- point 4length(p2)/6 of p2}
  \fmfi{fermion,label={\small $q_{N}$},label.side=right}{subpath (5length(p2)/6,4length(p2)/6) of p2}
  \fmfiv{decor.shape=circle,decor.filled=full,decor.size=2thick}{point 4length(p2)/6 of p2}
  \fmfiset{p4}{point 5length(pp2)/6 of pp2 -- point 5length(p2)/6 of p2}
  \fmfi{phantom,tension=1.0}{subpath (0,length(p4)) of p4}
  \fmfiv{decor.shape=circle,decor.filled=full,decor.size=2thick}{point length(p4) of p4}
  \fmfi{fermion,label={\small $q_{1}$},label.side=left}{subpath (5length(p2)/6,6length(p2)/6) of p2}
  \fmfiset{p5}{point 5length(pp1)/6 of pp1 -- point 5length(p1)/6 of p1}
  \fmfi{phantom,tension=1.0}{subpath (0,length(p5)) of p5}
  \fmfiv{decor.shape=circle,decor.filled=full,decor.size=2thick}{point length(p5) of p5}
  \fmfiset{p3}{point length(p5) of p5 .. a3 .. point length(p4) of p4}
  \fmfi{fermion,rubout,label={\small $q_{N+1}$},label.side=right}{subpath (0,length(p3)) of p3}
  \fmfv{label={\small Class of topologies $\calt_{3}$},label.dist=0.20w,label.angle=-115}{vv4}
\end{fmfgraph*}}
\caption{\small Diagram picturing a class of 
planar topology with $N$ external vertices and zero internal vertex.}
\label{fig02} 
\end{figure}

\noindent
In figs.~\ref{fig01} and~\ref{fig02}, the dots represent trees whatever they are which contain the propagator whose labels are missing on the picture. 
For the class of topologies $\calt_{1}$, the different sets $S_k$ are equal to $S_1 = \{1, \dots, i\}$, $S_2 = \{ i+1, \dots, N+2\}$ and $S_3 = \{N+3\}$, for $\calt_2$, they are given by $S_1 = \{1, \dots, i\}$, $S_2 = \{ i+1, \dots, N+1\}$ and $S_3 = \{N+2\}$ while for $\calt_{3}$, they are equal to $S_1 = \{1, \dots, i\}$, $S_2 = \{ i+1, \dots, N\}$ and $S_3 = \{N+1\}$.
The key idea is that, in order to be able to build a $\bB$ matrix which is homogeneous in the parameters $u_i$, the sets $S_i$ $i=1,2,3$ must contain at least one label of a $n$-vector $\hr_i$ whose linear combination of the external momenta $p_k$, as defined in eq.~(\ref{eqdecomp0}), is zero. If we choose $\br_1 = \hr_k$ with $k \in S_1$, $\br_2 = \hr_l$ with $l \in S_2$, then the discussion comes down to know if a $n$-vector $\hr_j$ with $j \in S_3$ which is a linear combination of only $\br_1$ and $\br_2$ can be built. It is clear that taking $\br_1 = \hr_{i}$ and $\br_2 = \hr_{i+1}$, $\hr_{N+3}$ (resp.~$\hr_{N+2}$) in the class of topologies $\calt_{1}$ (resp.~$\calt_{2}$) is equal to $\br_1 + \br_2$ and thus, for these classes of topologies, it exists a $n$-vector $\hr$ whose label belongs to the set $S_3$ which has this property. On the contrary, for topology $\calt_{3}$, there is no way to build the $n$-vector $\hr_{N+1}$ as a linear combination of $\br_1$ and $\br_2$ only because on the two vertices connecting three internal lines, there is an external $n$-momentum connection.
In this case, all the Feynman parameters $\tau_k$ with $k \in S_3$ will appear in the entries of the $\bB$ matrix. And thus, after the re-parameterisation in terms of the parameters $u_i$ and $\rho_j$, some entries of this matrix will not be homogeneous of degree $1$ in the parameter $u_i$.
The same conclusion can be reached by studying the class of non planar topologies.
To end up the discussion, the condition for which the factorisation of $\det(A)$ does not hold for the vector $V$ and the scalar $C$ for a given diagram is that it has two four-leg vertices connecting three internal lines. \\

\noindent
In the next sections, we present a set of scalar topologies having a number of internal lines $I$ less than or equal to five\footnote{For the purpose of this article, we do not consider two-loop one-point functions, because the generalised one-loop function associated is trivial, in the sense that it does not contain extra integration.}. For a fixed value of $I$, a detailed calculation is given only for the primary topologies, they are parent topologies from where child topologies can be generated. This is due to the fact that in eq.~(\ref{eqG1a}), the complexity of the integrand depends only on $I$ and not on the number of external legs.
Thus the way to generate other topologies having the same structure as an initial $N$-point topology is to change the three-leg vertices into four-leg one. But it appears two kinds of three-leg vertices: the one connecting three internal legs and the one connecting two internal lines and one external line. Changing the first kind of three-leg vertex into a four-leg vertex amounts to add an extra external leg to the initial topology leading to a $N+1$ topology whose kinematics cannot be guessed from the one of the initial topology without performing the explicit calculation. On the contrary, changing the second kind of three-leg vertex into a four-leg vertex leads also to a new $N+1$ topology but with a kinematics which is a trivial extension of the one of the initial topology.
These latter topologies will just be mentioned in the article.\\

\noindent
In the rest of the article, we will come across typical integrals as
\[
  \int_0^1 d\rho \, \rho^{-1+\varepsilon} \, f(\rho)
\]
where $f$ is a regular function as $\rho \rightarrow 0$.
Up to terms of order of $O(\varepsilon^2)$, the distribution $\rho^{-1+\varepsilon}$ can be replaced by
\begin{align}
  \rho^{-1+\varepsilon} \equiv \frac{1}{\varepsilon} \, \delta(\rho) + \frac{1}{(\rho)_+} + \varepsilon \, \left( \frac{\ln(\rho)}{\rho} \right)_+
  \label{eqdistrib0}
\end{align}
The $(+)$ distributions are defined as usual
\begin{align}
  \int_0^1 d\rho \, \frac{f(\rho)}{(\rho)_+} &\equiv \int_0^1 d\rho \, \frac{f(\rho)-f(0)}{\rho}
  \label{eqplusdist1}
\end{align}
and
\begin{align}
  \int_0^1 d\rho \, f(\rho) \, \left( \frac{\ln(\rho)}{\rho} \right)_+ &\equiv \int_0^1 d\rho \, \frac{(f(\rho)-f(0)) \, \ln(\rho)}{\rho}
  \label{eqplusdist2}
\end{align}

\section{Topologies with $I=3$}\label{secIeq3}

\begin{figure}[h]
\centering
\parbox[c][45mm][t]{40mm}{%
\begin{fmfgraph*}(40,40)
  \fmfstraight
  \fmfleftn{i}{1} \fmfrightn{o}{1}
  \fmfset{arrow_len}{3mm}
  \fmf{fermion,label={\small $p$},label.side=right}{i1,v2}
  \fmf{fermion,label={\small $-p$},label.side=left}{o1,v3}
  \fmf{phantom,tension=0.2,left}{v2,v3}
  \fmf{phantom,tension=0.2,left}{v3,v2}
  \fmfposition
  \fmfipath{p}
  \fmfipath{pp}
  \fmfiset{p}{vpath(__v3,__v2)}
  \fmfiset{pp}{vpath(__v2,__v3)}
  \fmfi{fermion,label={\small $q_2$}}{subpath (0,length(p)) of p}
  \fmfi{fermion,label={\small $q_1$}}{subpath (length(pp),0) of pp}
  \fmffreeze
  \fmf{fermion,label={\small $q_3$},label.side=left}{v2,v3}
\end{fmfgraph*}
}
\caption{\small Topology $\calt_{22\_0\_2\_1}$ ($e111|e|$).}
\label{fig1} 
\end{figure}
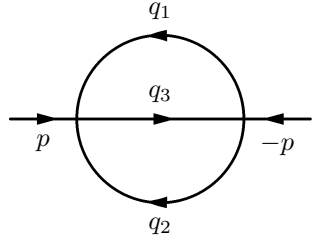

There is in fact only one topology depicted in fig.~\ref{fig1} having three internal propagators. From the discussion of sect.~\ref{gen_topo}, this topology has zero three-leg vertex and two four-leg vertices and the superficial degree of divergence is $2$.

\vspace{1cm}

\noindent
For this topology, we parameterise the internal momenta in the following way
\begin{align}
  q_1 = k_1 + \hr_1 \label{eqdefq1} \\
  q_2 = k_2 + \hr_2 \label{eqdefq2} \\
  q_3 = k_1 + k_2 + \hr_3 \label{eqdefq3}
\end{align}
Thus the sets $S_k$, $k=1\dots 3$ are given by
\begin{equation}
  S_1 = \left\{ 1 \right\} , \quad S_2 = \left\{ 2 \right\} \quad \text{and} \quad S_3 = \left\{ 3 \right\}
  \label{eqdefsets}
\end{equation}
which fills the matrix $A$, cf.~eq.~(\ref{eqdefA0}).
Furthermore we choose $\br_1 = \hr_1$ and $\br_2 = \hr_2$.
$T$ is a one dimensional vector $T = [p]$ and the matrices $\bB$ and $\Gamma$ are given by
\begin{align}
  \bB &= \tau_3 \, \left[
  \begin{array}{c}
    1 \\
    1
  \end{array}
\right]
  \label{eqdefBbar1t}
\end{align}
and
\begin{align}
  \Gamma &= \left[
  \begin{array}{c}
     \tau_3
  \end{array}
\right]
  \label{eqdefGamma1t}
\end{align}
Expanding the right-hand side of eq.~(\ref{eqdefpolycalf}) gives
\begin{align}
  \calf(\{\tau_k\}) &= - p^2 \, \tau_1 \, \tau_2 \, \tau_3 + (\tau_1 \, \tau_2 + \tau_1 \, \tau_3 + \tau_2 \, \tau_3) \, (\tau_1 \, m_1^2 + \tau_2 \, m_2^2 + \tau_3 \, m_3^2)
  \label{eqdefcaf1}
\end{align}

\noindent
There is no need to introduce the $\rho_i$ parameters in this case, obviously $\rho_i=\tau_i$ for $i=1,2,3$. 
Then we set $\tau_1 = \rho \, \xi$ and $\tau_2 = \rho \, ( 1- \xi)$, the two-loop two-point function then becomes
\begin{align}
  ^{(2)}I_{2}^{n}\left( \kappa_a \, ; \calt_{22\_0\_2\_1} \right)  &= (4 \, \pi)^{-4 + 2 \, \varepsilon} \, \Gamma(-1 + 2 \, \varepsilon) \int_0^1 d \rho \int_0^1 d \xi \, \rho^{-1+\varepsilon} \, \left( 1 - \rho + \rho \, \xi \, (1-\xi) \right)^{-3 + 3 \, \varepsilon} \notag \\
  &\quad {} \times \left( \calft(\rho,\xi) - i \, \lambda \right)^{1 - 2 \, \varepsilon}
  \label{eqdefT22n2}
\end{align}
with
\begin{align}
  \calft(\rho,\xi) &=  - p^2 \, \rho \, (1-\rho) \, \xi \, (1-\xi) + (1 - \rho + \rho \, \xi \, (1-\xi)) \, (\rho \, \xi \, m_1^2 + \rho \, (1-\xi) \, m_2^2 + (1-\rho) \, m_3^2 )
  \label{eqdefFbar}
\end{align}
and $\kappa_a = \{p^2,m_1^2,m_2^2,m_3^2\}$.
Eq.~(\ref{eqdefT22n2}) represents a two dimensional integration of a ``generalised one-loop'' tadpole function in a space-time of dimension $4 - 4 \, \varepsilon$, the function $\calft(\rho,\xi)$ plays the role of the mass of the effective particle which propagates through the tadpole.\\

\noindent
The details of the computation have been put in appendix~\ref{det_cal_t22021} to facilitate the reading. Summing the different parts and gathering terms proportional to $\ln(\calftp(\rho,\xi) - i \, \lambda)$, where $\calftp(\rho,\xi) = \calft(\rho,\xi)/(1 - \rho + \rho \, \xi \, (1-\xi))$, yields
\begin{align}
\hspace{2em}&\hspace{-2em}^{(2)}I_2^n\left( \kappa_a; \calt_{22\_0\_2\_1} \right) \notag \\
&= (4 \, \pi)^{-4 + 2 \, \varepsilon} \, \frac{\Gamma(1 + 2 \, \varepsilon)}{(-1 + 2 \, \varepsilon) \, 2 \, \varepsilon} \, \left\{ \frac{1}{\varepsilon} \, \sum_{i=1}^3 m_i^2 + \sum_{i=1}^3 m_i^2 + \frac{1}{2} \, (-p^2) - 2 \, \sum_{i=1}^3 m_i^2 \, \ln(m_i^2 - i \, \lambda) \right. \notag \\
&\quad {} + \varepsilon \left[ 2 \, \sum_{i=1}^3 m_i^2 \, \ln^2(m_i^2 - i \, \lambda) + m_3^2 \, \left( 1 + 3 \, I_1 + I_3 + I_4  \right) + (-p^2) \, \left( I_6 + I_7 \right) \right. \notag \\
&\quad {} + \sum_{i=1}^2 m_i^2 \, \left( 1 + \frac{2}{\xi^+ - \xi^-} \, \left( \dilog(\xi^-) - \dilog(\xi^+) \right) - \frac{\pi^2}{6} \right) \notag \\
&\quad {} - 2 \, \sum_{i=1}^2 m_i^2 \, \ln(m_i^2 - i \, \lambda) + 2 \, m_1^2 \, \dilog\left( \frac{m_1^2 - m_2^2}{m_1^2 - i \, \lambda} \right) + 2 \, m_2^2 \, \dilog\left( \frac{m_2^2 - m_1^2}{m_2^2 - i \, \lambda} \right) \notag \\
&\quad {}  -2 \, m_3^2 \, \int_0^1 d \xi \int_0^1 \frac{d \rho}{\rho} \, \left[ (1-\rho) \, \ln(\calft(\rho,\xi) - i \, \lambda) - \ln(m_3^2 - i \, \lambda) \right] \notag \\
&\quad {}  - 2 \, \int_0^1 d \xi \int_0^1 d \rho \, \left(  \frac{\calftp(\rho,\xi) \, \ln(\calftp(\rho,\xi)- i \, \lambda)}{\rho \, (1-\rho+\rho \, \xi \, (1-\xi))^2} - \frac{(1-\rho) \, m_3^2 \,  \ln(\calftp(\rho,\xi)- i \, \lambda)}{\rho} \right. \notag \\
&\qquad \qquad \qquad \qquad \qquad {} - \left. \left. \left. \frac{(\xi \, m_1^2 + (1-\xi) \, m_2^2) \,\ln(\xi \, m_1^2 + (1-\xi) \, m_2^2 - i \, \lambda)}{(1 - \rho + \rho\, \xi \, (1-\xi))^2} \right) \right] \right\}
  \label{eqdeftot1}
\end{align}
where $\xi^{\pm}$ are the roots of the polynomial $\xi^2-\xi+1$ and the integrals $I_j$ $j=1,7$ are given in the appendix~\ref{intsecpar}. 
From eq.~(\ref{eqdeftot1}), the divergent parts are obviously symmetric under the exchange of two masses. This is also the case for the finite part even if it is difficult to read it from the last equation. This symmetry can be seen from eq.~(\ref{eqdefcaf1}), it is the way the Feynman parameters $\tau_i$ are re-parameterised which breaks this symetry. The fact that the coefficient of the $\varepsilon^{-2}$ term is proportional to masses can be understood as follows. The diagram depicted in fig.~\ref{fig1} has divergent subdiagrams which are one-loop bubbles. Shrinking one of them to a point leads to a tadpole which is proportional to a mass.
\\

\noindent
On a practical point of view, the two double integrals on the variables $\rho$ and $\xi$ appearing in the eq.~(\ref{eqdeftot1}) are computed numerically. Nevertheless, an improvement can be made here by noticing that $\calft(\rho,\xi)$ is only quadratic in $\rho$, thus the $\rho$ integration could be performed analytically leaving only one dimensional integral to be computed numerically.

\section{Topologies with $I=4$}\label{secIeq4}

\begin{figure}[h]
\centering
\parbox[c][35mm][t]{30mm}{%
\begin{fmfgraph*}(35,30)
  \fmfstraight
  \fmfleftn{i}{1} \fmfrightn{o}{1}
  \fmfset{arrow_len}{3mm}
  \fmf{fermion,label={\small $p$},label.side=right}{i1,v2}
  \fmf{fermion,label={\small $-p$},label.side=left}{o1,v3}
  \fmf{phantom,tension=0.2,left}{v2,v3}
  \fmf{phantom,tension=0.2,left}{v3,v2}
  \fmfposition
  \fmfipath{p}
  \fmfipath{pp}
  \fmfiset{p}{vpath(__v3,__v2)}
  \fmfiset{pp}{vpath(__v2,__v3)}
  \fmfi{fermion,label={\small $q_2$}}{subpath (0,length(p)) of p}
  \fmfi{fermion,label={\small $q_1$}}{subpath (length(pp)/2,0) of pp}
  \fmfi{fermion,label={\small $q_3$}}{subpath (length(pp)/2,length(pp)) of pp}
  \fmfi{fermion,label={\small $q_4$},left}{subpath (length(pp)/2,0) of pp rotated 180 shifted (0.72w,1.133w)}
\end{fmfgraph*}
}
\caption{\small Topology $\calt_{22\_2\_1\_1}$ ($e112|2|e|$).}
\label{figure2} 
\end{figure}
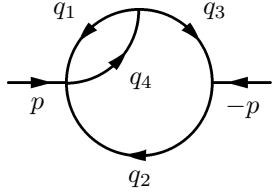

These topologies have a superficial degree of divergence $w(G) = 0$. There is only one primary topology: topology $\calt_{22\_2\_1\_1}$ depicted in fig.~\ref{figure2}, the others can be built out of it as will be shown later on.

\subsection{Topology $\calt_{22\_2\_1\_1}$}\label{t22211}

\vspace{1cm}

\noindent
The internal momenta are parameterised in the following way
\begin{align}
  q_1 &= k_1 + \hr_1 \notag \\
  q_2 &= k_2 + \hr_2 \notag \\
  q_3 &= k_2 + \hr_3 \notag \\
  q_4 &= k_1 + k_2 + \hr_4
  \label{eqdefqint0}
\end{align}
bringing about the sets $S_i$
\begin{equation}
  S_1 = \{1\} \quad S_2 = \{2,3\} \quad S_3 = \{4\} 
  \label{eqdefsets2t}
\end{equation}
and about the matrix $A$, cf. eq.~(\ref{eqdefA0}).
We choose $\br_1 = \hr_1$ and $\br_2 = \hr_3$. This choice leads to
\begin{align}
  \bB &= \left[
  \begin{array}{c}
    0 \\
    - \tau_2
  \end{array}
\right]
  \label{eqdefBbar2t}
\end{align}
and
\begin{align}
  \Gamma &= [ \tau_2]
  \label{eqdefGamma2t}
\end{align}
As in sec.~\ref{secIeq3}, $T$ is a one dimensional vector whose element is $p$.
Using eq.~(\ref{eqdefpolycalf}), we get
\begin{align}
  \calf(\{\tau_k\}) &= - p^2 \, \tau_2 \, (\tau_1 \, \tau_3 + \tau_1 \, \tau_4 + \tau_3 \, \tau_4) + (\tau_1 \, (\tau_2 + \tau_3) + \tau_1 \, \tau_4 + \tau_4 \, (\tau_2 + \tau_3)) \notag \\
  &\quad {} \times (\tau_1 \, m_1^2 + \tau_2 \, m_2^2 + \tau_3 \, m_3^2 + \tau_4 \, m_4^2)
  \label{eqdefcalf2t}
\end{align}
Parameterising the Feynman parameters $\tau_i$ in the following way
\begin{align}
  \tau_1 &= \rho_1  & \tau_2 &= \rho_2 \, u \notag \\
  \tau_3 &= \rho_2 \, (1-u) & \tau_4 &= \rho_3
  \label{eqreparam2t}
\end{align}
and performing the change of variable $\rho_1 = \rho \, \xi$ and $\rho_3 = \rho \, (1 - \xi)$, then the two-loop scalar amplitude becomes
\begin{align}
  ^{(2)}I_{2}^{n}\left( \kappa_b \, ; \calt_{22\_2\_1\_1} \right) &= - (4 \, \pi)^{-4 + 2 \, \varepsilon} \, \Gamma(2 \, \varepsilon) \int_0^1 d \rho \, \int_0^1 d \xi \, \left( 1 - \rho + \rho \, \xi (1-\xi) \right)^{-2+3 \, \varepsilon} \, \rho^{-1 + \varepsilon} \, (1-\rho) \notag \\
  &\quad {} \times \int_0^1 du \left( \calft(u,\rho,\xi) - i \, \lambda \right)^{- 2 \, \varepsilon}
  \label{eqUV6}
\end{align}
where $\kappa_b = \{p^2,m_1^2,m_2^2,m_3^2,m_4^2\}$ and
\begin{align}
  \calft(u,\rho,\xi) &= u^2 \, \tG_b - 2 \, \tV_b \, u -\tC_b
  \label{eqUV7}
\end{align}
with
\begin{align}
  \tG_b &=  p^2 \, (1-\rho)^2  \label{eqdefGt2t} \\
  \tV_b &=  \frac{1}{2} \, (1-\rho) \, (1 - \rho + \rho \, \xi \,(1-\xi)) \, (p^2 + m_3^2 - m_2^2) \label{eqdefVt2t} \\
  \tC_b &=  - (1 - \rho + \rho \, \xi \, (1-\xi)) \, \left( \rho \, \xi \, m_1^2 + (1-\rho) \, m_3^2 + \rho \, (1-\xi) \, m_4^2 \right) \label{eqdefCt2t}
\end{align}
With the parameterisation (\ref{eqreparam2t}), all the non zero entries of the matrix $\bB$ are homogeneous in $u$, thus the factorisation of $1 - \rho + \rho \, \xi \, (1-\xi)$ for the coefficients of $\tV_b$ and $\tC_b$ in the polynomial of eq.~(\ref{eqUV7}).\\ 

\noindent
After a partial fraction decomposition, eq.~(\ref{eqUV6}) can be re-written as
\begin{align}
  \hspace{2em}&\hspace{-2em}^{(2)}I_{2}^{n}\left( \kappa_b \, ; \calt_{22\_2\_1\_1} \right) \notag \\
  &= - (4 \, \pi)^{-4 + 2 \, \varepsilon} \, \Gamma(2 \, \varepsilon) \int_0^1 d \rho \, \int_0^1 d \xi \, \left[ \rho^{-1 + \varepsilon} \, G_1(\rho,\xi) + (1-\rho+\rho \, \xi \, (1-\xi))^{-2 + 3 \, \varepsilon} \, G_2(\rho,\xi) \right]
  \label{eqUV8}
\end{align}
with
\begin{align}
  G_1(\rho,\xi) &= (1-\rho+\rho \, \xi \, (1-\xi))^{3 \, \varepsilon} \, (1-\rho) \, \int_0^1 du \left( \calft(u,\rho,\xi) - i \, \lambda \right)^{- 2 \, \varepsilon} \label{eqUVG1} \\
  G_2(\rho,\xi) &= (1 - \xi \, (1-\xi)) \, \rho^{\varepsilon} \, (1-\rho) \, (2 - \rho \, (1 - \xi \, (1-\xi)) \, \int_0^1 du \left( \calft(u,\rho,\xi) - i \, \lambda \right)^{- 2 \, \varepsilon} \label{eqUVG2}
\end{align}
Notice that only the first term of eq.~(\ref{eqUV8}) diverges. Indeed, there is a global factor $1-\rho$ in $G_1$ and $G_2$ and thus $G_2(1,\xi) = 0$ which implies that the second term is finite.\\

\noindent
Let us start with the first term of eq. (\ref{eqUV8}) and for that let us introduce
\begin{align}
  M_1 &= \int_0^1 d\xi \, \int_0^1 d\rho \, \rho^{-1 + \varepsilon} \, G_1(\rho,\xi)
  \label{eqUVdefT1p0}
\end{align}
The function $G_1(\rho,\xi)$ can be expanded around $\varepsilon = 0$ 
\[
  G_1(\rho,\xi) = G_1^{(0)}(\rho,\xi) + \varepsilon \, G_1^{(1)}(\rho,\xi) + \varepsilon^2 \, G_1^{(2)}(\rho,\xi)
\]
with
\begin{align}
  G_1^{(0)}(\rho,\xi) &= 1- \rho \label{eqUVG10} \\
  G_1^{(1)}(\rho,\xi) &= (1-\rho) \, \left[ 3 \, \ln\left( 1-\rho + \rho \, \xi \, (1-\xi) \right) - 2 \, \int_0^1 du \, \ln\left( \calft(u,\rho,\xi) - i \, \lambda \right) \right] \label{eqUVG11} \\
G_1^{(2)}(\rho,\xi) &= (1-\rho) \, \left[ \frac{9}{2} \, \ln^2\left( 1-\rho+\rho \, \xi \, (1-\xi) \right) + 2 \, \int_0^1 du \, \ln^2\left( \calft(u,\rho,\xi) - i \, \lambda \right) \right. \notag \\
&\qquad \qquad \qquad {} - \left. 3 \, \ln\left( 1-\rho+\rho \, \xi \, (1-\xi) \right) \, \int_0^1 du \, \ln\left( \calft(u,\rho,\xi) - i \, \lambda \right) \right]
  \label{eqUVG12}
\end{align}
Using eq.~(\ref{eqdistrib0}) and the definition of the $+$ distributions (eqs.~(\ref{eqplusdist1}) and (\ref{eqplusdist2})), this leads to
\begin{align}
  M_1 &= \int_0^1 d\xi \left[  \frac{1}{\varepsilon} \, \int_0^1 du \left( \calft(u,0,\xi) - i \, \lambda \right)^{- 2 \, \varepsilon} - 1 + \varepsilon + \varepsilon \, \int_0^1 \frac{d \rho}{\rho} \left( G_1^{(1)}(\rho,\xi) - G_1^{(1)}(0,\xi) \right) \right]
  \label{eqUVdefT1p1}
\end{align}
From eq. (\ref{eqUV7}), one can see that $\calft(u,0,\xi)$ is a function of $u$ only
\begin{equation}
  \calft(u,0,\xi) = u^2 \, p^2 - u \, (p^2 + m_3^2 - m_2^2) + m_3^2 \equiv \calh(u)
  \label{eqUVdefKp1}
\end{equation}
so eq. (\ref{eqUVdefT1p1}) becomes
\begin{align}
M_1 &= \frac{1}{\varepsilon} \, \int_0^1 du \left( \calh(u) - i \, \lambda \right)^{- 2 \, \varepsilon} - 1  + \varepsilon + 3 \, \varepsilon \, I_1 \notag \\
&\quad {} - 2 \, \varepsilon \, \int_0^1 d\xi \, \int_0^1 \frac{d\rho}{\rho} \, \int_0^1 du \, \left[ (1-\rho) \, \ln\left( \calft(u,\rho,\xi) - i \, \lambda \right) - \ln\left( \calh(u) - i \, \lambda \right) \right]
  \label{eqUVdefT1p2}
\end{align}
with $I_1$ is given in appendix~(\ref{intsecpar}).\\

\noindent
For the second term in eq.~(\ref{eqUV8}), we have to study
\begin{align}
  M_2 &= \int_0^1 d\xi \, \int_0^1 d\rho \, (1-\rho+\rho \, \xi \, (1-\xi) )^{-2 + 3 \, \varepsilon} \, G_2(\rho,\xi)
  \label{eqUVdefT2p}
\end{align}
This term does not diverge and so expanding around $\varepsilon=0$ and keeping the relevant terms yields
\begin{align}
  M_2 &= I_2 + \varepsilon \, \left[ 3 \,I_3 + I_4 - 2 \, \int_0^1 d\xi \, \int_0^1 d\rho \, \frac{(1 - \xi \, (1-\xi)) \, (1-\rho) \, (2 - \rho \, (1 - \xi \, (1-\xi)))}{(1 - \rho + \rho \, \xi \, (1-\xi))^2} \, \right. \notag \\
&\qquad \qquad \qquad \qquad \qquad \qquad \qquad \qquad {} \times \left. \int_0^1 du \, \ln\left( \calft(u,\rho,\xi) - i \, \lambda \right) \right]
  \label{eqUVdefT2p1}
\end{align}
where the integrals $I_2$, $I_3$ and $I_4$ are given in appendix~\ref{intsecpar}. Then summing up the results for $M_1$ and $M_2$ and gathering terms proportional to $\ln\left( \calft(u,\rho,\xi) - i \, \lambda \right)$ leads to the scalar two-loop amplitude
\begin{align}
  \hspace{2em}&\hspace{-2em}^{(2)}I_{2}^{n}\left( \kappa_b \, ; \calt_{22\_2\_1\_1} \right) \notag \\
  &= - (4 \, \pi)^{-4 + 2 \, \varepsilon} \, \Gamma(1 + 2 \, \varepsilon) \, \left\{\frac{1}{2 \, \varepsilon^2} + \frac{1}{\varepsilon} \, \left[ \frac{1}{2} - \wtI_{2}^{\,(1)}(\Sigma_{(1)};\tG_b,\tV_b,\tC_b,0,\xi) \right] \right. \notag \\ 
  &\qquad \qquad \qquad \qquad \qquad \qquad {} + \wtI_{2}^{\,(2)}(\Sigma_{(1)};\tG_b,\tV_b,\tC_b,0,\xi) 
  + \frac{1}{2} \, \left[ 1 + 3 \, I_1 +  3 \,I_3 +  I_4 \right]  \notag \\
  &\qquad \qquad \qquad \qquad \qquad \qquad {} - \int_0^1 d\xi \, \int_0^1 \frac{d\rho}{\rho} \, \left[ \frac{(1-\rho)}{(1 - \rho + \rho \, \xi \, (1-\xi))^2} \, \wtI_{2}^{\,(1)}(\Sigma_{(1)};\tG_b,\tV_b,\tC_b,\rho,\xi) \right. \notag \\
  &\qquad \qquad \qquad \qquad \qquad \qquad \qquad \qquad \qquad \qquad {} - \left. \left. \wtI_{2}^{\,(1)}(\Sigma_{(1)};\tG_b,\tV_b,\tC_b,0,\xi) \vphantom{\frac{(1-\rho)}{(1 - \rho + \rho \, \xi \, (1-\xi))^2}} \right] \right\}
  \label{eqUVresfin2}
\end{align}
Note that $\wtI_{2}^{\,(1)}(\Sigma_{(1)};\tG_b,\tV_b,\tC_b,0,\xi)$ does not depend on $\xi$ (cf.~eq.~(\ref{eqUVdefKp1})). In eq.~(\ref{eqUVresfin2}), the ''generalised one-loop functions'' $\wtI_{2}^{\,(k)}$ are computed analytically leaving two integrations to be performed numerically. The functions $\wtI_{2}^{\,(1)}$ and $\wtI_{2}^{\,(2)}$ can be expressed in terms of dilogarithms and logarithms as shown in appendix~\ref{ol2Oe}.

\subsection{Topology $\calt_{23\_1\_2\_1}$}\label{t23121}

\begin{figure}[h]
\centering
\parbox[c][35mm][t]{30mm}{%
  \begin{fmfgraph*}(35,30)
  \fmfstraight
  \fmfleftn{i}{1} \fmfrightn{o}{2}
  \fmfset{arrow_len}{3mm}
  \fmf{fermion,label={\small $p_1$}}{i1,v2}
  \fmf{fermion,label={\small $p_2$}}{o1,v3}
  \fmf{fermion,label={\small $p_3$}}{o2,v4}
  \fmf{phantom,tension=0.5}{v2,v3p,v2p,v3}
  \fmf{phantom,tension=0.5}{v4,v1p,v4p,v2}
  \fmffreeze
  \fmf{fermion,label={\small $q_4$},label.side=right,right=0.5}{v2,v4}
  \fmf{fermion,rubout,label={\small $q_2$},label.side=left}{v4,v3}
  \fmf{fermion,label={\small $q_3$},label.side=left}{v3,v2}
  \fmf{fermion,label={\small $q_1$},right=0.5}{v4,v2}
\end{fmfgraph*}
}
\caption{\small Topology $\calt_{23\_1\_2\_1}$ ($e112|e2|e|$).}
\label{figure3} 
\end{figure}
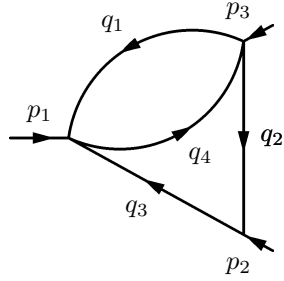

This is the topology, depicted in fig.~\ref{figure3}, obtained by inserting an external leg to the internal three-leg vertex of fig.~\ref{figure2}.

\vspace{1cm}

\noindent
As already explained, the determination of the new matrix $\tG$, the new vector $\tV$ and the new scalar $\tC$ for this topology will be done explicitly, they cannot be extracted from those of the primary topology $\calt_{22\_2\_1\_1}$.
The internal momenta are parameterised as follows
\begin{align}
  q_1 &= k_1 + \hr_1 \notag \\
  q_2 &= k_2 + \hr_2 \notag \\
  q_3 &= k_2 + \hr_3 \notag \\
  q_4 &= k_1 + k_2 + \hr_4
  \label{eqdefqint3t}
\end{align}
leading to the sets $S_i$
\begin{equation}
  S_1 = \{1\} \quad S_2 = \{2,3\} \quad S_3 = \{4\} 
  \label{eqdefsets3t}
\end{equation}
and thus to the matrix $A$ through eq.~(\ref{eqdefA0}).
We choose $\br_1 = \hr_1$ and $\br_2 = \hr_2$. This choice leads to
\begin{align}
  \bB &= \left[
  \begin{array}{c c}
    0 & \tau_4 \\
    \tau_3 & \tau_4
  \end{array}
\right]
  \label{eqdefBbar3t}
\end{align}
and
\begin{align}
  \Gamma &= \left[
  \begin{array}{c c}
    \tau_3 & 0 \\
    0 & \tau_4
  \end{array}
\right]
  \label{eqdefGamma3t}
\end{align}
The vector $T$ is taken to be
\begin{align}
  T &= \left[
  \begin{array}{c}
    p_2 \\
    p_1 + p_2
  \end{array}
\right]
  \label{eqdefT3t}
\end{align}
Using eq.~(\ref{eqdefpolycalf}), we get
\begin{align}
  \calf(\{\tau_k\}) &= - p_2^2 \, \tau_2 \, (\tau_1 \, \tau_4 + \tau_3 \, \tau_4 + \tau_1 \, \tau_3) - p_1^2 \, \tau_1 \, \tau_4 \, (\tau_2 + \tau_3) - 2 \, p_1 \cdot p_2 \, \tau_1 \, \tau_2 \, \tau_4 \notag \\
  &\quad {} + (\tau_1 \, (\tau_2 + \tau_3) + \tau_1 \, \tau_4 + \tau_4 \, (\tau_2 + \tau_3)) \, (\tau_1 \, m_1^2 + \tau_2 \, m_2^2 + \tau_3 \, m_3^2 + \tau_4 \, m_4^2)
  \label{eqdefcalf3t}
\end{align}

\noindent
The Feynman parameters $\tau_i$ are parameterised in the following way
\begin{align}
  \tau_1 &= \rho_1  & \tau_2 &= \rho_2 \, u \notag \\
  \tau_3 &= \rho_2 \, (1-u) & \tau_4 &= \rho_3
  \label{eqreparam3t}
\end{align}
Then, the following change of variables is performed $\rho_1 = \rho \, \xi$,  $\rho_3 = \rho \, (1 - \xi)$ leading to $\rho_2 = 1 - \rho$ because the $\rho_i$'s sum to $1$. This yields
\begin{align}
  ^{(2)}I_{3}^{n}\left( \kappa_c \, ; \calt_{23\_1\_2\_1} \right) &= - (4 \, \pi)^{-4 + 2 \, \varepsilon} \, \Gamma(2 \, \varepsilon) \int_0^1 d \rho \, \int_0^1 d \xi \, \left( 1 - \rho + \rho \, \xi (1-\xi) \right)^{-2+3 \, \varepsilon} \, \rho^{-1 + \varepsilon} \, (1-\rho) \notag \\
  &\quad {} \times \int_0^1 du \left( \calft(u,\rho,\xi) - i \, \lambda \right)^{- 2 \, \varepsilon}
  \label{eqUV3t6}
\end{align}
where $\kappa_c = \{p_1^2,p_2^2,p_3^2,m_1^2,m_2^2,m_3^2,m_4^2\}$ and
\begin{align}
  \calft(u,\rho,\xi) &=  u^2 \, \tG_c - 2 \, \tV_c \, u -\tC_c
  \label{eqUV3t7}
\end{align}
with
\begin{align}
  \tG_c &=  p_2^2 \, (1-\rho)^2  \label{eqdefGt3t} \\
  \tV_c &=  \frac{1}{2} \, (1-\rho) \, \left[ (1 - \rho + \rho \, \xi \, (1-\xi)) \, (p_2^2 + m_3^2 - m_2^2) + 2 \, p_1 \cdot p_2 \, \rho \, \xi \, (1-\xi) \right] \label{eqdefVt3t} \\
  \tC_c &=   \rho \, (1-\rho) \, \xi \, (1-\xi) \, p_1^2 - (1 - \rho + \rho \, \xi \, (1-\xi)) \, \left( \rho \, \xi \, m_1^2 + (1-\rho) \, m_3^2 + \rho \, (1-\xi) \, m_4^2 \right) \label{eqdefCt3t}
\end{align}
As discussed previously, the new kinematics carried by $\tG_c$, $\tV_c$ and $\tC_c$ cannot be guessed from the kinematics of the primary topology $\calt_{22\_2\_1\_1}$. Indeed,
since the topology $\calt_{23\_1\_2\_1}$ has two four-leg vertices connecting three internal lines, the factorisation of $(1 - \rho + \rho \, \xi \, (1-\xi))$ for the coefficients of $\tV_c$ and $\tC_c$ in the polynomial of eq.~(\ref{eqUV3t7}) does not hold (cf.~sec.~\ref{secskel}).
Note that the kinematics of the primary topology (eqs.~(\ref{eqdefGt2t}), (\ref{eqdefVt2t}) and (\ref{eqdefCt2t})) can be recovered by letting $p_1=0$ because of the choice of the labelling of the internal legs in fig.~\ref{figure3}. Making the change of variable $u = 1-w$ in eq.~(\ref{eqUV3t6}) leads to a new polynomial in $w$ whose coefficients are
\begin{align}
  \tG_c^{\prime} &=  p_2^2 \, (1-\rho)^2 \label{eqdefGt3tp} \\
  \tV_c^{\prime} &=  \frac{1}{2} \, (1-\rho) \, \left[ (1 - \rho + \rho \, \xi \, (1-\xi)) \, (p_2^2 + m_2^2 - m_3^2) + 2 \, p_2 \cdot p_3 \, \rho \, \xi \, (1-\xi) \right] \label{eqdefVt3tp} \\
  \tC_c^{\prime} &= \rho \, (1-\rho) \, \xi \, (1-\xi) \, p_3^2 - (1 - \rho + \rho \, \xi \, (1-\xi)) \, \left( \rho \, \xi \, m_1^2 + (1-\rho) \, m_2^2 + \rho \, (1-\xi) \, m_4^2 \right)  \label{eqdefCt3tp}
\end{align}
Thus letting $p_3=0$ leads to the kinematics of the primary topology also but with $m_2 \leftrightarrow m_3$ which is not straightforward by looking at eqs.~(\ref{eqdefGt3t}), (\ref{eqdefVt3t}) and (\ref{eqdefCt3t}). This is a consequence of the mirror symmetry of the diagram depicted in fig.~\ref{figure3}, namely $p_1 \leftrightarrow p_3$ and $m_2 \leftrightarrow m_3$.
The two-loop three-point function is given by the right-hand side of eq.~(\ref{eqUVresfin2}) as it is with a kinematics described by eqs.~(\ref{eqdefGt3t}), (\ref{eqdefVt3t}) and (\ref{eqdefCt3t}).
\begin{align}
  \hspace{2em}&\hspace{-2em}^{(2)}I_{3}^{n}\left( \kappa_c \, ; \calt_{23\_1\_2\_1} \right) \notag \\
  &= - (4 \, \pi)^{-4 + 2 \, \varepsilon} \, \Gamma(1 + 2 \, \varepsilon) \, \left\{\frac{1}{2 \, \varepsilon^2} + \frac{1}{\varepsilon} \, \left[ \frac{1}{2} - \wtI_{2}^{\,(1)}(\Sigma_{(1)};\tG_c,\tV_c,\tC_c,0,\xi) \right] \right. \notag \\ 
  &\qquad \qquad \qquad \qquad \qquad \qquad {} + \wtI_{2}^{\,(2)}(\Sigma_{(1)};\tG_c,\tV_c,\tC_c,0,\xi) + \frac{1}{2} \, \left[ 1 + 3 \, I_1 +  3 \,I_3 +  I_4 \right] \notag \\  
  &\qquad \qquad \qquad \qquad \qquad \qquad {}- \int_0^1 d\xi \, \int_0^1 \frac{d\rho}{\rho} \, \left[ \frac{(1-\rho)}{(1 - \rho + \rho \, \xi \, (1-\xi))^2} \, \wtI_{2}^{\,(1)}(\Sigma_{(1)};\tG_c,\tV_c,\tC_c,\rho,\xi) \right. \notag \\
  &\qquad \qquad \qquad \qquad \qquad \qquad \qquad \qquad \qquad \qquad {} - \left. \left. \wtI_{2}^{\,(1)}(\Sigma_{(1)};\tG_c,\tV_c,\tC_c,0,\xi) \vphantom{\frac{(1-\rho)}{(1 - \rho + \rho \, \xi \, (1-\xi))^2}} \right] \right\}
  \label{eqUV3tresfin2}
\end{align}

\subsubsection{Other topologies with $I=4$}

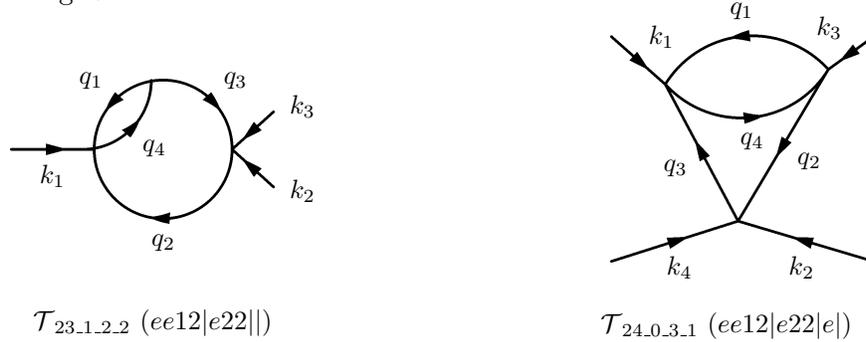
\begin{figure}[h]
\centering
\parbox[c][35mm][t]{40mm}{%
\begin{fmfgraph*}(35,30)
  \fmfstraight
  \fmfleftn{i}{1} \fmfrightn{o}{4}
  \fmfset{arrow_len}{3mm}
  \fmf{fermion,label={\small $k_1$}}{i1,v2}
  \fmf{fermion}{o2,v3}
  \fmf{fermion}{o3,v3}
  \fmfv{label={\small $k_2$}}{o2}
  \fmfv{label={\small $k_3$}}{o3}
  \fmf{phantom,tension=0.3,left}{v2,v3}
  \fmf{phantom,tension=0.3,left}{v3,v2}
  \fmfposition
  \fmfipath{p}
  \fmfipath{pp}
  \fmfiset{p}{vpath(__v3,__v2)}
  \fmfiset{pp}{vpath(__v2,__v3)}
  \fmfi{fermion,label={\small $q_2$}}{subpath (0,length(p)) of p}
  \fmfi{fermion,label={\small $q_1$}}{subpath (length(pp)/2,0) of pp}
  \fmfi{fermion,label={\small $q_3$}}{subpath (length(pp)/2,length(pp)) of pp}
  \fmfi{fermion,label={\small $q_4$},left}{subpath (length(pp)/2,0) of pp rotated 180 shifted (0.85w,1.12w)}
  \fmfv{label={\small $\calt_{23\_1\_2\_2}$ ($ee12|e22||$)},label.dist=0.65w,label.angle=-70}{v2}
\end{fmfgraph*}
}
\qquad \qquad \qquad \qquad \qquad
\parbox[c][35mm][t]{40mm}{%
  \begin{fmfgraph*}(35,30)
  \fmfstraight
  \fmfleftn{i}{2} \fmfrightn{o}{2}
  \fmfset{arrow_len}{3mm}
  \fmf{fermion,label={\small $k_1$}}{i2,v2}
  \fmf{fermion,label={\small $k_2$},label.side=left}{o1,v3}
  \fmf{fermion,label={\small $k_3$}}{o2,v4}
  \fmf{fermion,label={\small $k_4$}}{i1,v3}
  \fmf{phantom,tension=0.5}{v2,v3p,v2p,v3}
  \fmf{phantom,tension=0.5}{v4,v1p,v4p,v2}
  \fmf{fermion,tension=0.2,label={\small $q_2$},label.side=left}{v4,v3}
  \fmf{fermion,tension=0.2,label={\small $q_3$},label.side=left}{v3,v2}
  \fmffreeze
  \fmf{fermion,label={\small $q_4$},label.side=right,right=0.5}{v2,v4}
  \fmf{fermion,label={\small $q_1$},right=0.5}{v4,v2}
  \fmfv{label={\small $\calt_{24\_0\_3\_1}$ ($ee12|e22|e|$)},label.dist=0.9w,label.angle=-75}{v2}
\end{fmfgraph*}
}
\vspace{1cm}
\caption{\small New topologies generated from $\calt_{22\_2\_1\_1}$.}
\label{figure3p} 
\end{figure}

For the sake of completeness, let us mention two other topologies generated by changing the three-leg vertex having an external leg in topologies $\calt_{22\_2\_1\_1}$ and $\calt_{23\_1\_2\_1}$ into a four-leg vertex, they are depicted in fig.~\ref{figure3p}

\noindent
The amplitudes for these two topologies are still given by the right-hand side of eq.~(\ref{eqUVresfin2}). In these cases, no need to redo the explicit computation for the kinematics. The kinematics of $\calt_{23\_1\_2\_2}$ is the one of topology $\calt_{22\_2\_1\_1}$ with $p=k_1= -(k_2+k_3)$ and the kinematics of the topology $\calt_{24\_0\_3\_1}$ is the one of the topology $\calt_{23\_1\_2\_1}$ with $p_1=k_1$, $p_3=k_3$ and $p_2 = k_2+k_4$.

\section{Topologies with $I=5$}\label{secIeq5}

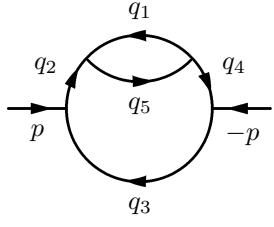
\begin{figure}[h]
\centering
\parbox[c][35mm][t]{30mm}{%
\begin{fmfgraph*}(35,30)
  \fmfstraight
  \fmfleftn{i}{1} \fmfrightn{o}{1}
  \fmfset{arrow_len}{3mm}
  \fmf{fermion,label={\small $p$}}{i1,v2}
  \fmf{fermion,label={\small $-p$},label.side=left}{o1,v3}
  \fmf{phantom,tension=0.2,left}{v2,v3}
  \fmf{phantom,tension=0.2,left}{v3,v2}
  \fmfposition
  \fmfipath{p}
  \fmfipath{pp}
  \fmfiset{p}{vpath(__v3,__v2)}
  \fmfiset{pp}{vpath(__v2,__v3)}
  \fmfi{fermion,label={\small $q_3$}}{subpath (0,length(p)) of p}
  \fmfi{fermion,label={\small $q_2$}}{subpath (0,length(pp)/4) of pp}
  \fmfi{fermion,label={\small $q_1$}}{subpath (3length(pp)/4,length(pp)/4) of pp}
  \fmfi{fermion,label={\small $q_4$}}{subpath (3length(pp)/4,length(pp)) of pp}
  \fmfi{fermion,label={\small $q_5$},left}{subpath (3length(pp)/4,length(pp)/4) of pp rotated 180 shifted (1.00w,1.237w)}
\end{fmfgraph*}
}
\caption{\small Topology $\calt_{22\_4\_0\_1}$ ($e12|e3|33||$).}
\label{figure5s} 
\end{figure}

For amplitudes having $I=5$, the superficial degree of divergence $w(G) = -2$ but some topologies can have UV divergent subdiagrams. There are two primary topologies, one having a UV divergent subdiagram $\calt_{22\_4\_0\_1}$ depicted in fig.~\ref{figure5s} and another one with a UV free topology $\calt_{22\_4\_0\_2}$ depicted in fig.~\ref{figure4}. 

\subsection{Topology $\calt_{22\_4\_0\_1}$}\label{t22401}

\vspace{1cm}

\noindent
The internal momenta are parameterised in the following way
\begin{align}
  q_1 &= k_1 + \hr_1 \notag \\
  q_2 &= k_2 + \hr_2 \notag \\
  q_3 &= k_2 + \hr_3 \notag \\
  q_4 &= k_2 + \hr_4 \notag \\
  q_5 &= k_1 + k_2 + \hr_5
  \label{eqdefqint5ts}
\end{align}
bringing about the sets $S_i$
\begin{equation}
  S_1 = \{1\} \quad S_2 = \{2,3,4\} \quad S_3 = \{5\} 
  \label{eqdefsets5ts}
\end{equation}
and about the matrix $A$ through eq.~(\ref{eqdefA0}).
We choose $\br_1 = \hr_1$ and $\br_2 = \hr_2$. This choice leads to
\begin{align}
  \bB &= \left[
  \begin{array}{c}
    0 \\
    -\tau_3
  \end{array}
\right]
  \label{eqdefBbar5ts}
\end{align}
and
\begin{align}
  \Gamma &= \left[
    \tau_3
\right]
  \label{eqdefGamma5ts}
\end{align}
The vector $T$ is taken to be
\begin{align}
  T &= \left[
    p 
\right]
  \label{eqdefT5ts}
\end{align}
Using eq.~(\ref{eqdefpolycalf}), we get for the polynomial $\calf(\{\tau_k\})$
\begin{align}
  \calf(\{\tau_k\}) &= - p^2 \, \tau_3 \, ( (\tau_1+\tau_5) \, (\tau_2+\tau_4) + \tau_1 \, \tau_5) + ( (\tau_1+\tau_5) \, (\tau_2+\tau_3+\tau_4) + \tau_1 \, \tau_5) \, \sum_{j=1}^{5} \tau_j \, m_j^2
  \label{eqdefcalf5ts}
\end{align}
Then, we parameterise the Feynman parameters $\tau_i$ in the following way
\begin{align}
  \tau_1 &= \rho_1  & \tau_2 &= \rho_2 \, u_1 \notag \\
  \tau_3 &= \rho_2 \, u_2 & \tau_4 &= \rho_2 \, (1-u_1-u_2) \notag \\
  \tau_5 &= \rho_3 & 
  \label{eqreparam5ts}
\end{align}
and performing the following change of variable $\rho_1 = \rho \, \xi$, $\rho_3 = \rho \, (1 - \xi)$ the two-loop scalar amplitude becomes
\begin{align}
  ^{(2)}I_{2}^{n}\left( \kappa_d \, ; \calt_{22\_4\_0\_1} \right) &= (4 \, \pi)^{-4 + 2 \, \varepsilon} \, \Gamma(1 + 2 \, \varepsilon) \int_0^1 d \rho \, \int_0^1 d \xi \, \left( 1 - \rho + \rho \, \xi (1-\xi) \right)^{-1+3 \, \varepsilon} \, \rho^{-1 + \varepsilon} \, (1-\rho)^2 \notag \\
  &\quad {} \times \int_{\Sigma_{(2)}} d u_1 \, d u_2 \left( \calft(u_1,u_2,\rho,\xi) - i \, \lambda \right)^{-1 - 2 \, \varepsilon}
  \label{eqUV5ts6}
\end{align}
where $\kappa_d = \{p^2,m_1^2,m_2^2,m_3^2,m_4^2,m_5^2\}$ and
\begin{align}
  \calft(u_1,u_2,\rho,\xi) &= U^T \cdot \tG_d \cdot U - 2 \, \tV^T_d \cdot U - \tC_d
  \label{eqUV5ts7}
\end{align}
with
\begin{align}
  \tG_d &= (1-\rho)^2 \, \left[
  \begin{array}{cc}
    0 & 0 \\
    0 & p^2
  \end{array}
\right] \label{eqdefaGt5ts} \\
\tV_d &= \frac{1}{2} \, (1 - \rho) \, (1 - \rho + \rho \, \xi \, (1-\xi)) \, \left[
\begin{array}{c}
  - m_2^2 + m_4^2 \\
  p^2 - m_3^2 + m_4^2
\end{array}
\right] \label{eqdefVt5ts} \\
\tC_d &= - (1 - \rho + \rho \, \xi \, (1-\xi)) \, \left( \rho  \, \xi \, m_1^2 + (1-\rho) \, m_4^2 + \rho \, (1-\xi) \, m_5^2 \right)
  \label{eqdefCt5ts}
\end{align}
\\

\noindent
We, then, proceed following the strategy developed for the other UV divergent diagrams.
After partial fraction decomposition, eq.~(\ref{eqUV5ts6}) can be re-written as
\begin{align}
  \hspace{2em}&\hspace{-2em}^{(2)}I_{2}^{n}\left( \kappa_d \, ; \calt_{22\_4\_0\_1} \right) \notag \\
  &= (4 \, \pi)^{-4 + 2 \, \varepsilon} \, \Gamma(1 + 2 \, \varepsilon) \int_0^1 d \rho \, \int_0^1 d \xi \, \left[ \rho^{-1 + \varepsilon} \, K_1(\rho,\xi) + (1-\rho+\rho \, \xi \, (1-\xi))^{-1 + 3 \, \varepsilon} \, K_2(\rho,\xi) \right]
  \label{eqUV5ts8}
\end{align}
with
\begin{align}
  K_1(\rho,\xi) &= (1-\rho+\rho \, \xi \, (1-\xi))^{3 \, \varepsilon} \, (1-\rho)^2 \, \int_{\Sigma_{(2)}} d u_1 \, d u_2 \, \left( \calft(u_1,u_2,\rho,\xi) - i \, \lambda \right)^{-1 - 2 \, \varepsilon} \label{eqUV5tsG1} \\
  K_2(\rho,\xi) &= \rho^{\varepsilon} \, (1-\rho)^2 \, (1 - \xi \, (1-\xi)) \, \int_{\Sigma_{(2)}} d u_1 \, d u_2 \, \left( \calft(u_1,u_2,\rho,\xi) - i \, \lambda \right)^{-1 - 2 \, \varepsilon} \label{eqUV5tsG2}
\end{align}
Notice that only the first term of the eq.~(\ref{eqUV5ts8}) diverges. Indeed, there is a global factor $(1-\rho)^2$ in $K_1$ and $K_2$ and thus $K_2(1,\xi) = 0$ makes the second term finite.\\

\noindent
Let us focus on the first term of eq. (\ref{eqUV5ts8}) and for that let us introduce
\begin{align}
  N_1 &= \int_0^1 d\xi \, \int_0^1 d\rho \, \rho^{-1 + \varepsilon} \, K_1(\rho,\xi)
  \label{eqUVdefT1p5ts0}
\end{align}
Using eq.~(\ref{eqdistrib0}) and keeping terms up to order $\varepsilon^{0}$, $N_1$ becomes
\begin{align}
  N_1 &= \frac{1}{\varepsilon} \, \int_0^1 d \xi \, K_1(0,\xi) + \int_0^1 d \xi \, \int_0^1 \frac{d \rho}{\rho} \, \left( K_1(\rho,\xi) - K_1(0,\xi) \right) + O(\varepsilon)
  \label{eqUVdefT1p5ts1}
\end{align}
with
\begin{align}
  K_1(0,\xi) &= \int_{\Sigma_{(2)}} d u_1 \, d u_2 \, \left( \calft(u_1,u_2,0,\xi) - i \, \lambda \right)^{-1 - 2 \, \varepsilon}
  \label{eqUVdefT1p5ts2}
\end{align}
Note that $\calft(u_1,u_2,0,\xi)$ does not depend on $\xi$, it is given by
\begin{align}
  \calft(u_1,u_2,0,\xi) &=  U^T \cdot \left[
  \begin{array}{cc}
    0 & 0 \\
    0 &  p^2
  \end{array}
\right] \cdot U -  \left[
\begin{array}{c}
 - m_2^2 + m_4^2 \\
p^2 - m_3^2 + m_4^2
\end{array}
\right]^T \cdot U + m_4^2 
  \label{eqcalf0xi5ts}
\end{align}
\\

\noindent
For the second term of eq.~(\ref{eqUV5ts8}), we just have to take $\varepsilon=0$. Putting the two contributions together and gathering terms proportional to $1/\calft(u_1,u_2,\rho,\xi)$, the two-loop amplitude reads
\begin{align}
\hspace{2em}&\hspace{-2em}^{(2)}I_{2}^{n}\left( \kappa_d \, ; \calt_{22\_4\_0\_1} \right) \notag \\
&= (4 \, \pi)^{-4 + 2 \, \varepsilon} \, \Gamma(1 + 2 \, \varepsilon) \,  \left\{  \frac{1}{\varepsilon} \, \wtI_{3}^{\,(0)}(\Sigma_{(2)};\tG_d,\tV_d,\tC_d,0,\xi) - 2 \, \wtI_{3}^{\,(1)}(\Sigma_{(2)};\tG_d,\tV_d,\tC_d,0,\xi) \right. \notag \\
&\quad {} + \int_0^1 d \xi \, \int_0^1 \frac{d \rho}{\rho} \left[ \frac{(1-\rho)^2}{1 - \rho + \rho \, \xi \, (1-\xi)} \, \wtI_{3}^{\,(0)}(\Sigma_{(2)};\tG_d,\tV_d,\tC_d,\rho,\xi) \right. \notag \\ 
&\qquad \qquad \qquad \qquad \quad {} - \left. \left. \wtI_{3}^{\,(0)}(\Sigma_{(2)};\tG_d,\tV_d,\tC_d,0,\xi) \vphantom{\frac{(1-\rho)^2}{1 - \rho + \rho \, \xi \, (1-\xi)}} \right] \right\}
  \label{eqUV5ts9}
\end{align}
The ``generalised one-loop three-point functions'' appearing in eq.~(\ref{eqUV5ts9}) are a bit special because the matrix $\tG_d$ is not invertible. The general case for the computation of ``generalised one-loop three- and four-point functions'' where $\det(G)=0$ has been treated in the Appendix C of ref.~\cite{paper1}. In this case, the one-loop three-point function boils down to a difference of one-loop two-point functions as can be seen from eq.~(2.25) of ref.~\cite{paper1} where the coefficients $b_i$ and $B$ are given by eqs.~(C.45) and (C.46) of the same reference, namely
\begin{align}
  \wtI_{3}^{\,(k)}(\Sigma_{(2)};\tG_d,\tV_d,\tC_d,\rho,\xi) 
  &= \frac{1}{2 \, (k+1) \, \tV_{1 \, d}} \, \left[ \wtI_{2}^{\, k+1}(\Sigma_{(1)}; \tG_{22 \, d}, \tV_{2 \, d},\tC_d,\rho,\xi) \right. \notag \\ 
  &\qquad \qquad \qquad \qquad {} - \left. \wtI_{2}^{\, k+1}(\Sigma_{(1)}; \tG_{22 \, d}, \tV_{2 \, d} - \tV_{1 \, d},\tC_d + 2 \, \tV_{1 \, d},\rho,\xi) \right] 
  \label{eqspeccase0}
\end{align}
where $\tG_{ij \, d}$ (resp.~$\tV_{i \, d}$) denotes the component $i,j$ of the matrix $\tG_d$ (resp.~the component $i$ of the vector $\tV_d$).

\subsection{Topology $\calt_{23\_3\_1\_1}$}\label{t23311}

This topology is built by inserting an external leg to one three-leg vertex connecting three internal lines in the topology $\calt_{22\_4\_0\_1}$. The kinematics will be computed explicitly.

\vspace{1cm}

\begin{figure}[h]
\centering
\parbox[c][35mm][t]{30mm}{%
  \begin{fmfgraph*}(30,30)
  \fmfstraight
  \fmfleftn{i}{1} \fmfrightn{o}{2}
  \fmfset{arrow_len}{3mm}
  \fmf{fermion,label={\small $p_1$}}{i1,v2}
  \fmf{fermion,label={\small $p_2$}}{o1,v3}
  \fmf{fermion,label={\small $p_3$}}{o2,v4}
  \fmf{phantom,tension=0.5}{v2,v3p,v2p,v3}
  \fmf{phantom,tension=0.5}{v4,v1p,v4p,v2}
  \fmffreeze
  \fmf{fermion,label={\small $q_5$},label.side=right,right=0.5}{v2,v1p}
  \fmf{fermion,rubout,label={\small $q_3$},label.side=left}{v4,v3}
  \fmf{fermion,label={\small $q_2$},label.side=left}{v3,v2}
  \fmf{fermion,label={\small $q_4$},label.side=left}{v1p,v4}
  \fmf{fermion,label={\small $q_1$},right=0.5}{v1p,v2}
\end{fmfgraph*}
}
\caption{\small Topology $\calt_{23\_3\_1\_1}$ ($e112|3|e3|e|$).}
\label{figure5} 
\end{figure}
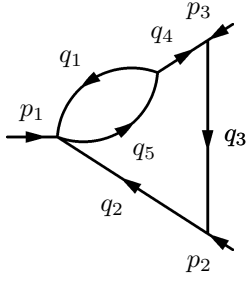

\noindent
The internal momenta are parameterised in the following way
\begin{align}
  q_1 &= k_1 + \hr_1 \notag \\
  q_2 &= k_2 + \hr_2 \notag \\
  q_3 &= k_2 + \hr_3 \notag \\
  q_4 &= k_2 + \hr_4 \notag \\
  q_5 &= k_1 + k_2 + \hr_5
  \label{eqdefqint5t}
\end{align}
leading to the sets $S_i$
\begin{equation}
  S_1 = \{1\} \quad S_2 = \{2,3,4\} \quad S_3 = \{5\} 
  \label{eqdefsets5t}
\end{equation}
and to the matrix $A$ through eq.~(\ref{eqdefA0}).
We choose $\br_1 = \hr_1$ and $\br_2 = \hr_2$. This choice leads to
\begin{align}
  \bB &= \left[
  \begin{array}{cc}
    \tau_5 & 0 \\
    \tau_4+\tau_5 & \tau_3
  \end{array}
\right]
  \label{eqdefBbar5t}
\end{align}
and
\begin{align}
  \Gamma &= \left[
  \begin{array}{cc}
    \tau_4+\tau_5 & 0 \\
    0 & \tau_3
  \end{array}
\right]
  \label{eqdefGamma5t}
\end{align}
The vector $T$ is taken to be
\begin{align}
  T &= \left[
  \begin{array}{c}
    p_1 \\
    -p_2
  \end{array}
\right]
  \label{eqdefT5t}
\end{align}
Using eq.~(\ref{eqdefpolycalf}), parameterising the Feynman parameters $\tau_i$ in the following way
\begin{align}
  \tau_1 &= \rho_1  & \tau_2 &= \rho_2 \, u_1 \notag \\
  \tau_3 &= \rho_2 \, u_2 & \tau_4 &= \rho_2 \, (1-u_1-u_2) \notag \\
  \tau_5 &= \rho_3 & 
  \label{eqreparam5t}
\end{align}
and performing the following change of variable $\rho_1 = \rho \, \xi$, $\rho_3 = \rho \, (1 - \xi)$ the two-loop scalar amplitude becomes
\begin{align}
  ^{(2)}I_{3}^{n}\left( \kappa_e \, ; \calt_{23\_3\_1\_1} \right) &= (4 \, \pi)^{-4 + 2 \, \varepsilon} \, \Gamma(1 + 2 \, \varepsilon) \int_0^1 d \rho \, \int_0^1 d \xi \, \left( 1 - \rho + \rho \, \xi (1-\xi) \right)^{-1+3 \, \varepsilon} \, \rho^{-1 + \varepsilon} \, (1-\rho)^2 \notag \\
  &\quad {} \times \int_{\Sigma_{(2)}} d u_1 \, d u_2 \left( \calft(u_1,u_2,\rho,\xi) - i \, \lambda \right)^{-1 - 2 \, \varepsilon}
  \label{eqUV5t6}
\end{align}
where $\kappa_e = \{p_1^2,p_2^2,p_3^2,m_1^2,m_2^2,m_3^2,m_4^2,m_5^2\}$ and
\begin{align}
  \calft(u_1,u_2,\rho,\xi) &= U^T \cdot \tG_e \cdot U - 2 \, \tV^T_e \cdot U - \tC_e
  \label{eqUV5t7}
\end{align}
with
\begin{align}
  \tG_e &= (1-\rho)^2 \, \left[
  \begin{array}{cc}
    p_1^2 & 1/2 \, (p_3^2 + p_1^2 - p_2^2) \\
    1/2 \, (p_3^2 + p_1^2 - p_2^2) & p_3^2
  \end{array}
\right] \label{eqdefaGt5t} \\
\tV_e &= \frac{1}{2} \, (1 - \rho) \, (1 - \rho + \rho \, \xi \, (1-\xi)) \, \left[
\begin{array}{c}
  p_1^2 - m_2^2 + m_4^2 \\
  p_3^2 - m_3^2 + m_4^2
\end{array}
\right] \label{eqdefVt5t} \\
\tC_e &= - (1 - \rho + \rho \, \xi \, (1-\xi)) \, \left( \rho  \, \xi \, m_1^2 + (1-\rho) \, m_4^2 + \rho \, (1-\xi) \, m_5^2 \right)
  \label{eqdefCt5t}
\end{align}
The coefficients $\tG_d$, $\tV_d$ and $\tC_d$ of the preceding subsection can be recover by setting $p_1=0$ and $p_2=p_3=p$.
\\

\noindent
Note that due to the choice of the $n$-momenta $\br_1$ and $\br_2$ the matrix $\bB$ is not explicitly homogeneous of degree $1$ in the variables $u_1$ and $u_2$. Nevertheless the factorisation of $1 - \rho + \rho \, \xi \, (1-\xi)$ for the coefficients $\tV_e$ and $\tC_e$ holds. Using the transformation given by eq.~(A.12) of Appendix A of ref.~\cite{letter}, we can introduce a matrix $\bB^{(0)}$ given by
\begin{align}
  \bB^{(0)} &\equiv A \cdot \Delta \notag \\
  &= \left[
  \begin{array}{cc}
    \tau_1 + \tau_5 & \tau_5 \\
    \tau_5 & \tau_2 + \tau_3 + \tau_4 + \tau_5
  \end{array}
\right] \cdot \left[
\begin{array}{cc}
  \alpha_1 & \alpha_2 \\
  \beta_1 & \beta_2
\end{array}
\right] 
  \label{eqdefbB05t}
\end{align}
Setting $\alpha_1 = \alpha_2 = \beta_2 = 0$ and $\beta_1=1$, we end with
\begin{align}
  \bB^{\prime} &\equiv \bB - \bB^{(0)} \notag \\
  &= \left[
  \begin{array}{cc}
    0 & 0 \\
    -(\tau_2+\tau_3) & \tau_3
  \end{array}
\right]
  \label{eqdefbBprime5t}
\end{align}
The new matrix $\bB^{\prime}$ is explicitly homogeneous of degree $1$ in the variables $u_1$ and $u_2$. With the help of eq.~(A.16) of ref.~\cite{letter}, the new matrix $\Gamma^{\prime}$ reads
\begin{align}
  \Gamma^{\prime} &\equiv \Delta^T \cdot A \cdot \Delta - \Delta^T \cdot \bB - \bB^T \cdot \Delta + \Gamma \notag \\
  &= \left[
  \begin{array}{cc}
    \tau_2+\tau_3 & - \tau_3 \\
    -\tau_3 & \tau_3
  \end{array}
\right]
  \label{eqdefGammaprime5t}
\end{align}
This is an example on how the transformation given by eq.~(A.12) of ref.~\cite{letter} works.\\

\noindent
Since this topology is a ``child'' of the topology $\calt_{22\_4\_0\_1}$, the two-loop amplitude is given by the right-hand side of eq.~(\ref{eqUV5ts9}) but with different coefficients for the polynomial in $u_1$ and $u_2$
\begin{align}
\hspace{2em}&\hspace{-2em}^{(2)}I_{3}^{n}\left( \kappa_e \, ; \calt_{23\_3\_1\_1} \right) \notag \\
&= (4 \, \pi)^{-4 + 2 \, \varepsilon} \, \Gamma(1 + 2 \, \varepsilon) \,  \left\{  \frac{1}{\varepsilon} \, \wtI_{3}^{\,(0)}(\Sigma_{(2)};\tG_e,\tV_e,\tC_e,0,\xi) - 2 \, \wtI_{3}^{\,(1)}(\Sigma_{(2)};\tG_e,\tV_e,\tC_e,0,\xi) \right. \notag \\
&\quad {} + \int_0^1 d \xi \, \int_0^1 \frac{d \rho}{\rho} \left[ \frac{(1-\rho)^2}{1 - \rho + \rho \, \xi \, (1-\xi)} \, \wtI_{3}^{\,(0)}(\Sigma_{(2)};\tG_e,\tV_e,\tC_e,\rho,\xi) \right. \notag \\ 
&\qquad \qquad \qquad \qquad \quad {} - \left. \left. \wtI_{3}^{\,(0)}(\Sigma_{(2)};\tG_e,\tV_e,\tC_e,0,\xi) \vphantom{\frac{(1-\rho)^2}{1 - \rho + \rho \, \xi \, (1-\xi)}} \right] \right\}
  \label{eqUV5t9}
\end{align}
Some comments are in order. Firstly, the coefficient of the divergent term is proportional to the one-loop three-point function obtained by shrinking the bubble in fig.~\ref{figure5} as it should be.
Secondly, the generalised one-loop three-point function $\wtI_{3}^{\,(0)}(\Sigma_{(2)};\tG_d,\tV_d,\tC_d,\rho,\xi)$ coincide with the genuine one-loop three-point function with no restriction on the kinematics, it is given in ref.~\cite{paper1}. But the generalised one-loop three-point function of order $\varepsilon$, $\wtI_{3}^{\,(1)}(\Sigma_{(2)};\tG_d,\tV_d,\tC_d,0,\xi)$ with $\det(\tG_d) \ne 0$  is new. Its analytical formula as well as the derivation to get it are given in appendix~\ref{new_integral}.

\subsection{Topology $\calt_{24\_2\_2\_1}$}\label{t24221}

This topology is built by inserting an external leg to the three-leg vertex connecting three internal lines in the topology $\calt_{23\_3\_1\_1}$. The kinematics will be computed explicitly.

\vspace{1cm}

\begin{figure}[h]
\centering
\parbox[c][35mm][t]{35mm}{%
  \begin{fmfgraph*}(35,35)
  \fmfstraight
  \fmfleftn{i}{2} \fmfrightn{o}{2}
  \fmfset{arrow_len}{3mm}
  \fmf{fermion,label={\small $p_1$}}{i2,v2}
  \fmf{fermion,label={\small $p_2$}}{i1,v1}
  \fmf{fermion,label={\small $p_3$}}{o1,v3}
  \fmf{fermion,label={\small $p_4$}}{o2,v4}
  \fmf{phantom,tension=0.5}{v2,v4}
  \fmf{fermion,tension=0.5,label={\small $q_4$},label.side=left}{v4,v3}
  \fmf{fermion,tension=0.5,label={\small $q_3$},label.side=left}{v3,v1}
  \fmf{fermion,tension=0.5,label={\small $q_2$},label.side=left}{v1,v2}
  \fmffreeze
  \fmf{fermion,label={\small $q_5$},label.side=right,right=0.5}{v2,v4}
  \fmf{fermion,label={\small $q_1$},right=0.5}{v4,v2}
\end{fmfgraph*}
}
\caption{\small Topology $\calt_{24\_2\_2\_1}$ ($e112|e3|e3|e|$).}
\label{figure7} 
\end{figure}
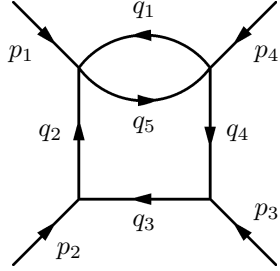

\noindent
The internal momenta are parameterised in the following way
\begin{align}
  q_1 &= k_1 + \hr_1 \notag \\
  q_2 &= k_2 + \hr_2 \notag \\
  q_3 &= k_2 + \hr_3 \notag \\
  q_4 &= k_2 + \hr_4 \notag \\
  q_5 &= k_1 + k_2 + \hr_5
  \label{eqdefqint7t}
\end{align}
leading to the sets $S_i$
\begin{equation}
  S_1 = \{1\} \quad S_2 = \{2,3,4\} \quad S_3 = \{5\} 
  \label{eqdefsets7t}
\end{equation}
and to the matrix $A$ through eq.~(\ref{eqdefA0}).
We choose $\br_1 = \hr_1$ and $\br_2 = \hr_2$. This choice leads to
\begin{align}
  \bB &= \left[
  \begin{array}{ccc}
    \tau_5 & 0 & 0\\
    \tau_5 & \tau_4 & \tau_3
  \end{array}
\right]
  \label{eqdefBbar7t}
\end{align}
and
\begin{align}
  \Gamma &= \left[
  \begin{array}{ccc}
    \tau_5 & 0  & 0 \\
    0 & \tau_4 & 0 \\
    0 & 0 & \tau_3
  \end{array}
\right]
  \label{eqdefGamma7t}
\end{align}
For this topology, the vector $T$ is chosen to be
\begin{align}
  T &= \left[
  \begin{array}{c}
    p_1 \\
    p_1+p_4 \\
    p_1+p_3+p_4
  \end{array}
\right]
  \label{eqdefT7t}
\end{align}

\noindent
Using eq.~(\ref{eqdefpolycalf}), and parameterising the Feynman parameters $\tau_i$ in the following way
\begin{align}
  \tau_1 &= \rho_1  & \tau_2 &= \rho_2 \, u_1 \notag \\
  \tau_3 &= \rho_2 \, u_2 & \tau_4 &= \rho_2 \, (1-u_1-u_2) \notag \\
  \tau_5 &= \rho_3 & 
  \label{eqreparam7t}
\end{align}
and performing the change of variable $\rho_1 = \rho \, \xi$ and $\rho_3 = \rho \, (1-\xi)$ as for the topology $\calt_{23\_3\_1\_1}$, this leads to
\begin{align}
  ^{(2)}I_{4}^{n}\left( \kappa_f \, ; \calt_{24\_2\_2\_1} \right) &= (4 \, \pi)^{-4 + 2 \, \varepsilon} \, \Gamma(1 + 2 \, \varepsilon) \int_0^1 d \rho \, \int_0^1 d \xi \, \left( 1 - \rho + \rho \, \xi (1-\xi) \right)^{-1+3 \, \varepsilon} \, \rho^{-1 + \varepsilon} \, (1-\rho)^2 \notag \\
  &\quad {} \times \int_{\Sigma_{(2)}} d u_1 \, d u_2 \left( \calft(u_1,u_2,\rho,\xi) - i \, \lambda \right)^{-1 - 2 \, \varepsilon}
  \label{eqUV7t6}
\end{align}
where $\kappa_f = \{p_1^2,p_2^2,p_3^2,p_4^2,s,t,m_1^2,m_2^2,m_3^2,m_4^2,m_5^2\}$ with $s = (p_1+p_2)^2$ and $t=(p_1+p_4)^2$.
For this topology, the function $\calft(u_1,u_2,\rho,\xi)$ is given by
\begin{align}
  \calft(u_1,u_2,\rho,\xi) &= U^T \cdot \tG_f \cdot U - 2 \, \tV_f \cdot U - \tC_f
  \label{eqUV7t7}
\end{align}
with
\begin{align}
  \tG_f &= \frac{1}{2} \, (1-\rho)^2 \, \left[
  \begin{array}{cc}
    2 \, t & t + p_3^2 - p_2^2 \\
    t + p_3^2 - p_2^2 & 2 \, p_3^2
  \end{array}
\right] \label{eqdefaGt7t} \\
\tV_f &= \frac{1}{2} \, (1 - \rho) \, \left[
\begin{array}{c}
  \rho \, \xi \, (1-\xi) \, (p_1^2 - p_4^2 - t) + (1 - \rho + \rho \, \xi \, (1-\xi)) \, (t - m_2^2 + m_4^2) \\
  \rho \, \xi \, (1-\xi) \, (s - p_4^2 - p_3^2) + (1 - \rho + \rho \, \xi \, (1-\xi))\, (p_3^2 - m_3^2 + m_4^2)
\end{array}
\right] \label{eqdefVt7t} \\
\tC_f &= \rho \, (1-\rho) \, \xi \, (1-\xi) \, p_4^2 - (1 - \rho + \rho \, \xi \, (1-\xi)) \, \left( \rho  \, \xi \, m_1^2 + (1-\rho) \, m_4^2 + \rho \, (1-\xi) \, m_5^2 \right)
  \label{eqdefCt7t}
\end{align}
Note that this case has two four-leg vertices connecting three internal lines thus there is no factorisation of $(1 - \rho + \rho \, \xi \, (1-\xi))$ in the components of the vector $\tV_f$ or in the scalar $\tC_f$.
This topology is similar to the topology $\calt_{23\_3\_1\_1}$, the only differences are the coefficients of the polynomial $\calf(u_1,u_2,\rho,\xi)$ in $u_1$ and $u_2$.
The two-loop four-point function is thus given by the right-hand side of eq.~(\ref{eqUV5ts9}) using the kinematics given by eqs.~(\ref{eqdefaGt7t}), (\ref{eqdefVt7t}) and (\ref{eqdefCt7t}) 
\begin{align}
\hspace{2em}&\hspace{-2em}^{(2)}I_{4}^{n}\left( \kappa_f \, ; \calt_{24\_2\_2\_1} \right) \notag \\
&= (4 \, \pi)^{-4 + 2 \, \varepsilon} \, \Gamma(1 + 2 \, \varepsilon) \,  \left\{  \frac{1}{\varepsilon} \, \wtI_{3}^{\,(0)}(\Sigma_{(2)};\tG_f,\tV_f,\tC_f,0,\xi) - 2 \, \wtI_{3}^{\,(1)}(\Sigma_{(2)};\tG_f,\tV_f,\tC_f,0,\xi) \right. \notag \\
&\quad {} + \int_0^1 d \xi \, \int_0^1 \frac{d \rho}{\rho} \left[ \frac{(1-\rho)^2}{1 - \rho + \rho \, \xi \, (1-\xi)} \, \wtI_{3}^{\,(0)}(\Sigma_{(2)};\tG_f,\tV_f,\tC_f,\rho,\xi) \right. \notag \\ 
&\qquad \qquad \qquad \qquad \quad {} - \left. \left. \wtI_{3}^{\,(0)}(\Sigma_{(2)};\tG_f,\tV_f,\tC_f,0,\xi) \vphantom{\frac{(1-\rho)^2}{1 - \rho + \rho \, \xi \, (1-\xi)}} \right] \right\}
  \label{eqUV7t8}
\end{align}
\\

\noindent
Note that the result of sec.~\ref{t23311} can be retrieved by setting $p_4 = 0$ ($t=p_1^2$ and $s=p_3^2$) in the latter equations.
In principle, setting $p_1=0$ would also lead to a kinematics of the same type as the one of sec.~\ref{t23311} but it is not easy to see that on eqs.~(\ref{eqdefaGt7t}), (\ref{eqdefVt7t}) and (\ref{eqdefCt7t}). Nevertheless, the transformation
\begin{align}
  u_1 &= 1 - w_1 - w_2 \notag \\
  u_2 &= w_2
  \label{eqtransf1}
\end{align}
lets invariant the integration volume. At the end of this transformation, the new polynomial in the integration variables $w_1$ and $w_2$ has the following structure
\begin{align}
  \calft(w_1,w_2,\rho,\xi) &= W^T \cdot \tG_f'' \cdot W - 2 \, \tV_f'' \cdot W - \tC_f''
  \label{eqUV7t7d}
\end{align}
with
\[
 W = \left[
\begin{array}{c}
  w_1 \\
  w_2
\end{array}
\right]
\]
and
\begin{align}
  \tG_f'' &= (1-\rho)^2 \, \frac{1}{2} \, \left[
  \begin{array}{cc}
    2 \, t & t + p_2^2 - p_3^2 \\
    t + p_2^2 - p_3^2 & 2 \, p_2^2
  \end{array}
\right] \label{eqdefaGt7td} \\
\tV_f'' &= \frac{1}{2} \, (1 - \rho) \, \left[
\begin{array}{c}
  \rho \, \xi \, (1-\xi) \, (p_4^2 - p_1^2 - t) + (1 - \rho + \rho \, \xi \, (1-\xi)) \, (t - m_4^2 + m_2^2) \\
  \rho \, \xi \, (1-\xi) \, (s - p_1^2 - p_2^2) + (1 - \rho + \rho \, \xi \, (1-\xi))\, (p_2^2 - m_3^2 + m_2^2)
\end{array}
\right] \label{eqdefVt7td} \\
\tC_f'' &= \rho \, (1-\rho) \, \xi \, (1-\xi) \, p_1^2 - (1 - \rho + \rho \, \xi \, (1-\xi)) \, \left( \rho  \, \xi \, m_1^2 + (1-\rho) \, m_2^2 + \rho \, (1-\xi) \, m_5^2 \right)
  \label{eqdefCt7td}
\end{align}
The last equations~(\ref{eqdefaGt7td}), (\ref{eqdefVt7td}) and (\ref{eqdefCt7td}) are more handy for the limit $p_1=0$. This is due to the mirror symmetry of the diagram depicted in fig.~\ref{figure7}, namely
\begin{align}
  p_1 &\leftrightarrow p_4 \\
  p_2 &\leftrightarrow p_3 \\
  m_2 &\leftrightarrow m_4
  \label{eqmirrsym1}
\end{align}

\subsubsection{Other UV divergent topologies with $I=5$}

As in sect.~\ref{secIeq4}, other UV divergent topologies can be generated by replacing the three-leg vertices connecting two internal legs with one external leg in the topologies $\calt_{23\_3\_1\_1}$ and $\calt_{24\_2\_2\_1}$ by four-leg vertices leading to the following topologies. They are presented in figs.~\ref{figure5sp}, \ref{figure5p} and \ref{figure7p}.
They correspond to a trivial extension of the kinematics from the one related to topologies $\calt_{23\_3\_1\_1}$ and $\calt_{24\_2\_2\_1}$.


\begin{figure}[h]
\centering
\parbox[c][40mm][t]{30mm}{%
\begin{fmfgraph*}(35,30)
  \fmfstraight
  \fmfleftn{i}{6} \fmfrightn{o}{1}
  \fmfset{arrow_len}{3mm}
  \fmf{fermion}{i3,v2}
  \fmf{fermion}{i4,v2}
  \fmfv{label={\small $p_1$}}{i3}
  \fmfv{label={\small $p_2$}}{i4}
  \fmf{fermion,label={\small $p_3$},label.side=left}{o1,v3}
  \fmf{phantom,tension=0.3,left}{v2,v3}
  \fmf{phantom,tension=0.3,left}{v3,v2}
  \fmfposition
  \fmfipath{p}
  \fmfipath{pp}
  \fmfiset{p}{vpath(__v3,__v2)}
  \fmfiset{pp}{vpath(__v2,__v3)}
  \fmfi{fermion,label={\small $q_3$}}{subpath (0,length(p)) of p}
  \fmfi{fermion,label={\small $q_2$}}{subpath (0,length(pp)/4) of pp}
  \fmfi{fermion,label={\small $q_1$}}{subpath (3length(pp)/4,length(pp)/4) of pp}
  \fmfi{fermion,label={\small $q_4$}}{subpath (3length(pp)/4,length(pp)) of pp}
  \fmfi{fermion,label={\small $q_5$},left}{subpath (3length(pp)/4,length(pp)/4) of pp rotated 180 shifted (0.84w,1.221w)}
  \fmfv{label={\small $\calt_{23\_3\_1\_3}$ ($ee12|e3|33||$)},label.dist=0.65w,label.angle=-70}{v2}
\end{fmfgraph*}
}
\qquad \qquad \qquad \qquad
\parbox[c][40mm][t]{30mm}{%
\begin{fmfgraph*}(35,30)
  \fmfstraight
  \fmfleftn{i}{6} \fmfrightn{o}{6}
  \fmfset{arrow_len}{3mm}
  \fmf{fermion}{i3,v2}
  \fmf{fermion}{i4,v2}
  \fmfv{label={\small $p_1$}}{i3}
  \fmfv{label={\small $p_2$}}{i4}
  \fmf{fermion}{o3,v3}
  \fmf{fermion}{o4,v3}
  \fmfv{label={\small $p_3$}}{o3}
  \fmfv{label={\small $p_4$}}{o4}
  \fmf{phantom,tension=0.4,left}{v2,v3}
  \fmf{phantom,tension=0.4,left}{v3,v2}
  \fmfposition
  \fmfipath{p}
  \fmfipath{pp}
  \fmfiset{p}{vpath(__v3,__v2)}
  \fmfiset{pp}{vpath(__v2,__v3)}
  \fmfi{fermion,label={\small $q_3$}}{subpath (0,length(p)) of p}
  \fmfi{fermion,label={\small $q_2$}}{subpath (0,length(pp)/4) of pp}
  \fmfi{fermion,label={\small $q_1$}}{subpath (3length(pp)/4,length(pp)/4) of pp}
  \fmfi{fermion,label={\small $q_4$}}{subpath (3length(pp)/4,length(pp)) of pp}
  \fmfi{fermion,label={\small $q_5$},left}{subpath (3length(pp)/4,length(pp)/4) of pp rotated 180 shifted (1.00w,1.237w)}
  \fmfv{label={\small $\calt_{24\_2\_2\_3}$ ($ee12|ee3|33||$)},label.dist=0.65w,label.angle=-70}{v2}
\end{fmfgraph*}
}
\caption{\small Trivial extensions from the primary topology $\calt_{22\_4\_0\_1}$.}
\label{figure5sp} 
\end{figure}


\begin{figure}[h]
\centering
\parbox[c][40mm][t]{30mm}{%
  \begin{fmfgraph*}(35,30)
  \fmfstraight
  \fmfleftn{i}{1} \fmfrightn{o}{6}
  \fmfset{arrow_len}{3mm}
  \fmf{fermion,label={\small $p_1$}}{i1,v2}
  \fmf{fermion}{o1,v3}
  \fmf{fermion}{o6,v4}
  \fmfv{label={\small $p_3$}}{o6}
  \fmfv{label={\small $p_2$}}{o1}
  \fmf{phantom,tension=0.5}{v2,v3p,v2p,v3}
  \fmf{phantom,tension=0.5}{v4,v1p,v4p,v2}
  \fmffreeze
  \fmf{fermion,label={\small $q_5$},label.side=right,right=0.5}{v2,v1p}
  \fmf{fermion,label={\small $q_1$},right=0.5}{v1p,v2}
  \fmf{fermion,label={\small $q_3$},label.side=left}{v4,v3}
  \fmf{fermion,label={\small $q_2$},label.side=left}{v3,v2}
  \fmf{fermion,label={\small $q_4$},label.side=left}{v1p,v4}
  \fmf{fermion}{o5,v4}
  \fmfv{label={\small $p_4$}}{o5}
  \fmfv{label={\small $\calt_{24\_2\_2\_4}$ ($e112|3|e3|ee|$)},label.dist=0.65w,label.angle=-70}{v2}
\end{fmfgraph*}
}
\qquad \qquad \qquad
\parbox[c][40mm][t]{30mm}{%
  \begin{fmfgraph*}(35,30)
  \fmfstraight
  \fmfleftn{i}{1} \fmfrightn{o}{6}
  \fmfset{arrow_len}{3mm}
  \fmf{fermion,label={\small $p_1$}}{i1,v2}
  \fmf{fermion}{o1,v3}
  \fmf{fermion}{o6,v4}
  \fmfv{label={\small $p_3$}}{o6}
  \fmfv{label={\small $p_2$}}{o1}
  \fmf{phantom,tension=0.5}{v2,v3p,v2p,v3}
  \fmf{phantom,tension=0.5}{v4,v1p,v4p,v2}
  \fmffreeze
  \fmf{fermion,label={\small $q_5$},label.side=right,right=0.5}{v2,v1p}
  \fmf{fermion,label={\small $q_1$},right=0.5}{v1p,v2}
  \fmf{fermion,label={\small $q_3$},label.side=left}{v4,v3}
  \fmf{fermion,label={\small $q_2$},label.side=left}{v3,v2}
  \fmf{fermion,label={\small $q_4$},label.side=left}{v1p,v4}
  \fmf{fermion}{o2,v3}
  \fmfv{label={\small $p_4$}}{o2}
  \fmfv{label={\small $\calt_{24\_2\_2\_5}$ ($e112|3|ee3|e|$)},label.dist=0.65w,label.angle=-70}{v2}
\end{fmfgraph*}
}
\qquad \qquad \qquad
\parbox[c][40mm][t]{30mm}{%
  \begin{fmfgraph*}(35,30)
  \fmfstraight
  \fmfleftn{i}{1} \fmfrightn{o}{6}
  \fmfset{arrow_len}{3mm}
  \fmf{fermion,label={\small $p_1$}}{i1,v2}
  \fmf{fermion}{o1,v3}
  \fmf{fermion}{o6,v4}
  \fmfv{label={\small $p_3$}}{o6}
  \fmfv{label={\small $p_2$}}{o1}
  \fmf{phantom,tension=0.5}{v2,v3p,v2p,v3}
  \fmf{phantom,tension=0.5}{v4,v1p,v4p,v2}
  \fmffreeze
  \fmf{fermion,label={\small $q_5$},label.side=right,right=0.5}{v2,v1p}
  \fmf{fermion,label={\small $q_1$},right=0.5}{v1p,v2}
  \fmf{fermion,label={\small $q_3$},label.side=left}{v4,v3}
  \fmf{fermion,label={\small $q_2$},label.side=left}{v3,v2}
  \fmf{fermion,label={\small $q_4$},label.side=left}{v1p,v4}
  \fmf{fermion}{o2,v3}
  \fmf{fermion}{o5,v4}
  \fmfv{label={\small $p_4$}}{o2}
  \fmfv{label={\small $p_5$}}{o5}
  \fmfv{label={\small $\calt_{25\_1\_3\_1}$ ($e112|3|ee3|ee|$)},label.dist=0.65w,label.angle=-70}{v2}
\end{fmfgraph*}
}
\caption{\small Trivial extensions from the primary topology $\calt_{23\_3\_1\_1}$.}
\label{figure5p} 
\end{figure}

\begin{figure}[h]
\centering
\parbox[c][45mm][t]{35mm}{%
  \begin{fmfgraph*}(40,35)
  \fmfstraight
  \fmfleftn{i}{2} \fmfrightn{o}{6}
  \fmfset{arrow_len}{3mm}
  \fmf{fermion,label={\small $p_1$}}{i2,v2}
  \fmf{fermion,label={\small $p_2$}}{i1,v1}
  \fmf{fermion,label={\small $p_4$}}{o6,v4}
  \fmf{fermion}{o1,v3}
  \fmfv{label={\small $p_3$}}{o1}
  \fmf{phantom,tension=0.5}{v2,v4}
  \fmf{fermion,tension=0.5,label={\small $q_4$},label.side=left}{v4,v3}
  \fmf{fermion,tension=0.5,label={\small $q_3$},label.side=left}{v3,v1}
  \fmf{fermion,tension=0.5,label={\small $q_2$},label.side=left}{v1,v2}
  \fmffreeze
  \fmf{fermion,label={\small $q_5$},label.side=right,right=0.5}{v2,v4}
  \fmf{fermion,label={\small $q_1$},right=0.5}{v4,v2}
  \fmf{fermion}{o2,v3}
  \fmfv{label={\small $p_5$}}{o2}
  \fmfv{label={\small $\calt_{25\_1\_3\_2}$ ($e12|ee3|e33|e|$)},label.dist=0.85w,label.angle=-70}{v2}
\end{fmfgraph*}
}
\qquad \qquad \qquad \qquad \qquad
\parbox[c][45mm][t]{35mm}{%
  \begin{fmfgraph*}(40,35)
  \fmfstraight
  \fmfleftn{i}{6} \fmfrightn{o}{6}
  \fmfset{arrow_len}{3mm}
  \fmf{fermion,label={\small $p_1$}}{i6,v2}
  \fmf{fermion}{i1,v1}
  \fmf{fermion,label={\small $p_4$}}{o6,v4}
  \fmf{fermion}{o1,v3}
  \fmfv{label={\small $p_2$}}{i1}
  \fmfv{label={\small $p_3$}}{o1}
  \fmf{phantom,tension=0.5}{v2,v4}
  \fmf{fermion,tension=0.5,label={\small $q_4$},label.side=left}{v4,v3}
  \fmf{fermion,tension=0.5,label={\small $q_3$},label.side=left}{v3,v1}
  \fmf{fermion,tension=0.5,label={\small $q_2$},label.side=left}{v1,v2}
  \fmffreeze
  \fmf{fermion,label={\small $q_5$},label.side=right,right=0.5}{v2,v4}
  \fmf{fermion,label={\small $q_1$},right=0.5}{v4,v2}
  \fmf{fermion}{o2,v3}
  \fmf{fermion}{i2,v1}
  \fmfv{label={\small $p_5$}}{o2}
  \fmfv{label={\small $p_6$}}{i2}
  \fmfv{label={\small $\calt_{26\_0\_4\_1}$ ($ee12|ee3|e33|e|$)},label.dist=0.85w,label.angle=-70}{v2}
\end{fmfgraph*}
}
\caption{\small  Trivial extensions from the topology $\calt_{24\_2\_2\_1}$.}
\label{figure7p} 
\end{figure}

\clearpage

\subsection{Topology $\calt_{22\_4\_0\_2}$}\label{t22402}

This is the second primary topology with $I=5$ mentioned at the beginning of sec.~\ref{secIeq5}, the corresponding amplitude is free of UV divergences.

\vspace{1cm}

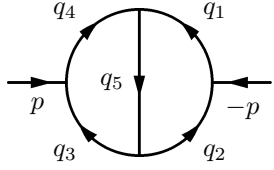
\begin{figure}[h]
\centering
\parbox[c][35mm][t]{30mm}{%
\begin{fmfgraph*}(35,30)
  \fmfstraight
  \fmfleftn{i}{1} \fmfrightn{o}{1}
  \fmfset{arrow_len}{3mm}
  \fmf{fermion,label={\small $p$},label.side=right}{i1,v2}
  \fmf{fermion,label={\small $-p$},label.side=left}{o1,v3}
  \fmf{phantom,tension=0.2,left}{v2,v3}
  \fmf{phantom,tension=0.2,left}{v3,v2}
  \fmfposition
  \fmfipath{p}
  \fmfipath{pp}
  \fmfiset{p}{vpath(__v3,__v2)}
  \fmfiset{pp}{vpath(__v2,__v3)}
  \fmfi{fermion,label={\small $q_2$}}{subpath (length(p)/2,0) of p}
  \fmfi{fermion,label={\small $q_3$}}{subpath (length(p)/2,length(p)) of p}
  \fmfi{fermion,label={\small $q_4$}}{subpath (0,length(pp)/2) of pp}
  \fmfi{fermion,label={\small $q_1$}}{subpath (length(pp),length(pp)/2) of pp}
  \fmfi{fermion,label={\small $q_5$}}{point length(pp)/2 of pp -- point length(p)/2 of p}
\end{fmfgraph*}
}
\caption{\small Topology $\calt_{22\_4\_0\_2}$ ($e12|23|3|e|$).}
\label{figure4} 
\end{figure}

\noindent
The internal momenta are parameterised in the follwing way
\begin{align}
  q_1 &= k_1 + \hr_1 \notag \\
  q_2 &= k_1 + \hr_2 \notag \\
  q_3 &= k_2 + \hr_3 \notag \\
  q_4 &= k_2 + \hr_4 \notag \\
  q_5 &= k_1 + k_2 + \hr_5
  \label{eqdefqint4t}
\end{align}
leading to the sets $S_i$
\begin{equation}
  S_1 = \{1,2\} \quad S_2 = \{3,4\} \quad S_3 = \{5\} 
  \label{eqdefsets4t}
\end{equation}
and, through eq.~(\ref{eqdefA0}), to the matrix $A$.
We choose $\br_1 = \hr_2$ and $\br_2 = \hr_3$. This choice leads to
\begin{align}
  \bB &= \left[
  \begin{array}{c}
    -\tau_1 \\
    \tau_4
  \end{array}
\right]
  \label{eqdefBbar4t}
\end{align}
and
\begin{align}
  \Gamma &= [ \tau_1+\tau_4]
  \label{eqdefGamma4t}
\end{align}
The vector $T$ is $T= [p]$.
Using eq.~(\ref{eqdefpolycalf}), we get
\begin{align}
  \calf(\{\tau_k\}) &= - p^2 \, \left[ (\tau_1+\tau_4) \, \tau_2 \, \tau_3 + (\tau_2+\tau_3) \, \tau_1 \, \tau_4 + \tau_5 \, (\tau_1+\tau_4) \, (\tau_2+\tau_3) \right] \notag \\
  &\quad {} + ( (\tau_1 + \tau_2) \, (\tau_3 + \tau_4 + \tau_5) + \tau_5 \, (\tau_3 + \tau_4)) \, (\tau_1 \, m_1^2 + \tau_2 \, m_2^2 + \tau_3 \, m_3^2 + \tau_4 \, m_4^2 + \tau_5 \, m_5^2)
  \label{eqdefcalf4t}
\end{align}
Parameterising the Feynman parameters $\tau_i$ in the following way
\begin{align}
  \tau_1 &= \rho_1 \, u_1  & \tau_2 &= \rho_1 \, (1-u_1) \notag \\
  \tau_3 &= \rho_2 \, (1-u_2) & \tau_4 &= \rho_2 \, u_2 \notag \\
  \tau_5 &= \rho_3 & 
  \label{eqreparam4t}
\end{align}
and performing the following change of variable $\rho_1 = \rho \, \xi$, $\rho_2 = \rho \, (1 - \xi)$ the two-loop scalar amplitude becomes
\begin{align}
  ^{(2)}I_{2}^{n}\left( \kappa_g \, ; \calt_{22\_4\_0\_2} \right) &= (4 \, \pi)^{-4 + 2 \, \varepsilon} \, \Gamma(1 + 2 \, \varepsilon) \int_0^1 d \rho \, \int_0^1 d \xi \, \left( 1 - \rho + \rho \, \xi (1-\xi) \right)^{-1+3 \, \varepsilon} \, \rho^{1 + \varepsilon} \, \xi \, (1-\xi) \notag \\
  &\quad {} \times \int_{K_{(2)}} d u_1 \, d u_2 \left( \calft(u_1,u_2,\rho,\xi) - i \, \lambda \right)^{-1 - 2 \, \varepsilon}
  \label{eqUV4t6}
\end{align}
where $\kappa_g = \{p^2,m_1^2,m_2^2,m_3^2,m_4^2,m_5^2\}$ and 
\begin{align}
  \calft(u_1,u_2,\rho,\xi) &= U^T \cdot \tG_g \cdot U - 2 \, \tV_g \cdot U - \tC_g
  \label{eqUV4t7}
\end{align}
with
\begin{align}
  \tG_g &= \rho \, p^2 \, \left[
  \begin{array}{cc}
    \xi^2 \, (1 - \rho \, \xi) & \xi \, (1-\xi) \, (1-\rho)  \\
    \xi \, (1-\xi) \, (1-\rho) & (1-\xi)^2 \, (1-\rho+\rho \, \xi)
  \end{array}
\right] \label{eqdefaGt4t} \\
\tV_g &= \frac{1}{2} \, \rho \, (1 - \rho + \rho \, \xi \, (1-\xi)) \, \left[
\begin{array}{c}
  \xi \, (p^2 - m_1^2 + m_2^2) \\
  (1-\xi) \, (p^2 - m_4^2 + m_3^2)
\end{array}
\right] \label{eqdefVt4t} \\
\tC_g &= - (1 - \rho + \rho \, \xi \, (1-\xi)) \, \left( \rho  \, \xi \, m_2^2 + \rho \, (1-\xi) \, m_3^2 + (1-\rho) \, m_5^2 \right)
  \label{eqdefCt4t}
\end{align}
It is clear from eq. (\ref{eqUV4t6}) that there is no UV divergence in this case, thus we can safely take $\varepsilon=0$ in this equation.
Note that the choice of the parameterisation defined in eq.~(\ref{eqreparam4t}) makes $\bB$ homogeneous of degree 1 with respect to the variables $u_i$, thus the factorisation of $1 - \rho + \rho \, \xi \, (1-\xi)$ for the coefficients of $\tV_g$ and $\tC_g$ in the polynomial of eq.~(\ref{eqUV4t7}).\\

\noindent
Thus the two-loop amplitude reads
\begin{align}
^{(2)}I_{2}^{4}\left( \kappa_g \, ; \calt_{22\_4\_0\_2} \right) &= (4 \, \pi)^{-4} \, \int_0^1 d \xi \, \int_0^1 d \rho \, \frac{\rho \, \xi \, (1-\xi)}{1 - \rho + \rho \, \xi \, (1-\xi)} \, \wtI_{3}^{\,(0)}(K_{(2)};\tG_g,\tV_g,\tC_g,\rho,\xi)
  \label{eqUV4t8}
\end{align}

\noindent
The ``generalised one-loop function'' $\wtI_{3}^{\,(0)}(K_{(2)};\tG_g,\tV_g,\tC_g,\rho,\xi)$ looks like a one-loop three-point function but the Feynman parameters runs through the unit square instead of the usual simplex. Its computation is presented in appendix \ref{ol3ptsq}.

\subsection{Topology $\calt_{23\_3\_1\_2}$}\label{t23312}

This topology is generated from the primary topology $\calt_{22\_4\_0\_2}$ by replacing one of the three-leg vertex connecting three internal legs by a four-leg vertex. Let us compute the related kinematics.

\vspace{1cm}

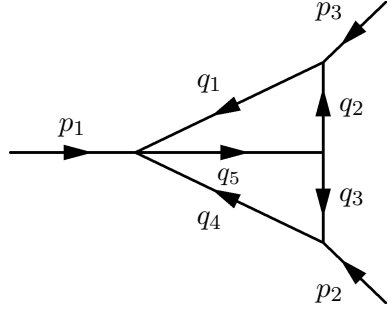
\begin{figure}[h]
\centering
\parbox[c][55mm][t]{40mm}{%
  \begin{fmfgraph*}(50,40)
  \fmfleftn{i}{1} \fmfrightn{o}{2}
  \fmf{fermion,tension=1.5,label=$p_1$}{i1,v1}
  \fmf{fermion,tension=1.5,label=$p_2$}{o1,v2}
  \fmf{fermion,tension=1.5,label=$p_3$}{o2,v3}
  \fmf{fermion,tension=0.5,label=$q_4$,label.side=left}{v2,v1}
  \fmf{fermion,tension=0.5,label=$q_3$,label.side=left}{v0,v2}
  \fmf{fermion,tension=0.5,label=$q_2$,label.side=right}{v0,v3}
  \fmf{fermion,tension=0.5,label=$q_1$,label.side=right}{v3,v1}
  \fmffreeze
  \fmf{fermion,label={\small $q_5$},label.side=right}{v1,v0}
\end{fmfgraph*}
}
\caption{\small Topology $\calt_{23\_3\_1\_2}$ ($e12|e23|3|e|$).}
\label{figure6} 
\end{figure}

\noindent
The internal momenta are parameterised in the following way
\begin{align}
  q_1 &= k_1 + \hr_1 \notag \\
  q_2 &= k_1 + \hr_2 \notag \\
  q_3 &= k_2 + \hr_3 \notag \\
  q_4 &= k_2 + \hr_4 \notag \\
  q_5 &= k_1 + k_2 + \hr_5
  \label{eqdefqint6t}
\end{align}
leading to the sets $S_i$
\begin{equation}
  S_1 = \{1,2\} \quad S_2 = \{3,4\} \quad S_3 = \{5\} 
  \label{eqdefsets6t}
\end{equation}
and to the matrix $A$ through eq.~(\ref{eqdefA0}).
We choose $\br_1 = \hr_2$ and $\br_2 = \hr_3$. This choice leads to
\begin{align}
  \bB &= \left[
  \begin{array}{cc}
    \tau_1 & 0 \\
    0 & \tau_4
  \end{array}
\right]
  \label{eqdefBbar6t}
\end{align}
and
\begin{align}
  \Gamma &= \left[
  \begin{array}{cc}
    \tau_1 & 0 \\
    0 & \tau_4
  \end{array}
\right]
  \label{eqdefGamma6t}
\end{align}
For this topology, the vector $T$ is chosen to be
\begin{align}
  T &= \left[
  \begin{array}{c}
    p_3 \\
    p_2
  \end{array}
\right]
  \label{eqdefT6t}
\end{align}

\noindent
With the help of eq.~(\ref{eqdefpolycalf}), parameterising the Feynman parameters $\tau_i$ in the following way
\begin{align}
  \tau_1 &= \rho_1 \, u_1  & \tau_2 &= \rho_1 \, (1-u_1) \notag \\
  \tau_3 &= \rho_2 \, (1-u_2) & \tau_4 &= \rho_2 \, u_2 \notag \\
  \tau_5 &= \rho_3 & 
  \label{eqreparam6t}
\end{align}
and performing the change of variable $\rho_1 = \rho \, \xi$, $\rho_2 = \rho \, (1 - \xi)$ which implies due to the constraint $\sum_{i=1}^3 \, \rho_i = 1$, $\rho_3 = 1 - \rho$, the two-loop amplitude becomes at $\varepsilon=0$
\begin{align}
  ^{(2)}I_{3}^{n}\left( \kappa_h \, ; \calt_{23\_3\_1\_2} \right) &= (4 \, \pi)^{-4} \, \int_0^1 d \rho \, \int_0^1 d \xi \, \frac{\rho \, \xi \, (1-\xi) }{1 - \rho + \rho \, \xi (1-\xi)} \, \int_{K_{(2)}} d u_1 \, d u_2 \left( \calft(u_1,u_2,\rho,\xi) - i \, \lambda \right)^{-1}
  \label{eqUV6t6}
\end{align}
where $\kappa_g = \{p_1^2,p_2^2,p_3^2,m_1^2,m_2^2,m_3^2,m_4^2,m_5^2\}$ and
\begin{align}
  \calft(u_1,u_2,\rho,\xi)  &= U^T \cdot \tG_h \cdot U - 2 \, \tV^T_h \cdot U - \tC_h
  \label{eqUV6t7}
\end{align}
with
\begin{align}
  \tG_h &= \frac{1}{2} \, \rho \, \left[
  \begin{array}{cc}
    2 \, \xi^2 \, (1 - \rho \, \xi) \, p_3^2 & \xi \, (1-\xi) \, (1-\rho) \, (p_3^2 + p_2^2 - p_1^2) \\
    \xi \, (1-\xi) \, (1-\rho) \, (p_3^2 + p_2^2 - p_1^2) & 2 \, (1-\xi)^2 \, (1-\rho+\rho \, \xi) \, p_2^2
  \end{array}
\right] \label{eqdefaGt6t} \\
\tV_h &= \frac{1}{2} \, \rho \, (1 - \rho + \rho \, \xi \, (1-\xi)) \, \left[
\begin{array}{c}
  \xi \, (p_3^2 - m_1^2 + m_2^2) \\
  (1-\xi) \, (p_2^2 - m_4^2 + m_3^2)
\end{array}
\right] \label{eqdefVt6t} \\
\tC_h &= - (1 - \rho + \rho \, \xi \, (1-\xi)) \, \left( \rho  \, \xi \, m_2^2 + \rho \, (1-\xi) \, m_3^2 + (1-\rho) \, m_5^2 \right)
  \label{eqdefCt6t}
\end{align}

\noindent
The two-loop amplitude is given by the right-hand side of eq.~(\ref{eqUV4t8}) but with a different kinematics
\begin{align}
^{(2)}I_{3}^{4}\left( \kappa_h \, ; \calt_{23\_3\_1\_2} \right) &= (4 \, \pi)^{-4} \, \int_0^1 d \xi \, \int_0^1 d \rho \, \frac{\rho \, \xi \, (1-\xi)}{1 - \rho + \rho \, \xi \, (1-\xi)} \, \wtI_{3}^{\,(0)}(K_{(2)};\tG_h,\tV_h,\tC_h,\rho,\xi)
  \label{eqUV6t8}
\end{align}
The results of the primary topology $\calt_{22\_4\_0\_2}$ can be recovered by setting $p_1=0$ which implies that $p_2=-p_3$ in eqs.~(\ref{eqdefaGt6t}), (\ref{eqdefVt6t}) and (\ref{eqdefCt6t}).

\subsection{Topology $\calt_{24\_2\_2\_2}$}\label{t24222}

This topology is generated from the primary topology $\calt_{22\_4\_0\_2}$ by replacing the two three-leg vertices connecting three internal legs by four-leg vertices. In this case also let us compute the related kinematics.

\vspace{1cm}

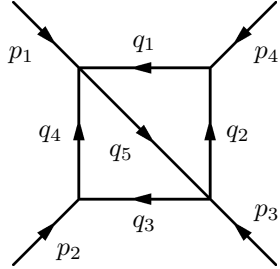
\begin{figure}[h]
\centering
\parbox[c][35mm][t]{35mm}{%
  \begin{fmfgraph*}(35,35)
  \fmfstraight
  \fmfleftn{i}{2} \fmfrightn{o}{2}
  \fmfset{arrow_len}{3mm}
  \fmf{fermion,label={\small $p_1$}}{i2,v2}
  \fmf{fermion,label={\small $p_2$}}{i1,v1}
  \fmf{fermion,label={\small $p_3$}}{o1,v3}
  \fmf{fermion,label={\small $p_4$}}{o2,v4}
  \fmf{fermion,tension=0.5,label={\small $q_1$},label.side=right}{v4,v2}
  \fmf{fermion,tension=0.5,label={\small $q_2$},label.side=right}{v3,v4}
  \fmf{fermion,tension=0.5,label={\small $q_4$},label.side=left}{v1,v2}
  \fmf{fermion,tension=0.5,label={\small $q_3$},label.side=left}{v3,v1}
  \fmffreeze
  \fmf{fermion,label={\small $q_5$},label.side=right}{v2,v3}
\end{fmfgraph*}
}
\caption{\small Topology $\calt_{24\_2\_2\_2}$ ($e12|e23|e3|e|$).}
\label{figure8} 
\end{figure}

\noindent
The internal momenta are parameterised in the following way
\begin{align}
  q_1 &= k_1 + \hr_1 \notag \\
  q_2 &= k_1 + \hr_2 \notag \\
  q_3 &= k_2 + \hr_3 \notag \\
  q_4 &= k_2 + \hr_4 \notag \\
  q_5 &= k_1 + k_2 + \hr_5
  \label{eqdefqint8t}
\end{align}
leading to the sets $S_i$
\begin{equation}
  S_1 = \{1,2\} \quad S_2 = \{3,4\} \quad S_3 = \{5\} 
  \label{eqdefsets8t}
\end{equation}
and, through eq.~(\ref{eqdefA0}), to the matrix $A$.
We choose $\br_1 = \hr_2$ and $\br_2 = \hr_3$. This choice leads to
\begin{align}
  \bB &= \left[
  \begin{array}{ccc}
    \tau_1 & 0 & \tau_5 \\
    0 & \tau_4 & \tau_5
  \end{array}
\right]
  \label{eqdefBbar8t}
\end{align}
and
\begin{align}
  \Gamma &= \left[
  \begin{array}{ccc}
    \tau_1 & 0 & 0 \\
    0 & \tau_4 & 0 \\
    0 & 0 & \tau_5
  \end{array}
\right]
  \label{eqdefGamma8t}
\end{align}
For this topology, we choose
\begin{align}
  T &= \left[
  \begin{array}{c}
    p_4 \\
    p_2 \\
    - p_3
  \end{array}
\right]
  \label{eqdefT8t}
\end{align}

\noindent
The use of eq.~(\ref{eqdefpolycalf}) and the following parameterisation of the Feynman parameters $\tau_i$
\begin{align}
  \tau_1 &= \rho_1 \, u_1  & \tau_2 &= \rho_1 \, (1-u_1) \notag \\
  \tau_3 &= \rho_2 \, (1-u_2) & \tau_4 &= \rho_2 \, u_2 \notag \\
  \tau_5 &= \rho_3 & 
  \label{eqreparam8t}
\end{align}
define the polynomial in $u_1$, $u_2$ in the integrand.
Furthermore the following change of variable is performed $\rho_1 = \rho \, \xi$, $\rho_2 = \rho \, (1 - \xi)$. Thus letting $\varepsilon \rightarrow 0$, the two-loop amplitude reads
\begin{align}
  ^{(2)}I_{4}^{4}\left( \kappa_i \, ; \calt_{24\_2\_2\_2} \right) &= (4 \, \pi)^{-4} \, \int_0^1 d \rho \, \int_0^1 d \xi \,\frac{\rho \, \xi \, (1-\xi) }{1 - \rho + \rho \, \xi (1-\xi)} \, \int_{K_{(2)}} d u_1 \, d u_2 \left( \calft(u_1,u_2,\rho,\xi) - i \, \lambda \right)^{-1}
  \label{eqUV8t6}
\end{align}
where $\kappa_i =\{p_1^2,p_2^2,p_3^2,p_4^2,s,t,m_1^2,m_2^2,m_3^2,m_4^2,m_5^2\}$ and $s=(p_1+p_2)^2$ and $t=(p_1+p_4)^2$. The polynomial in the variables $u_1$, $u_2$ is given by
\begin{align}
  \calft(u_1,u_2,\rho,\xi)  &= U^T \cdot \tG_i \cdot U - 2 \, \tV^T_i \cdot U - \tC_i
  \label{eqUV8t7}
\end{align}
with
\begin{align}
  \tG_i &= \frac{1}{2} \, \rho \, \left[
  \begin{array}{cc}
    2 \, \xi^2 \, (1 - \rho \, \xi) \, p_4^2 & \xi \, (1-\xi) \, (1-\rho) \, (p_2^2 + p_4^2 - u) \\
    \xi \, (1-\xi) \, (1-\rho) \, (p_2^2 + p_4^2 - u) & 2 \, (1-\xi)^2 \, (1-\rho+\rho \, \xi) \, p_2^2
  \end{array}
\right] \label{eqdefaGt8t} \\
\tV_i &= \frac{1}{2} \, \rho \, \left[
\begin{array}{c}
  \xi \, ( (1-\xi) \, (1-\rho) \, (s - p_3^2 - p_4^2) + (1 - \rho + \rho \, \xi \, (1-\xi))\, (p_4^2 - m_1^2 + m_2^2)) \\
  (1-\xi) \, ( \xi \, (1-\rho) \, (t - p_2^2 - p_3^2) + (1 - \rho + \rho \, \xi \, (1-\xi))\, (p_2^2 - m_4^2 + m_3^2))
\end{array}
\right] \label{eqdefVt8t} \\
\tC_i &= \rho \, (1-\rho) \, \xi \, (1-\xi) \, p_3^2 - (1 - \rho + \rho \, \xi \, (1-\xi)) \, \left( \rho  \, \xi \, m_2^2 + \rho \, (1-\xi) \, m_3^2 + (1-\rho) \, m_5^2 \right)
  \label{eqdefCt8t}
\end{align}
The Mandelstam variable $u$ is taken to be $u=(p_1+p_3)^2$. This topology, also, has two four-leg vertices connecting three internal lines, thus there is no factorisation of $1 - \rho + \rho \, \xi \, (1-\xi)$ in $\tV_i$ and $\tC_i$.\\

\noindent
The two-loop amplitude is still given by the right-hand side of eq.~(\ref{eqUV4t8}) but again with a different kinematics
\begin{align}
^{(2)}I_{4}^{4}\left( \kappa_i \, ; \calt_{24\_2\_2\_2} \right) &= (4 \, \pi)^{-4} \, \int_0^1 d \xi \, \int_0^1 d \rho \, \frac{\rho \, \xi \, (1-\xi)}{1 - \rho + \rho \, \xi \, (1-\xi)} \, \wtI_{3}^{\,(0)}(K_{(2)};\tG_i,\tV_i,\tC_i,\rho,\xi)
  \label{eqUV8t8}
\end{align}
The results of sec.~\ref{t23312} can be obtained by setting $p_3=0$, i.e. $s=p_4^2$, $t=p_2^2$ and $u=p_1^2$ ($p_4$ for this kinematics plays the role of $p_3$ for the kinematics related to topology $\calt_{23\_3\_1\_2}$). One could recover the results of sec.~\ref{t23312} by letting $p_1=0$, it is not straightforward to see that from eqs.~(\ref{eqdefaGt8t}), (\ref{eqdefVt8t}) and (\ref{eqdefCt8t}). It is better to use the transformation on the integration variables $U = L - W$ where
\[
  L = \left[
  \begin{array}{c}
    1 \\
    1
  \end{array}
\right]
\qquad W = \left[
\begin{array}{c}
  w_1 \\
  w_2
\end{array}
\right]
\]
This transformation lets the integration volume invariant. The polynomial in $u_1$ and $u_2$ becomes in terms of the new variables $w_1$ and $w_2$
\begin{align}
\calft(w_1,w_2,\rho,\xi)  &= W^T \cdot \tG_i \cdot W - 2 \, \left( L^T \cdot \tG_i - \tV^T_i \right) \cdot W - \left( \tC_i - L^T \cdot \tG_i \cdot L + 2 \, V^T_i \cdot L \right) \notag \\
&= W^T \cdot \tG_i''' \cdot W - 2 \, \tV_i''' \cdot W - \tC_i'''
  \label{eqnewtransf2}
\end{align}
with
\begin{align}
  \tG_i''' &= \tG_i \label{eqdefaGt8tt} \\
\tV_i''' &= \frac{1}{2} \, \rho \, \left[
\begin{array}{c}
  \xi \, ( (1-\xi) \, (1-\rho) \, (t - p_1^2 - p_4^2) + (1 - \rho + \rho \, \xi \, (1-\xi))\, (p_4^2 - m_2^2 + m_1^2)) \\
  (1-\xi) \, ( \xi \, (1-\rho) \, (s - p_1^2 - p_2^2) + (1 - \rho + \rho \, \xi \, (1-\xi))\, (p_2^2 - m_3^2 + m_4^2))
\end{array}
\right] \label{eqdefVt8tt} \\
\tC_i''' &= \rho \, (1-\rho) \, \xi \, (1-\xi) \, p_1^2 - (1 - \rho + \rho \, \xi \, (1-\xi)) \, \left( \rho  \, \xi \, m_1^2 + \rho \, (1-\xi) \, m_4^2 + (1-\rho) \, m_5^2 \right)
  \label{eqdefCt8tt}
\end{align}
This is more handy to use eqs~(\ref{eqdefaGt8tt}), (\ref{eqdefVt8tt}) and (\ref{eqdefCt8tt}) to take the limit $p_1 = 0$. This is related to the mirror symmetry of the diagram depicted in fig.~\ref{figure8}, namely
\begin{align}
  p_1 &\leftrightarrow p_3 \notag \\
  m_1 &\leftrightarrow m_2 \notag \\
  m_3 &\leftrightarrow m_4
  \label{eqmirrsymm2}
\end{align}
The kinematics of the sec.~\ref{t22402} can be recovered by letting $p_1=p_3=0$, $p_4=-p_2$, $s=p_2^2$, $t=p_2^2$ and $u=0$.

\subsubsection{Other topologies with $I=5$ free of UV divergences}

As in the preceding cases, some new topologies can be generated from the topologies $\calt_{22\_4\_0\_2}$, $\calt_{23\_3\_1\_2}$ and $\calt_{24\_2\_2\_2}$ by changing one or two three-leg vertices connecting one external leg and two internal lines into four-leg vertices. The kinematics related to these topologies are trivial extensions of those related to the starting topologies, the former will not be discussed further. For the sake of completeness, they are presented in figs.~\ref{figure4s}, \ref{figure6s} and \ref{figure8s}.

\begin{figure}[h]
\centering
\parbox[c][45mm][t]{30mm}{%
\begin{fmfgraph*}(35,30)
  \fmfstraight
  \fmfleftn{i}{6} \fmfrightn{o}{1}
  \fmfset{arrow_len}{3mm}
  \fmf{fermion}{i3,v2}
  \fmf{fermion}{i4,v2}
  \fmfv{label={\small $p_1$}}{i3}
  \fmfv{label={\small $p_2$}}{i4}
  \fmf{fermion,label={\small $p_3$}}{o1,v3}
  \fmf{phantom,tension=0.2,left}{v2,v3}
  \fmf{phantom,tension=0.2,left}{v3,v2}
  \fmfposition
  \fmfipath{p}
  \fmfipath{pp}
  \fmfiset{p}{vpath(__v3,__v2)}
  \fmfiset{pp}{vpath(__v2,__v3)}
  \fmfi{fermion,label={\small $q_2$}}{subpath (length(p)/2,0) of p}
  \fmfi{fermion,label={\small $q_3$}}{subpath (length(p)/2,length(p)) of p}
  \fmfi{fermion,label={\small $q_4$}}{subpath (0,length(pp)/2) of pp}
  \fmfi{fermion,label={\small $q_1$}}{subpath (length(pp),length(pp)/2) of pp}
  \fmfi{fermion,label={\small $q_5$}}{point length(pp)/2 of pp -- point length(p)/2 of p}
  \fmfv{label={\small $\calt_{23\_3\_1\_4}$ ($ee12|23|3|e|$)},label.dist=0.65w,label.angle=-70}{v2}
\end{fmfgraph*}
}
\qquad \qquad \qquad \qquad
\parbox[c][45mm][t]{30mm}{%
\begin{fmfgraph*}(35,30)
  \fmfstraight
  \fmfleftn{i}{6} \fmfrightn{o}{6}
  \fmfset{arrow_len}{3mm}
  \fmf{fermion}{i3,v2}
  \fmf{fermion}{i4,v2}
  \fmfv{label={\small $p_1$}}{i3}
  \fmfv{label={\small $p_2$}}{i4}
  \fmf{fermion}{o3,v3}
  \fmf{fermion}{o4,v3}
  \fmfv{label={\small $p_3$}}{o3}
  \fmfv{label={\small $p_4$}}{o4}
  \fmf{phantom,tension=0.3,left}{v2,v3}
  \fmf{phantom,tension=0.3,left}{v3,v2}
  \fmfposition
  \fmfipath{p}
  \fmfipath{pp}
  \fmfiset{p}{vpath(__v3,__v2)}
  \fmfiset{pp}{vpath(__v2,__v3)}
  \fmfi{fermion,label={\small $q_2$}}{subpath (length(p)/2,0) of p}
  \fmfi{fermion,label={\small $q_3$}}{subpath (length(p)/2,length(p)) of p}
  \fmfi{fermion,label={\small $q_4$}}{subpath (0,length(pp)/2) of pp}
  \fmfi{fermion,label={\small $q_1$}}{subpath (length(pp),length(pp)/2) of pp}
  \fmfi{fermion,label={\small $q_5$}}{point length(pp)/2 of pp -- point length(p)/2 of p}
  \fmfv{label={\small $\calt_{24\_2\_2\_6}$ ($ee12|23|3|ee|$)},label.dist=0.65w,label.angle=-70}{v2}
\end{fmfgraph*}
}
\caption{\small Topologies coming from $\calt_{22\_4\_0\_2}$.}
\label{figure4s} 
\end{figure}

\begin{figure}[h]
\centering
\parbox[c][55mm][t]{40mm}{%
  \begin{fmfgraph*}(50,40)
  \fmfleftn{i}{1} \fmfrightn{o}{6}
  \fmf{fermion,tension=1.5,label=$p_1$}{i1,v1}
  \fmf{fermion,tension=1.5}{o1,v2}
  \fmf{fermion,tension=1.5}{o6,v3}
  \fmfv{label=$p_2$}{o1}
  \fmfv{label=$p_3$}{o6}
  \fmf{fermion,tension=0.5,label=$q_4$,label.side=left}{v2,v1}
  \fmf{fermion,tension=0.5,label=$q_3$,label.side=left}{v0,v2}
  \fmf{fermion,tension=0.5,label=$q_2$,label.side=right}{v0,v3}
  \fmf{fermion,tension=0.5,label=$q_1$,label.side=right}{v3,v1}
  \fmffreeze
  \fmf{fermion,label={\small $q_5$},label.side=right}{v1,v0}
  \fmf{fermion,tension=1.5}{o5,v3}
  \fmfv{label=$p_4$}{o5}
  \fmfv{label={\small $\calt_{24\_2\_2\_7}$ ($e123|e2|3|ee|$)},label.dist=0.65w,label.angle=-70}{v1}
\end{fmfgraph*}
}
\qquad \qquad \qquad \qquad
\parbox[c][55mm][t]{40mm}{%
  \begin{fmfgraph*}(50,40)
  \fmfleftn{i}{1} \fmfrightn{o}{6}
  \fmf{fermion,tension=1.5,label=$p_1$}{i1,v1}
  \fmf{fermion,tension=1.5}{o1,v2}
  \fmf{fermion,tension=1.5}{o6,v3}
  \fmfv{label=$p_2$}{o1}
  \fmfv{label=$p_3$}{o6}
  \fmf{fermion,tension=0.5,label=$q_4$,label.side=left}{v2,v1}
  \fmf{fermion,tension=0.5,label=$q_3$,label.side=left}{v0,v2}
  \fmf{fermion,tension=0.5,label=$q_2$,label.side=right}{v0,v3}
  \fmf{fermion,tension=0.5,label=$q_1$,label.side=right}{v3,v1}
  \fmffreeze
  \fmf{fermion,label={\small $q_5$},label.side=right}{v1,v0}
  \fmf{fermion,tension=1.5}{o2,v2}
  \fmf{fermion,tension=1.5}{o5,v3}
  \fmfv{label=$p_4$}{o2}
  \fmfv{label=$p_5$}{o5}
  \fmfv{label={\small $\calt_{25\_1\_3\_3}$ ($e123|ee2|3|ee|$)},label.dist=0.65w,label.angle=-70}{v1}
\end{fmfgraph*}
}
\caption{\small Topologies coming from $\calt_{23\_3\_1\_2}$.}
\label{figure6s} 
\end{figure}
\begin{figure}[h]
\centering
\parbox[c][50mm][t]{35mm}{%
  \begin{fmfgraph*}(35,35)
  \fmfstraight
  \fmfleftn{i}{2} \fmfrightn{o}{6}
  \fmfset{arrow_len}{3mm}
  \fmf{fermion,label={\small $p_1$}}{i2,v2}
  \fmf{fermion,label={\small $p_2$}}{i1,v1}
  \fmf{fermion}{o1,v3}
  \fmf{fermion}{o6,v4}
  \fmfv{label={\small $p_3$}}{o1}
  \fmfv{label={\small $p_4$}}{o6}
  \fmf{fermion,tension=0.5,label={\small $q_1$},label.side=right}{v4,v2}
  \fmf{fermion,tension=0.5,label={\small $q_2$},label.side=right}{v3,v4}
  \fmf{fermion,tension=0.5,label={\small $q_4$},label.side=left}{v1,v2}
  \fmf{fermion,tension=0.5,label={\small $q_3$},label.side=left}{v3,v1}
  \fmffreeze
  \fmf{fermion,label={\small $q_5$},label.side=right}{v2,v3}
  \fmf{fermion}{o5,v4}
  \fmfv{label={\small $p_5$}}{o5}
  \fmfv{label={\small $\calt_{25\_1\_3\_4}$ ($e12|e23|e3|ee|$)},label.dist=0.60w,label.angle=-70}{v1}
\end{fmfgraph*}
}
\qquad \qquad \qquad \qquad
\parbox[c][50mm][t]{35mm}{%
  \begin{fmfgraph*}(35,35)
  \fmfstraight
  \fmfleftn{i}{6} \fmfrightn{o}{6}
  \fmfset{arrow_len}{3mm}
  \fmf{fermion}{i6,v2}
  \fmf{fermion}{i1,v1}
  \fmfv{label={\small $p_1$}}{i6}
  \fmfv{label={\small $p_2$}}{i1}
  \fmf{fermion}{o1,v3}
  \fmf{fermion}{o6,v4}
  \fmfv{label={\small $p_3$}}{o1}
  \fmfv{label={\small $p_4$}}{o6}
  \fmf{fermion,tension=0.5,label={\small $q_1$},label.side=right}{v4,v2}
  \fmf{fermion,tension=0.5,label={\small $q_2$},label.side=right}{v3,v4}
  \fmf{fermion,tension=0.5,label={\small $q_4$},label.side=left}{v1,v2}
  \fmf{fermion,tension=0.5,label={\small $q_3$},label.side=left}{v3,v1}
  \fmffreeze
  \fmf{fermion,label={\small $q_5$},label.side=right}{v2,v3}
  \fmf{fermion}{o5,v4}
  \fmf{fermion}{i2,v1}
  \fmfv{label={\small $p_5$}}{o5}
  \fmfv{label={\small $p_6$}}{i2}
  \fmfv{label={\small $\calt_{26\_0\_4\_2}$ ($ee12|e23|e3|ee|$)},label.dist=0.60w,label.angle=-70}{v1}
\end{fmfgraph*}
}
\caption{\small Topologies coming from $\calt_{24\_2\_2\_2}$.}
\label{figure8s} 
\end{figure}

\clearpage

\section{Numerical results}\label{numres}

The results are presented under the following way
\begin{align}
  ^{(2)}I_N^n &= (4 \, \pi)^{-4 + 2 \, \varepsilon} \, \Gamma(1 + 2 \, \varepsilon) \, \left[ \frac{a_{-2}}{\varepsilon^2} + \frac{a_{-1}}{\varepsilon} + a_{0} \right]
  \label{eqpresres0}
\end{align}
We keep unexpanded around $\varepsilon=0$ the overall factor $(4 \, \pi)^{-4 + 2 \, \varepsilon} \, \Gamma(1 + 2 \, \varepsilon)$. For each topology, two tables are presented. The first one gives five points of the phase space, they are chosen more or less randomly avoiding exceptional kinematics since the purpose of this publication is a ``proof of concept'' of the method. The second one gives, for each phase space point, the real and imaginary parts for the coefficients $a_{-2}$, $a_{-1}$ and $a_0$ with the errors related. These errors are the errors reported by the numerical integration package\footnote{The integration package is a homemade code using a two-dimensional adaptive Gauss-Kronrod method to integrate a complex function over the square.} used to performed the $\rho$ and $\xi$ integration. Note that the numerical values for the coefficients $a_{-2}$ and $a_{-1}$ come from the evaluation of analytical formulae, there is no numerical integration here. This explains that the errors for these coefficients are set to zero. \\

\noindent
To compute the numbers in the second table we built a {\tt Fortran} program to perform the numerical evaluation of the different formulae presented in this article. The results of this program are cross checked by another program built in {\tt Mathematica}~\cite{Mathematica}. In addition, for each phase space point, the coefficients $a_{-2}$, $a_{-1}$ and $a_0$ are also compared with the results of the {\tt SecDec} program~\cite{Borowka:2015mxa}. Our results are in perfect agreement with those of {\tt SecDec} within the error bars quoted by the three programs.
Note that we differ by a global $-1$ factor with {\tt SecDec} because of the use of a different measure for the loop integration.\\

\noindent
As it is, our {\tt Fortran} program is, in average, not faster than {\tt SecDec} for the same accuracy: there are less integrations and the number of dimensions per integration is also lower but the integrand is more complicated. It is difficult to give quantitative results because there are lot of options in {\tt SecDec} and we did not play with them and also our computations have been done without any optimisation and there is room for that. Indeed, our program is very fast for Euclidian phase space points and slower for phase points in the Minkowski region. In these last regions, the numerical integration is performed by brute force over the square, it could be improved by contour deformations or using Richardson extrapolation. We can also test with other integration packages. All these problems as well as quantitative comparisons with other programs will be postponed to a dedicated article when our program will become public.

\clearpage

\subsubsection*{Topology $\calt_{22\_0\_2\_1}$ ($e111|e|$)}

\vspace{0.5cm}

\noindent
{\footnotesize
  \hspace{0.90cm}
\begin{tabular}{|c|c|c|c|c|}
  \hline
  Kinem &$p^2$ & $m_1^2$ & $m_2^2$ & $m_3^2$ \\ \hline
  1 & 3. & 3.5 & 1. & 6. \\ \hline
  2 & 13. & 0.5 & 3. & 6. \\ \hline
  3 & 43. & 0.5 & 3. & 2. \\ \hline
  4 & -13. & 0.5 & 3. & 6. \\ \hline
  5 & -3. & 4.5 & 3. & 16. \\ \hline
\end{tabular}

\vspace{0.5cm}

\noindent
\hspace{1cm}
\begin{tabular}{|r|c c c c|}
  \hline
\multirow{3}{*}{1} & $a_{-2}$ = & -0.52500000000E+01 -  0.00000000000E+00\, I & $\pm$ &  0.00E+00 +  0.00E+00\, I \\
 & $a_{-1}$ = &  0.13522720510E+00 +  0.00000000000E+00\, I & $\pm$ &  0.00E+00 +  0.00E+00\, I \\
 & $a_{0}$ = & -0.29537026732E+01 +  0.00000000000E+00\, I & $\pm$ &  0.10E-04 +  0.00E+00\, I \\
\hline
\multirow{3}{*}{2} & $a_{-2}$ = & -0.47500000000E+01 -  0.00000000000E+00\, I & $\pm$ &  0.00E+00 +  0.00E+00\, I \\
 & $a_{-1}$ = &  0.26998200911E+01 +  0.00000000000E+00\, I & $\pm$ &  0.00E+00 +  0.00E+00\, I \\
 & $a_{0}$ = & -0.14244771397E+02 +  0.00000000000E+00\, I & $\pm$ &  0.27E-04 +  0.00E+00\, I \\
\hline
\multirow{3}{*}{3} & $a_{-2}$ = & -0.27500000000E+01 -  0.00000000000E+00\, I & $\pm$ &  0.00E+00 +  0.00E+00\, I \\
 & $a_{-1}$ = &  0.68355576368E+01 +  0.00000000000E+00\, I & $\pm$ &  0.00E+00 +  0.00E+00\, I \\
 & $a_{0}$ = & -0.14606436989E+02 +  0.20823098455E+02\, I & $\pm$ &  0.57E-03 +  0.54E-03\, I \\
\hline
\multirow{3}{*}{4} & $a_{-2}$ = & -0.47500000000E+01 -  0.00000000000E+00\, I & $\pm$ &  0.00E+00 +  0.00E+00\, I \\
 & $a_{-1}$ = & -0.38001799089E+01 +  0.00000000000E+00\, I & $\pm$ &  0.00E+00 +  0.00E+00\, I \\
 & $a_{0}$ = &  0.97794331166E+01 +  0.00000000000E+00\, I & $\pm$ &  0.21E-04 +  0.00E+00\, I \\
\hline
\multirow{3}{*}{5} & $a_{-2}$ = & -0.11750000000E+02 -  0.00000000000E+00\, I & $\pm$ &  0.00E+00 +  0.00E+00\, I \\
 & $a_{-1}$ = &  0.18425604707E+02 +  0.00000000000E+00\, I & $\pm$ &  0.00E+00 +  0.00E+00\, I \\
 & $a_{0}$ = & -0.18702875620E+02 +  0.00000000000E+00\, I & $\pm$ &  0.27E-04 +  0.00E+00\, I \\
\hline
\end{tabular}

\subsubsection*{Topology $\calt_{22\_2\_1\_1}$ ($e112|2|e|$)}

\vspace{0.5cm}

\noindent
  \hspace{0.90cm}
\begin{tabular}{|c|c|c|c|c|c|}
  \hline
  Kinem &$p^2$ & $m_1^2$ & $m_2^2$ & $m_3^2$ & $m_4^2$ \\ \hline
  1 & 3. & 3.5 & 1. & 4. & 1.5 \\ \hline
  2 & -3. & 3.5 & 1. & 4. & 1.5 \\ \hline
  3 & 30. & 3.5 & 1. & 4. & 1.5 \\ \hline
  4 & 20. & 3.5 & 1. & 4. & 1.5 \\ \hline
  5 & -30. & 3.5 & 1. & 4. & 1.5 \\ \hline
\end{tabular}

\vspace{0.5cm}

\noindent
\hspace{1cm}
\begin{tabular}{|r|c c c c|}
  \hline
\multirow{3}{*}{ 1} & $a_{-2}$ = & -0.50000000000E+00 +  0.00000000000E+00\, I & $\pm$ &  0.00E+00 +  0.00E+00\, I \\
 & $a_{-1}$ = &  0.95493937276E-01 -  0.00000000000E+00\, I & $\pm$ &  0.00E+00 +  0.00E+00\, I \\
 & $a_{0}$ = &  0.20043038078E+01 -  0.26620059104E-14\, I & $\pm$ &  0.11E-04 +  0.71E-16\, I \\
\hline
\multirow{3}{*}{ 2} & $a_{-2}$ = & -0.50000000000E+00 +  0.00000000000E+00\, I & $\pm$ &  0.00E+00 +  0.00E+00\, I \\
 & $a_{-1}$ = &  0.54138398520E+00 -  0.00000000000E+00\, I & $\pm$ &  0.00E+00 +  0.00E+00\, I \\
 & $a_{0}$ = &  0.12127813518E+01 +  0.18461994542E-14\, I & $\pm$ &  0.11E-04 +  0.18E-15\, I \\
\hline
\multirow{3}{*}{ 3} & $a_{-2}$ = & -0.50000000000E+00 +  0.00000000000E+00\, I & $\pm$ &  0.00E+00 +  0.00E+00\, I \\
 & $a_{-1}$ = &  0.33480410172E+00 -  0.25842663004E+01\, I & $\pm$ &  0.00E+00 +  0.00E+00\, I \\
 & $a_{0}$ = &  0.68789376397E+01 +  0.46589953606E+01\, I & $\pm$ &  0.11E-04 +  0.23E-07\, I \\
\hline
\multirow{3}{*}{ 4} & $a_{-2}$ = & -0.50000000000E+00 +  0.00000000000E+00\, I & $\pm$ &  0.00E+00 +  0.00E+00\, I \\
 & $a_{-1}$ = & -0.25963239148E+00 -  0.22708739066E+01\, I & $\pm$ &  0.00E+00 +  0.00E+00\, I \\
 & $a_{0}$ = &  0.64884335104E+01 +  0.21118973200E+01\, I & $\pm$ &  0.11E-04 +  0.91E-07\, I \\
\hline
\multirow{3}{*}{ 5} & $a_{-2}$ = & -0.50000000000E+00 +  0.00000000000E+00\, I & $\pm$ &  0.00E+00 +  0.00E+00\, I \\
 & $a_{-1}$ = &  0.14373846668E+01 -  0.00000000000E+00\, I & $\pm$ &  0.00E+00 +  0.00E+00\, I \\
 & $a_{0}$ = & -0.13844570957E+01 +  0.35709948415E-14\, I & $\pm$ &  0.11E-04 +  0.11E-15\, I \\
\hline
\end{tabular}

\subsubsection*{Topology $\calt_{23\_1\_2\_1}$ ($e112|e2|e|$)}

\vspace{0.5cm}

\noindent
  \hspace{0.90cm}
\begin{tabular}{|c|c|c|c|c|c|c|c|}
  \hline
  Kinem &$p_1^2$ & $p_2^2$ & $p_3^2$ & $m_1^2$ & $m_2^2$ & $m_3^2$ & $m_4^2$ \\ \hline
  1 & 3. & 1. & 2. & 3.5 & 1. & 4. & 1.5 \\ \hline
  2 & -3. & -1. & -2. & 3.5 & 1. & 4. & 1.5 \\ \hline
  3 & 30. & 10. & 2. & 3.5 & 1. & 4. & 1.5 \\ \hline
  4 & 30. & 50. & 70. & 3.5 & 1. & 4. & 1.5 \\ \hline
  5 & -30. & 50. & -70. & 3.5 & 1. & 4. & 1.5 \\ \hline
\end{tabular}

\vspace{0.5cm}

\noindent
\hspace{1cm}
\begin{tabular}{|r|c c c c|}
  \hline
\multirow{3}{*}{ 1} & $a_{-2}$ = & -0.50000000000E+00 +  0.00000000000E+00\, I & $\pm$ &  0.00E+00 +  0.00E+00\, I \\
 & $a_{-1}$ = &  0.27258872224E+00 -  0.00000000000E+00\, I & $\pm$ &  0.00E+00 +  0.00E+00\, I \\
 & $a_{0}$ = &  0.15863678305E+01 -  0.25959202695E-17\, I & $\pm$ &  0.11E-04 +  0.70E-17\, I \\
\hline
\multirow{3}{*}{ 2} & $a_{-2}$ = & -0.50000000000E+00 +  0.00000000000E+00\, I & $\pm$ &  0.00E+00 +  0.00E+00\, I \\
 & $a_{-1}$ = &  0.41779504852E+00 -  0.44408920985E-15\, I & $\pm$ &  0.00E+00 +  0.00E+00\, I \\
 & $a_{0}$ = &  0.15908919040E+01 +  0.71026301077E-15\, I & $\pm$ &  0.11E-04 +  0.10E-14\, I \\
\hline
\multirow{3}{*}{ 3} & $a_{-2}$ = & -0.50000000000E+00 +  0.00000000000E+00\, I & $\pm$ &  0.00E+00 +  0.00E+00\, I \\
 & $a_{-1}$ = & -0.13909645111E+01 -  0.94247779608E+00\, I & $\pm$ &  0.00E+00 +  0.00E+00\, I \\
 & $a_{0}$ = &  0.13573181032E+01 -  0.28931884091E+01\, I & $\pm$ &  0.11E-04 +  0.74E-07\, I \\
\hline
\multirow{3}{*}{ 4} & $a_{-2}$ = & -0.50000000000E+00 +  0.00000000000E+00\, I & $\pm$ &  0.00E+00 +  0.00E+00\, I \\
 & $a_{-1}$ = &  0.10240315487E+01 -  0.28162411290E+01\, I & $\pm$ &  0.00E+00 +  0.00E+00\, I \\
 & $a_{0}$ = &  0.71580644577E+01 +  0.70330999378E+01\, I & $\pm$ &  0.11E-04 +  0.50E-07\, I \\
\hline
\multirow{3}{*}{ 5} & $a_{-2}$ = & -0.50000000000E+00 +  0.00000000000E+00\, I & $\pm$ &  0.00E+00 +  0.00E+00\, I \\
 & $a_{-1}$ = &  0.10240315487E+01 -  0.28162411290E+01\, I & $\pm$ &  0.00E+00 +  0.00E+00\, I \\
 & $a_{0}$ = &  0.62611140151E+01 +  0.11876292035E+02\, I & $\pm$ &  0.11E-04 +  0.37E-07\, I \\
\hline
\end{tabular}

\subsubsection*{Topology $\calt_{22\_4\_0\_1}$ ($e12|e3|33||$)}

\vspace{0.5cm}

\noindent
  \hspace{0.90cm}
\begin{tabular}{|c|c|c|c|c|c|c|}
  \hline
  Kinem &$p^2$ & $m_1^2$ & $m_2^2$ & $m_3^2$ & $m_4^2$ & $m_5^2$ \\ \hline
  1 & 3. & 3.5 & 1. & 4. & 1.5 & 2.5 \\ \hline
  2 & -3. & 3.5 & 1. & 4. & 1.5 & 2.5 \\ \hline
  3 & 30. & 3.5 & 1. & 4. & 1.5 & 2.5 \\ \hline
  4 & 20. & 3.5 & 1. & 4. & 1.5 & 2.5 \\ \hline
  5 & -30. & 3.5 & 1. & 4. & 1.5 & 2.5 \\ \hline
\end{tabular}

\vspace{0.5cm}

\noindent
\hspace{1cm}
\begin{tabular}{|r|c c c c|}
  \hline
\multirow{3}{*}{ 1} & $a_{-2}$ = &  0.00000000000E+00 +  0.00000000000E+00\, I & $\pm$ &  0.00E+00 +  0.00E+00\, I \\
 & $a_{-1}$ = & -0.32585507695E+00 -  0.11102230246E-15\, I & $\pm$ &  0.00E+00 +  0.00E+00\, I \\
 & $a_{0}$ = &  0.60075212574E+00 +  0.65005240953E-14\, I & $\pm$ &  0.40E-08 +  0.47E-15\, I \\
\hline
\multirow{3}{*}{ 2} & $a_{-2}$ = &  0.00000000000E+00 +  0.00000000000E+00\, I & $\pm$ &  0.00E+00 +  0.00E+00\, I \\
 & $a_{-1}$ = & -0.20937204709E+00 +  0.17763568394E-14\, I & $\pm$ &  0.00E+00 +  0.00E+00\, I \\
 & $a_{0}$ = &  0.50189220962E+00 -  0.14624659782E-13\, I & $\pm$ &  0.40E-08 +  0.36E-14\, I \\
\hline
\multirow{3}{*}{ 3} & $a_{-2}$ = &  0.00000000000E+00 +  0.00000000000E+00\, I & $\pm$ &  0.00E+00 +  0.00E+00\, I \\
 & $a_{-1}$ = &  0.12708618658E+00 -  0.14089364262E+00\, I & $\pm$ &  0.00E+00 +  0.00E+00\, I \\
 & $a_{0}$ = &  0.75653203129E-01 +  0.71967244786E+00\, I & $\pm$ &  0.65E-04 +  0.66E-04\, I \\
\hline
\multirow{3}{*}{ 4} & $a_{-2}$ = &  0.00000000000E+00 +  0.00000000000E+00\, I & $\pm$ &  0.00E+00 +  0.00E+00\, I \\
 & $a_{-1}$ = &  0.18023485520E+00 -  0.25430822792E+00\, I & $\pm$ &  0.00E+00 +  0.00E+00\, I \\
 & $a_{0}$ = &  0.31124925008E+00 +  0.97655849662E+00\, I & $\pm$ &  0.59E-04 +  0.61E-04\, I \\
\hline
\multirow{3}{*}{ 5} & $a_{-2}$ = &  0.00000000000E+00 +  0.00000000000E+00\, I & $\pm$ &  0.00E+00 +  0.00E+00\, I \\
 & $a_{-1}$ = & -0.91118403196E-01 +  0.00000000000E+00\, I & $\pm$ &  0.00E+00 +  0.00E+00\, I \\
 & $a_{0}$ = &  0.31076522901E+00 +  0.28162073710E-13\, I & $\pm$ &  0.41E-08 +  0.45E-15\, I \\
\hline
\end{tabular}

\subsubsection*{Topology $\calt_{23\_3\_1\_1}$ ($e112|3|e3|e|$)}

\vspace{0.5cm}

\noindent
  \hspace{0.90cm}
\begin{tabular}{|c|c|c|c|c|c|c|c|c|}
  \hline
  Kinem &$p_1^2$ & $p_2^2$ & $p_3^2$ & $m_1^2$ & $m_2^2$ & $m_3^2$ & $m_4^2$ & $m_5^2$ \\ \hline
  1 & 3. & 1. & 2. & 3.5 & 1. & 4. & 1.5 & 3. \\ \hline
  2 & -3. & -1. & -2. & 3.5 & 1. & 4. & 1.5 & 3. \\ \hline
  3 & 30. & 10. & 2. & 3.5 & 1. & 4. & 1.5 & 3. \\ \hline
  4 & 30. & 50. & 70. & 3.5 & 1. & 4. & 1.5 & 3. \\ \hline
  5 & -30. & 50. & -70. & 3.5 & 1. & 4. & 1.5 & 3. \\ \hline
\end{tabular}

\vspace{0.5cm}

\noindent
\hspace{1cm}
\begin{tabular}{|r|c c c c|}
  \hline
\multirow{3}{*}{ 1} & $a_{-2}$ = &  0.00000000000E+00 +  0.00000000000E+00\, I & $\pm$ &  0.00E+00 +  0.00E+00\, I \\
 & $a_{-1}$ = &  0.36784617870E+00 +  0.44408920985E-15\, I & $\pm$ &  0.00E+00 +  0.00E+00\, I \\
 & $a_{0}$ = & -0.60872125164E+00 -  0.43899482229E-14\, I & $\pm$ &  0.10E-05 +  0.10E-05\, I \\
\hline
\multirow{3}{*}{ 2} & $a_{-2}$ = &  0.00000000000E+00 +  0.00000000000E+00\, I & $\pm$ &  0.00E+00 +  0.00E+00\, I \\
 & $a_{-1}$ = &  0.19871625204E+00 +  0.23551386880E-15\, I & $\pm$ &  0.00E+00 +  0.00E+00\, I \\
 & $a_{0}$ = & -0.51321142324E+00 +  0.50234552650E-14\, I & $\pm$ &  0.17E-06 +  0.13E-14\, I \\
\hline
\multirow{3}{*}{ 3} & $a_{-2}$ = &  0.00000000000E+00 +  0.00000000000E+00\, I & $\pm$ &  0.00E+00 +  0.00E+00\, I \\
 & $a_{-1}$ = & -0.11285841800E+00 +  0.22679298979E+00\, I & $\pm$ &  0.00E+00 +  0.00E+00\, I \\
 & $a_{0}$ = & -0.47819030237E+00 -  0.86257747352E+00\, I & $\pm$ &  0.38E-05 +  0.43E-05\, I \\
\hline
\multirow{3}{*}{ 4} & $a_{-2}$ = &  0.00000000000E+00 +  0.00000000000E+00\, I & $\pm$ &  0.00E+00 +  0.00E+00\, I \\
 & $a_{-1}$ = & -0.60679868141E-01 +  0.88993218087E-02\, I & $\pm$ &  0.00E+00 +  0.00E+00\, I \\
 & $a_{0}$ = &  0.16404161151E+00 -  0.32974950110E+00\, I & $\pm$ &  0.40E-04 +  0.41E-04\, I \\
\hline
\multirow{3}{*}{ 5} & $a_{-2}$ = &  0.00000000000E+00 +  0.00000000000E+00\, I & $\pm$ &  0.00E+00 +  0.00E+00\, I \\
 & $a_{-1}$ = &  0.41120805206E-01 +  0.49265819840E-01\, I & $\pm$ &  0.00E+00 +  0.00E+00\, I \\
 & $a_{0}$ = & -0.25654789543E+00 -  0.19159473560E+00\, I & $\pm$ &  0.58E-06 +  0.52E-06\, I \\
\hline
\end{tabular}

\subsubsection*{Topology $\calt_{24\_2\_2\_1}$ ($e112|e3|e3|e|$)}

\vspace{0.5cm}

\noindent
  \hspace{0.90cm}
\begin{tabular}{|c|c|c|c|c|c|c|c|c|c|c|c|}
  \hline
  Kinem &$p_1^2$ & $p_2^2$ & $p_3^2$ & $p_4^2$ & $s$ & $t$ & $m_1^2$ & $m_2^2$ & $m_3^2$ & $m_4^2$ & $m_5^2$ \\ \hline
  1 & 3. & 1. & 2. & 0. & 2. & 3. & 3.5 & 1. & 4. & 1.5 & 3. \\ \hline
  2 & -3. & -1. & -2. & 4. & -7.4 & -4.7 & 3.5 & 1. & 4. & 1.5 & 3. \\ \hline
  3 & 30. & 10. & 2. & 4. & 7.4 & -3.7 & 3.5 & 1. & 4. & 1.5 & 3. \\ \hline
  4 & 30. & 50. & 70. & 40. & 74. & -37. & 3.5 & 1. & 4. & 1.5 & 3. \\ \hline
  5 & -3. & -5. & 7. & 4. & 37. & -3.7 & 3.5 & 1. & 4. & 1.5 & 3. \\ \hline
\end{tabular}

\vspace{0.5cm}

\noindent
\hspace{1cm}
\begin{tabular}{|r|c c c c|}
  \hline
\multirow{3}{*}{ 1} & $a_{-2}$ = &  0.00000000000E+00 +  0.00000000000E+00\, I & $\pm$ &  0.00E+00 +  0.00E+00\, I \\
 & $a_{-1}$ = &  0.36784617870E+00 +  0.44408920985E-15\, I & $\pm$ &  0.00E+00 +  0.00E+00\, I \\
 & $a_{0}$ = & -0.60872125164E+00 -  0.46774890239E-14\, I & $\pm$ &  0.10E-05 +  0.10E-05\, I \\
\hline
\multirow{3}{*}{ 2} & $a_{-2}$ = &  0.00000000000E+00 +  0.00000000000E+00\, I & $\pm$ &  0.00E+00 +  0.00E+00\, I \\
 & $a_{-1}$ = &  0.18642213411E+00 -  0.49113375056E-16\, I & $\pm$ &  0.00E+00 +  0.00E+00\, I \\
 & $a_{0}$ = & -0.49747267449E+00 +  0.61108240588E-15\, I & $\pm$ &  0.18E-06 +  0.68E-15\, I \\
\hline
\multirow{3}{*}{ 3} & $a_{-2}$ = &  0.00000000000E+00 +  0.00000000000E+00\, I & $\pm$ &  0.00E+00 +  0.00E+00\, I \\
 & $a_{-1}$ = &  0.46030231987E+00 +  0.17201731529E+00\, I & $\pm$ &  0.00E+00 +  0.00E+00\, I \\
 & $a_{0}$ = & -0.55631917291E+00 +  0.63955372748E+00\, I & $\pm$ &  0.54E-05 +  0.64E-05\, I \\
\hline
\multirow{3}{*}{ 4} & $a_{-2}$ = &  0.00000000000E+00 +  0.00000000000E+00\, I & $\pm$ &  0.00E+00 +  0.00E+00\, I \\
 & $a_{-1}$ = & -0.38332203430E-01 +  0.56282775535E-01\, I & $\pm$ &  0.00E+00 +  0.00E+00\, I \\
 & $a_{0}$ = & -0.65962616212E-01 -  0.38606258129E+00\, I & $\pm$ &  0.88E-04 +  0.90E-04\, I \\
\hline
\multirow{3}{*}{ 5} & $a_{-2}$ = &  0.00000000000E+00 +  0.00000000000E+00\, I & $\pm$ &  0.00E+00 +  0.00E+00\, I \\
 & $a_{-1}$ = &  0.24132803188E+00 +  0.00000000000E+00\, I & $\pm$ &  0.00E+00 +  0.00E+00\, I \\
 & $a_{0}$ = & -0.39052167827E+00 +  0.17704866554E-01\, I & $\pm$ &  0.91E-06 +  0.98E-06\, I \\
\hline
\end{tabular}

\subsubsection*{Topology $\calt_{22\_4\_0\_2}$ ($e12|23|3|e|$)}

\vspace{0.5cm}

\noindent
  \hspace{0.90cm}
\begin{tabular}{|c|c|c|c|c|c|c|}
  \hline
  Kinem &$p^2$ & $m_1^2$ & $m_2^2$ & $m_3^2$ & $m_4^2$ & $m_5^2$ \\ \hline
  1 & 3. & 3.5 & 1. & 4. & 1.5 & 2.5 \\ \hline
  2 & -3. & 3.5 & 1. & 4. & 1.5 & 2.5 \\ \hline
  3 & 30. & 3.5 & 1. & 4. & 1.5 & 2.5 \\ \hline
  4 & 20. & 3.5 & 1. & 4. & 1.5 & 2.5 \\ \hline
  5 & -30. & 3.5 & 1. & 4. & 1.5 & 2.5 \\ \hline
\end{tabular}

\vspace{0.5cm}

\noindent
\hspace{1cm}
\begin{tabular}{|r|c c c c|}
  \hline
\multirow{3}{*}{ 1} & $a_{-2}$ = &  0.00000000000E+00 +  0.00000000000E+00\, I & $\pm$ &  0.00E+00 +  0.00E+00\, I \\
 & $a_{-1}$ = &  0.00000000000E+00 +  0.00000000000E+00\, I & $\pm$ &  0.00E+00 +  0.00E+00\, I \\
 & $a_{0}$ = &  0.41219914357E+00 -  0.39936184599E-18\, I & $\pm$ &  0.14E-04 +  0.22E-17\, I \\
\hline
\multirow{3}{*}{ 2} & $a_{-2}$ = &  0.00000000000E+00 +  0.00000000000E+00\, I & $\pm$ &  0.00E+00 +  0.00E+00\, I \\
 & $a_{-1}$ = &  0.00000000000E+00 +  0.00000000000E+00\, I & $\pm$ &  0.00E+00 +  0.00E+00\, I \\
 & $a_{0}$ = &  0.27683187689E+00 +  0.14649979575E-16\, I & $\pm$ &  0.88E-05 +  0.13E-15\, I \\
\hline
\multirow{3}{*}{ 3} & $a_{-2}$ = &  0.00000000000E+00 +  0.00000000000E+00\, I & $\pm$ &  0.00E+00 +  0.00E+00\, I \\
 & $a_{-1}$ = &  0.00000000000E+00 +  0.00000000000E+00\, I & $\pm$ &  0.00E+00 +  0.00E+00\, I \\
 & $a_{0}$ = & -0.26621927150E+00 +  0.27333087222E+00\, I & $\pm$ &  0.18E-04 +  0.22E-04\, I \\
\hline
\multirow{3}{*}{ 4} & $a_{-2}$ = &  0.00000000000E+00 +  0.00000000000E+00\, I & $\pm$ &  0.00E+00 +  0.00E+00\, I \\
 & $a_{-1}$ = &  0.00000000000E+00 +  0.00000000000E+00\, I & $\pm$ &  0.00E+00 +  0.00E+00\, I \\
 & $a_{0}$ = & -0.25535371870E+00 +  0.57326996026E+00\, I & $\pm$ &  0.35E-04 +  0.41E-04\, I \\
\hline
\multirow{3}{*}{ 5} & $a_{-2}$ = &  0.00000000000E+00 +  0.00000000000E+00\, I & $\pm$ &  0.00E+00 +  0.00E+00\, I \\
 & $a_{-1}$ = &  0.00000000000E+00 +  0.00000000000E+00\, I & $\pm$ &  0.00E+00 +  0.00E+00\, I \\
 & $a_{0}$ = &  0.12186850114E+00 -  0.20043296705E-17\, I & $\pm$ &  0.39E-05 +  0.23E-16\, I \\
\hline
\end{tabular}

\subsubsection*{Topology $\calt_{23\_3\_1\_2}$ ($e12|e23|3|e|$)}

\vspace{0.5cm}

\noindent
  \hspace{0.90cm}
\begin{tabular}{|c|c|c|c|c|c|c|c|c|}
  \hline
  Kinem &$p_1^2$ & $p_2^2$ & $p_3^2$ & $m_1^2$ & $m_2^2$ & $m_3^2$ & $m_4^2$ & $m_5^2$ \\ \hline
  1 & 3. & 1. & 2. & 3.5 & 1. & 4. & 1.5 & 3. \\ \hline
  2 & -3. & -1. & -2. & 3.5 & 1. & 4. & 1.5 & 3. \\ \hline
  3 & 30. & 10. & 2. & 3.5 & 1. & 4. & 1.5 & 3. \\ \hline
  4 & 30. & 50. & 70. & 3.5 & 1. & 4. & 1.5 & 3. \\ \hline
  5 & -30. & 50. & -70. & 3.5 & 1. & 4. & 1.5 & 3. \\ \hline
\end{tabular}

\vspace{0.5cm}

\noindent
\hspace{1cm}
\begin{tabular}{|r|c c c c|}
  \hline
\multirow{3}{*}{ 1} & $a_{-2}$ = &  0.00000000000E+00 +  0.00000000000E+00\, I & $\pm$ &  0.00E+00 +  0.00E+00\, I \\
 & $a_{-1}$ = &  0.00000000000E+00 +  0.00000000000E+00\, I & $\pm$ &  0.00E+00 +  0.00E+00\, I \\
 & $a_{0}$ = &  0.36054528076E+00 -  0.43707313538E-18\, I & $\pm$ &  0.12E-04 +  0.46E-18\, I \\
\hline
\multirow{3}{*}{ 2} & $a_{-2}$ = &  0.00000000000E+00 +  0.00000000000E+00\, I & $\pm$ &  0.00E+00 +  0.00E+00\, I \\
 & $a_{-1}$ = &  0.00000000000E+00 +  0.00000000000E+00\, I & $\pm$ &  0.00E+00 +  0.00E+00\, I \\
 & $a_{0}$ = &  0.28847143871E+00 -  0.20852434825E-16\, I & $\pm$ &  0.96E-05 +  0.28E-15\, I \\
\hline
\multirow{3}{*}{ 3} & $a_{-2}$ = &  0.00000000000E+00 +  0.00000000000E+00\, I & $\pm$ &  0.00E+00 +  0.00E+00\, I \\
 & $a_{-1}$ = &  0.00000000000E+00 +  0.00000000000E+00\, I & $\pm$ &  0.00E+00 +  0.00E+00\, I \\
 & $a_{0}$ = &  0.10414467534E+01 +  0.31275711261E+00\, I & $\pm$ &  0.10E-03 +  0.67E-04\, I \\
\hline
\multirow{3}{*}{ 4} & $a_{-2}$ = &  0.00000000000E+00 +  0.00000000000E+00\, I & $\pm$ &  0.00E+00 +  0.00E+00\, I \\
 & $a_{-1}$ = &  0.00000000000E+00 +  0.00000000000E+00\, I & $\pm$ &  0.00E+00 +  0.00E+00\, I \\
 & $a_{0}$ = & -0.14315144501E+00 +  0.61247141825E-01\, I & $\pm$ &  0.13E-04 +  0.14E-04\, I \\
\hline
\multirow{3}{*}{ 5} & $a_{-2}$ = &  0.00000000000E+00 +  0.00000000000E+00\, I & $\pm$ &  0.00E+00 +  0.00E+00\, I \\
 & $a_{-1}$ = &  0.00000000000E+00 +  0.00000000000E+00\, I & $\pm$ &  0.00E+00 +  0.00E+00\, I \\
 & $a_{0}$ = &  0.61860272485E-01 +  0.12027275741E+00\, I & $\pm$ &  0.27E-04 +  0.22E-04\, I \\
\hline
\end{tabular}

\subsubsection*{Topology $\calt_{24\_2\_2\_2}$ ($e12|e23|e3|e|$)}

\vspace{0.5cm}

\noindent
  \hspace{0.90cm}
\begin{tabular}{|c|c|c|c|c|c|c|c|c|c|c|c|}
  \hline
  Kinem &$p_1^2$ & $p_2^2$ & $p_3^2$ & $p_4^2$ & $s$ & $t$ & $m_1^2$ & $m_2^2$ & $m_3^2$ & $m_4^2$ & $m_5^2$ \\ \hline
  1 & 3. & 1. & 0. & 2. & 2. & 1. & 3.5 & 1. & 4. & 1.5 & 3. \\ \hline
  2 & -3. & -1. & -2. & 4. & -7.4 & -4.7 & 3.5 & 1. & 4. & 1.5 & 3. \\ \hline
  3 & 30. & 10. & 2. & 4. & 7.4 & -3.7 & 3.5 & 1. & 4. & 1.5 & 3. \\ \hline
  4 & 30. & 50. & 70. & 40. & 74. & -37. & 3.5 & 1. & 4. & 1.5 & 3. \\ \hline
  5 & -3. & -5. & 7. & 4. & 37. & -3.7 & 3.5 & 1. & 4. & 1.5 & 3. \\ \hline
\end{tabular}

\vspace{0.5cm}

\noindent
\hspace{1cm}
\begin{tabular}{|r|c c c c|}
  \hline
\multirow{3}{*}{ 1} & $a_{-2}$ = &  0.00000000000E+00 +  0.00000000000E+00\, I & $\pm$ &  0.00E+00 +  0.00E+00\, I \\
 & $a_{-1}$ = &  0.00000000000E+00 +  0.00000000000E+00\, I & $\pm$ &  0.00E+00 +  0.00E+00\, I \\
 & $a_{0}$ = &  0.36054528065E+00 +  0.15531097836E-18\, I & $\pm$ &  0.12E-04 +  0.62E-18\, I \\
\hline
\multirow{3}{*}{ 2} & $a_{-2}$ = &  0.00000000000E+00 +  0.00000000000E+00\, I & $\pm$ &  0.00E+00 +  0.00E+00\, I \\
 & $a_{-1}$ = &  0.00000000000E+00 +  0.00000000000E+00\, I & $\pm$ &  0.00E+00 +  0.00E+00\, I \\
 & $a_{0}$ = &  0.33207668117E+00 +  0.00000000000E+00\, I & $\pm$ &  0.14E-04 +  0.00E+00\, I \\
\hline
\multirow{3}{*}{ 3} & $a_{-2}$ = &  0.00000000000E+00 +  0.00000000000E+00\, I & $\pm$ &  0.00E+00 +  0.00E+00\, I \\
 & $a_{-1}$ = &  0.00000000000E+00 +  0.00000000000E+00\, I & $\pm$ &  0.00E+00 +  0.00E+00\, I \\
 & $a_{0}$ = &  0.94941617014E+00 +  0.21434949328E+00\, I & $\pm$ &  0.51E-04 +  0.87E-05\, I \\
\hline
\multirow{3}{*}{ 4} & $a_{-2}$ = &  0.00000000000E+00 +  0.00000000000E+00\, I & $\pm$ &  0.00E+00 +  0.00E+00\, I \\
 & $a_{-1}$ = &  0.00000000000E+00 +  0.00000000000E+00\, I & $\pm$ &  0.00E+00 +  0.00E+00\, I \\
 & $a_{0}$ = & -0.17095069704E+00 +  0.10309141582E+00\, I & $\pm$ &  0.56E-05 +  0.61E-05\, I \\
\hline
\multirow{3}{*}{ 5} & $a_{-2}$ = &  0.00000000000E+00 +  0.00000000000E+00\, I & $\pm$ &  0.00E+00 +  0.00E+00\, I \\
 & $a_{-1}$ = &  0.00000000000E+00 +  0.00000000000E+00\, I & $\pm$ &  0.00E+00 +  0.00E+00\, I \\
 & $a_{0}$ = &  0.42930745092E+00 +  0.15170110537E-01\, I & $\pm$ &  0.13E-04 +  0.11E-06\, I \\
\hline
\end{tabular}
}

\clearpage

\section{Summary}

This article can be seen as a proof of concept of the validity of the novel framework of mixed analytical/numerical approach, introduced in ref. \cite{letter}, for the computation of scalar two-loop multi-leg Feynman diagrams appearing in scalar theories with three- and four-leg vertices. This method allows us to express any scalar two-loop $N$-point function in $n$ dimensions as a sum of double integrals of some generalised scalar one-loop type functions multiplied by some weight functions. In the current work, we limited ourselves to topologies with five internal propagators at most, where all the internal masses are taken to be real. To show the generality of our method, we considered several $N$-point two-loop functions, where some have global and/or sub-leading UV divergences. More complicated topologies will be considered in future publications. The studied topologies are classified in four categories. The topologies of each category are generated by inserting some extra external legs to the master topology. 
The first class (with $I=3$) contain only one topology, denoted $\calt_{22\_0\_2\_1}$, it corresponds to the two-loop two-point function with three internal lines and two external legs, which is globally UV divergent. The second class (with $I=4$) contains four globally UV divergent  topologies generated by the master topology $\calt_{22\_2\_1\_1}$, the later one corresponds to the two-loop two-point function with four internal lines and two external legs. The third and the fourth category (with $I=5$) have two master topologies, one of them, denoted $\calt_{22\_4\_0\_1}$, contains a sub-leading UV divergence and the other one, denoted $\calt_{22\_4\_0\_2}$, is finite. Both master topologies correspond to two-loop functions with five internal lines and two external legs. One can generate from the divergent topology nine other topologies, and from the finite one eight other topologies belonging to the same class. The topologies generated by the same master topology have the same formula with different coefficients for the polynomial $\tilde{\mathcal{F}}$. 
We implemented the two-dimensional integral representation of nine of these topologies belonging to the four classes in two different codes, one is written in {\tt Fortran} and the other one is written in {\tt Mathematica}, to perform the numerical integration over the two remaining variables ($\rho$ and $\xi$). We showed that our results are in good agreement with the public code {\tt SecDec}, which is based on sector decomposition method. This paper is accompanied with many appendices which are dedicated to the fully analytical calculation of the divergent parts and some related finite contributions of these topologies. Furthermore, we presented the full analytical calculation of the one-loop two-point function at $\mathcal{O}(\varepsilon)$ (Appendix \ref{ol2Oe}), the one-loop three-point function integrated over a square (Appendix \ref{ol3ptsq}) and the one-loop three point function at $\mathcal{O}(\varepsilon)$ (Appendix \ref{new_integral}). \\

\noindent
This method opens a new window towards the automation of two-loop computation, since traditional reduction methods can be applied to reduce tensor integrals to scalar integrals and scalar integrals to scalar integrals with less propgators. In principle, it is competitive with
the fully numerical approaches since it keeps only two numerical integrations for every two-loop amplitude in a systematic way.
It worthwhile to mention that, in the case where a sum of two-loop diagrams have to be computed, the numerical integration can be easily factorised out and performed once on the sum of the integrands which could lead to an extra gain in cpu time and precision. It is hard to do in other numerical methods.
Further developments can be carried out in different directions.
An immediate extension of this work would be to consider complex masses. Almost all the results are there. Indeed, all the new integrals presented in the appendix~\ref{ol2Oe} and \ref{new_integral} trivially extend to complex masses and the results for the ``generalised one-loop functions'' are given in ref.~\cite{paper2}.  These formulae for complex mass cases as well as numerical applications are not shown here mainly because this article contains already a lot of information. They will be postpone to other publications.
One could, inside this scalar theory, tackle amplitudes whose number of internal lines is greater than five. For $I=6$, the ``generalised one-loop function'' associated is a `` generalised one-loop four-point function'', the formulae for an extended kinematics being given in ref.~\cite{paper1,paper2,paper3} (see also \cite{Denner:2010tr}). While for $I=7$, we have to deal with a ``generalised one-loop five-point functions'' which are a sum of ``generalised one-loop four-point functions'' as can be seen with standard one-loop reduction methods.
Another direction would be to extend the method for theories including particles with spin $1/2$ or $1$, gauge theories for example. In this case, at the end of the re-parameterisation the numerator will be a polynomial in the parameters $u_i$ as well as in the parameters $\rho$ and $\xi$. To each monomial, will correspond a ``generalised one-loop functions'' with a numerator which can be reduced to ``generalised scalar integrals'' -- we think that any reduction method works, certainly the one proposed in \cite{Binoth:2005ff} will do. The extra dependance on $\rho$ and $\xi$ will be taken into account in the numerical integration. One has to care about the proliferation of terms in the integrand which may slow down the numerical integration.
Lastly, a third direction would be to consider two-loop amplitudes with IR divergences, some internal scalar particles are massless, where some extra work may have to be done on the ``generalised one-loop functions'' in order to push further the expansion around $\varepsilon=0$. Our method works in principle but it would be good to have a proof of concept in this case too. Along these lines, the IR cases with photons or gluons could be also investigated. \\


\section*{Acknowledgements}

This article is the continuation of a work initiated by Prof. Shimizu who passed away too early, we keep in memory a physicist who pushed his projects wholeheartedly.
We would like to thank P. Aurenche for his support during this project and for a careful reading of the manuscript.

\appendix

\section{Details of the computation for topology $\calt_{22\_0\_2\_1}$}\label{det_cal_t22021}

This appendix presents the gory details for the intermediate computation leading to eq.~(\ref{eqdeftot1}).
By splitting the factor $\calft(\rho,\xi)$ and using partial fraction decomposition, eq.~(\ref{eqdefT22n2}) can be written as
\begin{align}
  ^{(2)}I_{2}^{n} &= (4 \, \pi)^{-4 + 2 \, \varepsilon} \, \frac{\Gamma(1 + 2 \, \varepsilon)}{(-1 + 2 \, \varepsilon) \, 2 \, \varepsilon} \, \int_0^1 d \rho \int_0^1 d \xi \, \notag \\
&\quad {} \times \left[H_1(\rho,\xi) \, \rho^{-1+\varepsilon} + H_2(\rho,\xi) \, \left( 1 - \rho + \rho \, \xi \, (1-\xi) \right)^{-3 + 3 \, \varepsilon} \right. \notag \\
&\quad {} + \left. (H_3(\rho,\xi) + H_4(\rho,\xi)) \, \left( 1 - \rho + \rho \, \xi \, (1-\xi) \right)^{-2 + 3 \, \varepsilon}  \right] 
  \label{eqdefT22n4}
\end{align}
where
\begin{align}
  H_1(\rho,\xi) &= (1 - \rho + \rho \, \xi \, (1 - \xi))^{3 \, \varepsilon} \, (1-\rho) \, m_3^2 \, \left( \calft(\rho,\xi) - i \, \lambda \right)^{- 2 \, \varepsilon} \label{eqdefH1} \\
  H_2(\rho,\xi) &= \rho^{\varepsilon} \, (- p^2) \, (1-\rho) \, \xi \, (1-\xi) \, \left( \calft(\rho,\xi) - i \, \lambda \right)^{- 2 \, \varepsilon} \label{eqdefH2} \\
  H_3(\rho,\xi) &= \rho^{\varepsilon} \, (\xi \, m_1^2 + (1-\xi) \, m_2^2) \, \left( \calft(\rho,\xi) - i \, \lambda \right)^{- 2 \, \varepsilon} \label{eqdefH3} \\
  H_4(\rho,\xi) &= \rho^{\varepsilon} \, (1 - \xi \, (1-\xi)) \, (2 - \rho \, (1 - \xi \, (1-\xi))) \notag \\
  &\quad {} \times (1-\rho) \, m_3^2 \, \left( \calft(\rho,\xi) - i \, \lambda \right)^{- 2 \, \varepsilon} \label{eqdefH4}
\end{align}

\noindent
The first term in eq.~(\ref{eqdefT22n4}) diverges at $\rho=0$. 
Using eq.~(\ref{eqdistrib0}) and expanding the coefficient $H_1$ around $\varepsilon=0$: $H_1(\rho,\xi) = H_1^{(0)}(\rho,\xi) + \varepsilon \, H_1^{(1)}(\rho,\xi) + \varepsilon^2 \, H_1^{(2)}(\rho,\xi)$, the $\rho$ integration is now of the type
\begin{align}
  T_1 &= \int_0^1 d \rho \, H_1(\rho,\xi) \, \rho^{-1+\varepsilon} \notag \\
  &= \frac{1}{\varepsilon} \, H_1(0,\xi) + \int_0^1 \frac{d \rho}{\rho} \left( H_1^{(0)}(\rho,\xi) - H_1^{(0)}(0,\xi) \right) \notag \\
  &\quad {} + \varepsilon \,\int_0^1 \frac{d \rho}{\rho} \left( H_1^{(1)}(\rho,\xi) - H_1^{(1)}(0,\xi) \right) \notag \\
  &\quad {} + \varepsilon \, \int_0^1 d \rho \, \frac{\ln(\rho)}{\rho} \left( H_1^{(0)}(\rho,\xi) - H_1^{(0)}(0,\xi) \right) 
  \label{eqInt1t0}
\end{align}
with
\begin{equation}
\begin{aligned}
  \calft(0,\xi) &= m_3^2 \\
  H_1(0,\xi) &= \left( m_3^2 - i \, \lambda \right)^{1 - 2 \, \varepsilon} \\
  H_1^{(0)}(\rho,\xi) &= (1-\rho) \, m_3^2 \\
  H_1^{(1)}(\rho,\xi) &= H_1^{(0)}(\rho,\xi) \, \left[ 3 \, \ln\left( 1-\rho + \rho \, \xi \, (1-\xi) \right) - 2 \, \ln \left( \calft(\rho,\xi) - i \, \lambda \right) \right]
\end{aligned}
  \label{eqcoeffh1}
\end{equation}
Injecting the results of eqs.~(\ref{eqcoeffh1}) in eq.~(\ref{eqInt1t0}), we end with
\begin{align}
T_1 &= m_3^2 \, \left\{ \frac{1}{\varepsilon} \, \left( m_3^2 - i \, \lambda \right)^{- 2 \, \varepsilon} - 1 + \varepsilon + 3 \, \varepsilon \, I_1 \right. \notag \\
&\qquad \quad {} - \left.  2 \, \varepsilon \, \int_0^1 d \xi \, \int_0^1 \frac{d \rho}{\rho} \, \left[ (1-\rho) \, \ln\left( \calft(\rho,\xi) - i \, \lambda \right) - \ln( m_3^2 - i \, \lambda) \right]  \right\}
  \label{eqdefT11}
\end{align}
where $I_1$ is given in appendix~\ref{intsecpar}.
\\

\noindent
The second term is finite despite the factor $(1-\rho+\rho \, \xi \, (1-\xi))^{-3+3 \, \varepsilon}$, indeed the coefficient $H_2$ contains a factor $(1-\rho) \, \xi \, (1-\xi)$ which regularises the divergence. Care has to be taken because the denominator vanishes at $\rho=1$ and $\xi=0$ or $\xi=1$. For those points, $\calft(\rho,\xi)$ also vanishes but in $H_2$, it is raised to a negative power $- 2 \, \varepsilon$. Thus it is better to write
\begin{align}
  T_2 &\equiv \int_0^1 d \xi \int_0^1 d \rho \, H_2(\rho,\xi) \, (1 - \rho + \rho \, \xi \, (1-\xi))^{-3 + 3 \, \varepsilon} \notag \\
  &=  \int_0^1 d \xi \int_0^1 d \rho \, H_2^{\prime}(\rho,\xi) \, (1 - \rho + \rho \, \xi \, (1-\xi))^{-3 + \varepsilon}
  \label{eqdefT20}
\end{align}
with 
\begin{align}
  H_2^{\prime}(\rho,\xi) &= (-p^2) \, \rho^{\varepsilon} \, (1-\rho) \, \xi \, (1-\xi) \left( \calftp(\rho,\xi) - i \, \lambda \right)^{- 2 \, \varepsilon}
  \label{eqdefH2prime}
\end{align}
where
\begin{align}
  \calftp(\rho,\xi) &= \frac{\calft(\rho,\xi)}{1-\rho + \rho \, \xi \, (1-\xi)}
  \label{eqdefcalftp}
\end{align}
Note that $1-\rho + \rho \, \xi \, (1-\xi)$ is positive when the variables $\rho$ and $\xi$ run in the unit square and vanishes at the points $(\rho=1, \quad \xi=0)$ and $(\rho=1, \quad \xi=1)$. For these points,
$\calftp(1,0)$ and $\calftp(1,1)$ do not vanish because $\calft(1,\xi) = \xi \, (1-\xi) \, (\xi \, m_1^2 + (1-\xi) \, m_2^2)$.
It is enough to expand eq.~(\ref{eqdefT20}) around $\varepsilon=0$ and we get
\begin{align}
T_2 &= (-p^2) \, \left\{ I_5 + \varepsilon \left[ I_6 + I_7 - 2 \, \int_0^1 d \xi \int_0^1 d \rho \, \frac{(1-\rho) \, \xi \, (1-\xi) \, \ln\left( \calftp(\rho,\xi) - i \, \lambda \right)}{(1 - \rho + \rho \, \xi \, (1-\xi))^3} \right] \right\}
  \label{eqdefT21}
\end{align}
with the integrals $I_5$, $I_6$ and $I_7$ given in the appendix~\ref{intsecpar}.\\

\noindent
The third term diverges at $\rho=1$ and $\xi=0$ or $\xi=1$. The same care as for the second term has to be taken here. Indeed, $H_3(\rho,\xi)$ is not a regular function, thus we re-write the third term as
\begin{align}
  T_3 &\equiv \int_0^1 d \xi \int_0^1 d \rho \, H_3(\rho,\xi) \, (1 - \rho + \rho \, \xi \, (1-\xi))^{-2 + 3 \, \varepsilon} \notag \\
  &=  \int_0^1 d \xi \int_0^1 d \rho \, H_3^{\prime}(\rho,\xi) \, (1 - \rho + \rho \, \xi \, (1-\xi))^{-2 + \varepsilon}
  \label{eqdefT31}
\end{align}
where
\begin{align}
  H_3^{\prime}(\rho,\xi) &= \rho^{\varepsilon} \, (\xi \, m_1^2 + (1-\xi) \, m_2^2) \, \left( \calftp(\rho,\xi) - i \, \lambda \right)^{- 2 \, \varepsilon}
  \label{eqdefH3prime}
\end{align}
Now, $H_3^{\prime}(\rho,\xi)$ is a regular function except may be for exceptional kinematics. So the third term $T_3$ can be written as
\begin{align}
  T_3 &= \int_0^1 d \xi \int_0^1 d \rho \, \left( H_3^{\prime}(\rho,\xi) - H_3^{\prime}(1,\xi) \right)\, (1 - \rho + \rho \, \xi \, (1-\xi))^{-2 + \varepsilon} \notag \\
  &\quad {} + \int_0^1 d \xi \, H_3^{\prime}(1,\xi) \, \int_0^1 d \rho \, (1-\rho + \rho \, \xi \, (1-\xi))^{-2 + \varepsilon}
  \label{eqdefT32}
\end{align}
The first term of eq.~(\ref{eqdefT32}) is finite whereas the second one diverges. Let us focus on this second term. Integrating easily over $\rho$, we end for this term with
\begin{align}
  T^{\prime}_3 &\equiv \int_0^1 d \xi \, H_3^{\prime}(1,\xi) \, \int_0^1 d \rho \, (1-\rho + \rho \, \xi \, (1-\xi))^{-2 + \varepsilon} \notag \\
  &= \frac{1}{1-\varepsilon} \, \int_0^1 d \xi \, \frac{1}{1 - \xi \, (1-\xi)} \, \left( \xi \, m_1^2 + (1-\xi) \, m_2^2 \right)^{1 - 2 \, \varepsilon} \, \left[ \xi^{-1 + \varepsilon} \, (1-\xi)^{\varepsilon} + (1-\xi)^{-1+\varepsilon} \, \xi^{\varepsilon} - 1 \right]
  \label{eqdefTp31}
\end{align}
The integral over $\xi$ of the second term in the square bracket can be obtained from the integral of first term by changing $m_1 \leftrightarrow m_2$. The third term in the square bracket integrated over $\xi$ leads to finite terms. Thus, let us focus on the integral of the first term in the square bracket which is of the type
\begin{align}
  V_1 &= \int_0^1 d \xi \, F(\xi) \, \left( \frac{1}{\varepsilon} \, \delta(\xi) + \frac{1}{(\xi)_+} + \varepsilon \, \left( \frac{\ln(\xi)}{\xi} \right)_+ \right)
  \label{eqtmpTp31}
\end{align}
with\begin{align}
  F(\xi) &= \frac{1}{1 - \xi \, (1-\xi)} \, (\xi \, m_1^2 + (1-\xi) \, m_2^2)^{1 - 2 \, \varepsilon} \, (1-\xi)^{\varepsilon} \notag \\
  &= F^{(0)}(\xi) + \varepsilon \, F^{(1)}(\xi)
  \label{eqtmpTp32}
\end{align}
Using the definition of the Dirac and the ``plus'' distributions, eq.~(\ref{eqtmpTp31}) becomes
\begin{align}
  V_1 &= \frac{1}{\varepsilon} \, F(0) + \int_0^1 \frac{d \xi}{\xi} \, \left( F^{(0)}(\xi) - F^{(0)}(0) \right) + \varepsilon \, \int_0^1 \frac{d \xi}{\xi} \, \left( F^{(1)}(\xi) - F^{(1)}(0) \right) \notag \\
  &\quad {} + \varepsilon \, \int_0^1 d \xi \, \frac{\ln(\xi)}{\xi} \, \left( F^{(0)}(\xi) - F^{(0)}(0) \right)
  \label{eqdefTp32}
\end{align}
Playing with the fact that $\int_0^1 d \xi F(\xi) \, G(\xi \, (1-\xi)) = 1/2 \, \int_0^1 d \xi (F(\xi) + F(1-\xi)) \, G(\xi \, (1-\xi))$, we get for $V_1$
  \begin{align}
    V_1 &= \frac{1}{\varepsilon} \, \left( m_2^2 - i \lambda \right)^{1-2 \, \varepsilon} + \left( m_1^2 - \frac{1}{2} \, m_2^2 \right) \, \int_0^1 \frac{d \xi}{1- \xi \, (1-\xi)} + \varepsilon \, \int_0^1 \frac{d \xi}{\xi} \, \ln(1-\xi) \, (\xi \, m_1^2 + (1-\xi) \, m_2^2) \notag \\
    &\quad {} + \varepsilon \, \int_0^1 d \xi \, \ln(1-\xi) \, \frac{(1-\xi)}{1 - \xi \, (1-\xi)} \, (\xi \, m_1^2 + (1-\xi) \, m_2^2) \notag \\
    &\quad {} - 2 \, \varepsilon \, \int_0^1 \frac{d \xi}{\xi} \, (\xi \, m_1^2 + (1-\xi) \, m_2^2) \, \ln \left( 1 + \xi \, \frac{m_1^2 - m_2^2}{m_2^2 - i \, \lambda} \right) \notag \\ 
    &\quad {} - 2 \, \varepsilon \, \int_0^1 d \xi \, (\xi \, m_1^2 + (1-\xi) \, m_2^2) \, \frac{1-\xi}{1 - \xi \, (1-\xi)} \, \ln \left( 1 + \xi \, \frac{m_1^2 - m_2^2}{m_2^2 - i \, \lambda} \right) \notag \\ 
  &\quad {} - 2 \, \varepsilon \, \ln(m_2^2 - i \, \lambda) \, \left( m_1^2 - \frac{1}{2} \, m_2^2 \right) \, \int_0^1 \frac{d \xi}{1 - \xi \, (1-\xi)} + \varepsilon \, \int_0^1 d \xi \, \ln(\xi) \, \frac{m_1^2 - \xi \, m_2^2}{1 - \xi \, (1-\xi)}
    \label{eqdefTp33}
  \end{align}
  As already said, the integration of the second term of eq.~(\ref{eqdefTp31}) can be extracted from the result of eq.~(\ref{eqdefTp33}) by changing $m_1$ into $m_2$ and vice versa. The integration of the third term is just given by
  \begin{align}
    V_3 &= \int_0^1 d \xi \, \frac{\xi \, m_1^2 + (1-\xi) \, m_2^2}{1 - \xi \, (1-\xi)} \, \left[ 1 - 2 \, \varepsilon \, \ln \left( \xi \, m_1^2 + (1-\xi) \, m_2^2 - i \, \lambda \right) \right]
    \label{eqtmpTp33}
  \end{align}
Putting everything together and after some algebra, we get for $T^{\prime}_3$
\begin{align}
T^{\prime}_3 &= \frac{1}{1-\varepsilon} \, \left\{ \frac{1}{\varepsilon} \left( (m_1^2 - i \, \lambda)^{1 - 2 \, \varepsilon} + (m_2^2 - i \, \lambda)^{1 - 2 \, \varepsilon} \right) \right. \notag \\
&\quad {} + \varepsilon \left[ (m_1^2 + m_2^2) \, \int_0^1 \frac{d \xi}{\xi} \, \ln(1-\xi) + (m_1^2 + m_2^2) \, \int_0^1 d \xi \, \frac{\ln(\xi)}{1 - \xi \, (1-\xi)} \right. \notag \\
&\qquad {} - \left. \left. 2 \, m_1^2 \, \int_0^1 \frac{d \xi}{\xi} \, \ln\left( 1 + \frac{\xi \, (m_2^2-m_1^2)}{m_1^2 - i \, \lambda} \right)- 2 \, m_2^2 \, \int_0^1 \frac{d \xi}{\xi} \, \ln\left( 1 + \frac{\xi \, (m_1^2-m_2^2)}{m_2^2 - i \, \lambda} \right)   \right]\right\}
  \label{eqdefTp34}
\end{align}
The remaining integration over $\xi$ can be easily performed. Adding the finite part and using the result given by eq.~(\ref{eqUVdefI8p3}), the third term $T_3$ reads
\begin{align}
  T_3 &= \frac{1}{\varepsilon} \, \left( m_{1}^{2} + m_{2}^{2} \right) + (m_1^2 + m_2^2) - 2 \, m_1^2 \, \ln(m_1^2 - i \, \lambda) - 2 \, m_2^2 \, \ln(m_2^2 - i \, \lambda) \notag \\
&\quad {} + \varepsilon \left[ (m_1^2 + m_2^2) \, \left( 1 +\frac{2}{\xi^+ - \xi^-} \, (\dilog(\xi^-) - \dilog(\xi^+)) - \frac{\pi^2}{6} \right) \right. \notag \\
&\qquad \quad {} - 2 \, m_1^2 \, \ln(m_1^2 - i \, \lambda) - 2 \, m_2^2 \, \ln(m_2^2 - i \, \lambda) + 2 \, m_1^2 \, \ln^2(m_1^2 - i \, \lambda) + 2 \, m_2^2 \, \ln^2(m_2^2 - i \, \lambda) \notag \\
&\qquad \quad {} + 2 \, m_1^2 \, \dilog\left( \frac{m_1^2 - m_2^2}{m_2^2 - i \, \lambda} \right) + 2 \, m_2^2 \, \dilog\left( \frac{m_2^2 - m_1^2}{m_1^2 - i \, \lambda} \right) \notag \\
&\qquad \quad {} - \left. 2 \, \int_0^1 d \xi \int_0^1 d \rho \, (\xi \, m_1^2 + (1-\xi) \, m_2^2) \, 
\frac{\ln(\calftp(\rho,\xi) - i \, \lambda) - \ln(\xi \, m_1^2 + (1-\xi) \, m_2^2)}{(1 - \rho + \rho \, \xi \, (1-\xi))^2} \right]
  \label{eqdefT33}
\end{align}
The definition of the quantities $\xi^{\pm}$ is given in the appendix~\ref{intsecpar}.
\\

\noindent
The fourth term does not diverge either, it is enough to make an expansion around $\varepsilon=0$. Nevertheless, it is safer to re-write this term as
\begin{align}
  T_4 &\equiv \int_0^1 d \xi \, \int_0^1 d \rho \, H_4(\rho,\xi) \, (1 - \rho + \rho \, \xi \, (1-\xi))^{-2 + 3 \, \varepsilon} \notag \\
  &= \int_0^1 d \xi \int_0^1 d \rho \,H_4^{\prime}(\rho,\xi) \, (1 - \rho + \rho \, \xi \, (1-\xi))^{-2 +  \varepsilon} 
  \label{eqdefT41}
\end{align}
where
\begin{align}
  H_4^{\prime}(\rho,\xi) &= m_3^2 \, \rho^{\varepsilon} \, (1-\xi \, (1-\xi)) \, (1-\rho) \, (2 - \rho \, (1 - \xi \, (1-\xi))) \left( \calftp(\rho,\xi) - i \, \lambda \right)^{- 2 \, \varepsilon}
  \label{eqdefH4prime}
\end{align}
The expansion around $\varepsilon=0$ yields
\begin{align}
  T_4 &= m_3^2 \left\{ I_2 - 2 \, \varepsilon \, \int_0^1 d \xi \int_0^1 d \rho \, \frac{(1-\xi \, (1-\xi)) \, (1-\rho) \, (2 - \rho \, (1 - \xi \, (1-\xi))) \, \ln(\calftp(\rho,\xi) - i \, \lambda)}{(1- \rho \, (1-\xi \, (1-\xi)))^2} \right. \notag \\
  &\qquad \qquad {} + \left. \varepsilon \, (I_3 + I_4) \vphantom{\frac{(1-\xi \, (1-\xi)) \, (1-\rho) \, (2 - \rho \, (1 - \xi \, (1-\xi))) \, \ln(\calftp(\rho,\xi) - i \, \lambda)}{(1- \rho \, (1-\xi \, (1-\xi)))^2}} \right\}
  \label{eqdefT42}
\end{align}
where $I_2$, $I_3$ and $I_4$ are given in appendix~\ref{intsecpar}.

\section{Integrals appearing in globally UV divergent topologies}\label{intsecpar}

In this appendix, different integrals appearing in the UV globally divergent topologies are computed. Let us introduce the two roots of the polynomial $\xi^2 - \xi + 1$:
\begin{align}
  \xi^{\pm} &= \frac{1}{2} \pm i \, \frac{\sqrt{3}}{2}
  \label{eqUVdefroot}
\end{align}
which obey to the following relations
\begin{align}
  \xi^{\pm} &= e^{\pm i \, \pi/3} \notag \\
  \xi^+ &= 1 - \xi^-
  \label{eqUVproroot}
\end{align}
Let us also define $B = 1 - \xi \, (1 - \xi))$ throughout this appendix.

\subsection{$I_1$}

\begin{align}
  I_1 &= \int_0^1 d\xi \, \int_0^1 \frac{d \rho}{\rho} \, (1-\rho) \, \ln(1-\rho \, B)
  \label{eqUVintI1}
\end{align}
The $\rho$ integration can be performed easily:
\begin{align}
  I_1 &= \int_0^1 d\xi \, \left\{ - \dilog(1 - \xi \, (1-\xi)) + \frac{1}{1 - \xi \, (1-\xi)} \left[ \xi \, (1-\xi) \, \ln(\xi \, (1-\xi)) + 1 - \xi \, (1-\xi) \right] \right\}
  \label{eqUVintI11}
\end{align}
The first integral is done in appendix~\ref{intspe}, the other ones are not difficult to perform and finally we get
\begin{align}
  I_1 &= 3 + \frac{2}{\xi^+ - \xi^-} \, \left( \dilog(\xi^-) - \dilog(\xi^+) \right) - \int_0^1 d\xi \, \dilog(1 - \xi \, (1-\xi)) 
  \label{eqUVintI12}
\end{align}

\subsection{$I_2$}

This integral can be re-written as:
\begin{align}
  I_2 &\equiv \int_0^1 d\xi \, B \, \int_0^1 d\rho \, \frac{(1-\rho) \, (2-\rho \, B)}{(1 - \rho \, B)^2} \notag \\
  &= \int_0^1 d\xi \, B \, \int_0^1 d\rho \, \left[ \frac{1}{1 - \rho \, B} + \frac{\rho \, (B-1)}{(1- \rho \, B)^2} + 1 +  \frac{\rho \, (B-1)}{1- \rho \, B} \right] 
  \label{eqUVdefI2p1}
\end{align}
Integrating by part the second and the fourth terms in the squared brackets leads after the $\rho$ integration to
\begin{align}
  I_2 &= - \int_0^1 d\xi \, \ln\left( \xi \, (1-\xi) \right) = 2
  \label{eqUVdefI2p2}
\end{align}

\subsection{$I_3$}

\begin{align}
  I_3 &= \int_0^1 d\xi B \, \int_0^1  d\rho \, \frac{(1-\rho) \, (2 - \rho \, B) \, \ln(1 - \rho \, B)}{(1 - \rho \, B)^2} \\
&= \int_0^1 d\xi B \, \int_0^1  d\rho \, \left[ \frac{\ln(1- \rho \, B)}{1 - \rho \, B} + (B-1) \, \frac{\rho \, \ln(1- \rho \, B)}{(1 - \rho \, B)^2} + \ln(1- \rho \, B) \right. \notag \\
&\qquad \qquad \qquad \qquad \quad {} + \left. (B-1) \, \frac{\rho \, \ln(1- \rho \, B)}{1 - \rho \, B} \right] 
  \label{eqUVdefI3p1}
\end{align}
Integrating by part the second and the fourth terms in the squared brackets leads after the $\rho$ integration to
\begin{align}
  I_3 &= - \int_0^1 d\xi \, \left\{ 2 + \frac{1}{2} \, \ln^2\left( \xi \, (1-\xi) \right) + \frac{1 + \xi \, (1-\xi)}{1 - \xi \, (1-\xi)} \, \ln\left( \xi \, (1-\xi) \right) \right\} \\
  &= \frac{\pi^2}{6} - \frac{4}{\xi^+ - \xi^-} \, \left( \dilog\left( \xi^- \right) - \dilog\left( \xi^+ \right) \right) - 8
  \label{eqUVdefI3p2}
\end{align}

\subsection{$I_4$}

\begin{align}
  I_4 &= \int_0^1 d\xi \, B \, \int_0^1  d\rho \, \frac{(1-\rho) \, (2 - \rho \, B) \, \ln(\rho)}{(1 - \rho \, B)^2} \\
  &= \int_0^1 d\xi \, B \, \int_0^1  d\rho \, \left[ \frac{\ln(\rho)}{1 - \rho \, B} + (B-1) \, \frac{\rho \, \ln(\rho)}{(1 - \rho \, B)^2} + \ln(\rho) + (B-1) \, \frac{\rho \, \ln(\rho)}{1 - \rho \, B} \right] 
  \label{eqUVdefI4p1}
\end{align}
Integrating by part the second and the fourth terms in the squared brackets leads after the $\rho$ integration to
\begin{align}
  I_4 &= - \int_0^1 d\xi \, \left\{ \dilog\left( 1 -  \xi \, (1-\xi) \right) + 1 + \frac{\xi \, (1-\xi)}{1 - \xi \, (1-\xi)} \, \ln\left( \xi \, (1-\xi) \right) \right\} \\
  &= -  \int_0^1 d\xi \,\dilog\left( 1 -  \xi \, (1-\xi) \right) - 3 - \frac{2}{\xi^+ - \xi^-} \, \left( \dilog\left( \xi^- \right) - \dilog\left( \xi^+ \right) \right)
  \label{eqUVdefI4p2}
\end{align}

\subsection{$I_5$}

\begin{align}
  I_5 &= \int_0^1 d \xi \int_0^1 d \rho \, \frac{(1-\rho) \, \xi \, (1-\xi)}{(1 - \rho \, B)^3} \notag \\
  &= \int_0^1 d \xi \, \frac{1-B}{2 \, B^2} \, \left[ 2 \, \left( (1-B)^{-1} - 1  \right) - (1-B) \left( (1-B)^{-2} - 1 \right)  \right] \notag \\
  &= \frac{1}{2}
  \label{eqUVdefI5p1}
\end{align}

\subsection{$I_6$}

\begin{align}
  I_6 &= \int_0^1 d \xi \int_0^1 d \rho  \, \frac{(1-\rho) \, \xi \, (1-\xi) \, \ln(\rho)}{(1-\rho \, B)^3} \notag \\
  &= \int_0^1 d \xi \frac{1-B}{B} \, \int_0^1 d \rho \left[ \frac{\ln(\rho)}{(1-\rho \, B)^2} - (1-B) \, \frac{\ln(\rho)}{(1-\rho \, B)^3} \right]
  \label{eqUVdefI6p1}
\end{align}
Let us introduce $n$ a strictly positive integer ($n=2,3$) and integrating by part over the $\rho$ variable leads to
\begin{align}
  \int_0^1 d \rho \, \ln(\rho) \, (1 - \rho \, B)^{-n} &= - \frac{1}{(n-1) \, B} \, \int_0^1 d \rho \, \frac{\left( 1 - (1-\rho \, B)^{n-1} \right)}{\rho} \, (1-\rho \, B)^{-n+1} \notag \\
  &= \left\{
  \begin{array}{cc}
    \frac{1}{B} \, \ln(1-B) & \text{for $n=2$} \\
    \frac{1}{2 \, B} \, \left( 1 + \ln(1-B) - (1-B)^{-1} \right) & \text{for $n=3$}
  \end{array}
\right.
  \label{eqUVtmpI6p}
\end{align}
With this last result, the integral $I_6$ can be put under the following form
\begin{align}
  I_6 &= \frac{1}{2} \left\{ 2 \, \int_0^1 d \xi \, \frac{\ln(\xi)}{(1 - \xi \, (1-\xi))^2} - 2 \, \int_0^1 d \xi \, \ln(\xi) + \int_0^1 d \xi \, \frac{1}{1 - \xi \, (1-\xi)} - 1 \right\}
  \label{eqUVdefI6p2}
\end{align}
Then, using $\int_0^1 d x \, \ln(x)/(x-A)^2 = 1/A \, \ln( (A-1)/A) )$ where $A$ is a complex number with a non zero imaginary part, and the properties of the roots $\xi^+$ and $\xi^-$ yields
\begin{align}
  I_6 &= - \frac{2}{(\xi^+ - \xi^-)^3} \, \left[ \dilog(\xi^-) - \dilog(\xi^+) \right] + \frac{1}{2}
  \label{eqUVdefI6p3}
\end{align}

\subsection{$I_7$}

\begin{align}
  I_7 &= \int_0^1 d \xi \int_0^1 d \rho \, \frac{(1-\rho) \, \xi \, (1-\xi) \, \ln(1 - \rho \, B)}{(1-\rho \, B)^3} \notag \\
  &= \int_0^1 d \xi \, \frac{1-B}{B} \, \int_0^1 d \rho \left[ \frac{\ln(1-\rho \, B)}{(1-\rho \, B)^2} - (1-B) \, \frac{\ln(1-\rho \, B)}{(1 - \rho \, B)^3} \right]
  \label{eqUVdefI7p1}
\end{align}
For any integer $n > 1$, one can show that
\begin{align}
  \int_0^1 d \rho \, \frac{\ln(1-\rho \, B)}{(1 - \rho \, B)^n} &= \frac{1}{B \, (n-1)} \left[ \frac{\ln(1-B)}{(1-B)^{n-1}} + \frac{1}{(n-1) \, (1-B)^{n-1}} - \frac{1}{n-1} \right]
  \label{eqtmpI7p}
\end{align}
With this result, the $\rho$ integration can be easily performed leading to
\begin{align}
  I_7 &= \frac{1}{2} \left\{ 2 \, \int_0^1 d \xi \, \frac{\ln(\xi)}{(1 - \xi \, (1-\xi))^2} + \int_0^1 d \xi \, \frac{1}{1 - \xi \, (1-\xi)} + \frac{1}{2} \right\}
  \label{eqUVdefI7p2}
\end{align}
The integrals over $\xi$ are of the same type as those for $I_6$, thus using the same techniques we readily get
\begin{align}
  I_7 &=  - \frac{2}{(\xi^+ - \xi^-)^3} \, \left[ \dilog(\xi^-) - \dilog(\xi^+) \right] + \frac{1}{4}
  \label{eqUVdefI7p3}
\end{align}

\subsection{$I_8$}

\begin{align}
  I_8 &= \int_0^1 d \xi \int_0^1 d \rho \, \frac{(\xi \, m_1^2 + (1-\xi)\, m_2^2) \, \ln(\rho)}{(1- \rho \, B)^2}
  \label{eqUVdefI8p1}
\end{align}
The $\rho$ integration is performed using eq.~(\ref{eqUVtmpI6p}), yielding
\begin{align}
  I_8 &= \int_0^1 d \xi \, \frac{\xi \, m_1^2 + (1-\xi) \, m_2^2}{1 - \xi \, (1-\xi)} \, \ln(\xi \, (1-\xi))
  \label{eqUVdefI8p2}
\end{align}
The $\xi$ integration can be readily performed leading to
\begin{align}
  I_8 &= (m_1^2 + m_2^2) \, \frac{1}{\xi^+ - \xi^-} \, \left( \dilog(\xi^-) - \dilog(\xi^+) \right)
  \label{eqUVdefI8p3}
\end{align}

\section{Integral over $\dilog\left( 1 -  \xi \, (1-\xi) \right)$}\label{intspe}

In several places, the following integral appears
\begin{align}
  L &= \int_0^1 d\xi \,\dilog\left( 1 -  \xi \, (1-\xi) \right)
  \label{eqUVintspe1}
\end{align}
This integral is difficult to perform directly, nevertheless one can study the integral
\begin{align}
L_t &= \int_0^1 d\xi \, \int_0^1 \frac{d\rho}{\rho} \, \ln\left( 1 - \rho \, (1 - \xi \, (1-\xi))) \right)
  \label{eqUVintsub1}
\end{align}
Performing the $\rho$ integration first leads to
\begin{align}
  L_t &= - L
  \label{eqUVintsub2}
\end{align}
But now, the integral $L_t$ can be computed by integrating firstly on $\xi$. Let us note the two roots of the polynomial $-\rho \, \xi^2 + \rho \, \xi + 1 - \rho$ by
\begin{align}
  \barxi^{\pm} &= \frac{1}{2} \mp \frac{1}{2} \, \frac{\sqrt{\rho \, (4 - 3 \, \rho)}}{\rho}
  \label{eqUVrootofLt}
\end{align}
these roots obey $1 - \barxi^+ = \barxi^-$, thus $L_t$ becomes
\begin{align}
  L_t &= \int_0^1 \frac{d\rho}{\rho} \, \int_0^1 d\xi \, \ln\left( \rho \, (\barxi^- - \xi) \, (\xi - \barxi^+) \right)
  \label{eqUVintsub3}
\end{align}
Since the argument of the logarithm is a product of positive factors, it can be split and the $\xi$ integration is then performed easily, yielding
\begin{align}
L_t &= \int_0^1 \frac{d\rho}{\rho} \, \left\{ \ln(\rho) + \left( 1 - \frac{\sqrt{\rho \, (4 - 3 \, \rho)}}{\rho} \right) \, \ln\left[ \frac{1}{2} \, \left(\frac{\sqrt{\rho \, (4 - 3 \, \rho)}}{\rho}-1\right) \right] \right. \notag \\
&\quad {} + \left. \left( 1 + \frac{\sqrt{\rho \, (4 - 3 \, \rho)}}{\rho} \right) \, \ln\left[ \frac{1}{2} \, \left(\frac{\sqrt{\rho \, (4 - 3 \, \rho)}}{\rho}+1\right) \right] - 2 \right\}
  \label{eqUVintsub4}
\end{align}
To integrate over $\rho$, we use the Euler third change of variable: $\rho = 4/(t^2+3)$ which rationalises the square root and with some algebra, eq.~(\ref{eqUVintsub4}) becomes
\begin{align}
  L_t &= \int_1^{\infty} \frac{2 \, t \, dt}{t^2+3} \, \left[ \ln\left( \frac{t^2-1}{t^2+3} \right) + t \, \ln\left( \frac{t+1}{t-1} \right) - 2 \right]
  \label{eqUVintsub5}
\end{align}
Then, the integral over $t$ is split into three parts
\begin{align}
  L_t^{(1)} &= \int_1^{\infty} \frac{2 \, t \, dt}{t^2+3} \, \ln\left( \frac{t^2-1}{t^2+3} \right)
  \label{eqUVintLt1} \\
  L_t^{(2)} &= 2 \, \int_1^{\infty} dt \, \left[ \ln\left( \frac{t+1}{t-1} \right) - \frac{2 \, t}{t^2+3} \right] \label{eqUVintLt2} \\
  L_t^{(3)} &= -6 \, \int_1^{\infty} \frac{dt}{t^2+3} \, \ln\left( \frac{t+1}{t-1} \right) \label{eqUVintLt3} 
\end{align}
Let us then compute the different parts.\\

\noindent
For the first integral $L_t^{(1)}$ we make the following change of variable $t = 1/\sqrt{v}$, this leads to
\begin{align}
  L_T^{(1)} &= \int_0^1 \frac{dv}{v \, (1 + 3 \, v)} \, \ln\left( \frac{1-v}{1+3 \, v} \right)
  \label{eqUVintLt11}
\end{align}
A partial fraction decomposition and a splitting of the logarithm because the numerator and the denominator of its argument are positive yields
\begin{align}
  L_t^{(1)} &= \int_0^1 dv \, \frac{\ln(1-v)}{v} - \int_0^1 dv \, \frac{\ln(1+3 \, v)}{v} - 3 \, \int_0^1 dv \, \frac{\ln(1-v)}{1 + 3 \, v} + 3 \, \int_0^1 dv \, \frac{\ln(1 + 3 \, v)}{1 + 3 \, v}
  \label{eqUVintLt12}
\end{align}
All the integrals can be easily performed and we end with
\begin{align}
  L_t^{(1)} &= \dilog(-3) + \dilog\left( \frac{3}{4} \right) - \frac{\pi^2}{6} + 2 \, \ln^2(2)
  \label{eqUVintLt13}
\end{align}
\\

\noindent
For the second one $L_t^{(2)}$, the integrals over the two terms, taken separately, in the squared brackets of eq.~(\ref{eqUVintLt12}) diverge at $t = \infty$ but the integral of their sum is convergent. Thus, let us introduce a large $t$ cut-off $\Lambda$ such that
\begin{align}
  L_t^{(2)} &= \lim_{\Lambda \rightarrow \infty} \, M(\Lambda)
  \label{eqUVintLt20}
\end{align}
with
\begin{align}
  M(\Lambda) &= 2 \, \int_1^{\Lambda} dt \, \ln\left(  \frac{t+1}{t-1} \right) - 4 \, \int_1^{\Lambda} \frac{t \, dt}{t^2+3}
  \label{eqUVdefMLambda}
\end{align}
The logarithm can be split into two others and for the last integral of eq.~(\ref{eqUVdefMLambda}), we set $t=\sqrt{u}$, this leads to
\begin{align}
  M(\Lambda) &= 2 \, \left[  \int_1^{\Lambda} dt \, \ln(t+1) - \int_1^{\Lambda} dt \, \ln(t-1) - \int_1^{\Lambda^2} \frac{du}{u+3} \right]
  \label{eqUVdefMLambda1}
\end{align}
The integrals can be easily performed, the $\ln(\Lambda)$ terms vanish and we end with
\begin{align}
  M(\Lambda) &= 2 \, \left[ \left( \Lambda + 1 \right) \, \ln\left( 1 + \frac{1}{\Lambda} \right) - \left( \Lambda - 1 \right) \, \ln\left( 1 - \frac{1}{\Lambda} \right) - \ln\left( 1 + \frac{3}{\Lambda^2} \right)\right]
  \label{eqUVdefMLambda2}
\end{align}
Thus, taking the limit $\Lambda \rightarrow \infty$, we get
\begin{align}
  L_t^{(2)} &= 4
  \label{eqUVintLt21}
\end{align}
\\

\noindent
Lastly, for the third integral, we set $t = 1/u$, and we have to compute
\begin{align}
  L_t^{(3)} &= - 2 \, \int_0^1 \frac{du}{u^2 + 1/3} \, \ln\left( \frac{1+u}{1-u} \right)
  \label{eqUVintLt30}
\end{align}
A partial fraction decomposition and a splitting of the logarithm leads to
\begin{align}
  L_t^{(3)} &= i \, \sqrt{3} \, \left[ \int_0^1 du \, \frac{\ln(u+1)}{u - i/\sqrt{3}} - \int_0^1 du \, \frac{\ln(1-u)}{u - i/\sqrt{3}} - \int_0^1 du \, \frac{\ln(u+1)}{u + i/\sqrt{3}} + \int_0^1 du \, \frac{\ln(1-u)}{u + i/\sqrt{3}} \right]
  \label{eqUVintLt31}
\end{align}
Then, a change of variable is performed in each integral in such way that after this change of variable the argument of the logarithm is just the new integration variable $v$
\begin{align}
L_t^{(3)} &= i \, \sqrt{3} \, \left[ \int_1^2 dv \, \frac{\ln(v)}{v - (1 + i/\sqrt{3})}  + \int_0^1 dv \, \frac{\ln(v)}{v - (1 - i/\sqrt{3})} - \int_1^2 dv \, \frac{\ln(v)}{v - (1 - i/\sqrt{3})} \right. \notag \\
&\qquad \qquad {} - \left. \int_0^1 dv \, \frac{\ln(v)}{v - (1 + i/\sqrt{3})} \right]
  \label{eqUVintLt32}
\end{align}
The integration between $1$ and $2$ can be safely split because the pole is far away from the real axis: $\int_1^2 dv = \int_0^2 dv - \int_0^1 dv$ and then for the integrals between $0$ and $2$ we set $v=2 \, t$. After having performed the integrals, we end with
\begin{align}
L_t^{(3)} &= i \, \sqrt{3} \, \left\{ \ln(2) \, \left[ \ln\left( \frac{1 - i/\sqrt{3}}{-1 - i/\sqrt{3}} \right) - \ln\left( \frac{1 + i/\sqrt{3}}{-1 + i/\sqrt{3}} \right) \right] + \dilog\left( \frac{2}{1 + i/\sqrt{3}} \right) \right. \notag \\
&\qquad \qquad {} - \left. \dilog\left( \frac{2}{1- i/\sqrt{3}} \right) - 2 \, \dilog\left( \frac{1}{1 + i/\sqrt{3}} \right) + 2 \, \dilog\left( \frac{1}{1- i/\sqrt{3}} \right) \right\}
  \label{eqUVintLt33}
\end{align}
\\

\noindent
So putting things together, and expressing all the arguments of the dilogarithms in terms of $\xi^+$ and $\xi^-$, the following result is obtained
\begin{align}
  L &= \frac{\pi^2}{6} - \dilog((\xi^+ - \xi^-)^2) - \dilog\left( \left[\frac{1}{2} - \xi^+ \right] \, \left[ \frac{1}{2} - \xi^- \right] \right) - 2 \, \ln^2(2) - 4  \notag \\
  &\quad {}  - i \, \sqrt{3} \, \left[  \vphantom{\frac{1+\xi^-}{2}} \ln(2) \, \left( \ln(-\xi^-) - \ln(-\xi^+) \right) + \dilog(1+\xi^-) - \dilog(1+\xi^+) \right. \notag \\
&\qquad \qquad \quad {} - \left. 2 \, \dilog\left( \frac{1+\xi^-}{2} \right) + 2 \, \dilog\left( \frac{1+\xi^+}{2} \right) \right]
  \label{eqUVintspe2}
\end{align}

\section{One-loop two-point function at $O(\varepsilon)$ \label{ol2Oe}}

In this appendix, the computation of the one-loop two-point function at order $O(\varepsilon^0)$ and $O(\varepsilon^1)$ is presented. These functions appear in the $\varepsilon$ expansion of the UV divergent topologies $\calt_{22\_2\_1\_1}$ and $\calt_{23\_1\_2\_1}$. Let us remind their expression\footnote{As the $\rho$ and $\xi$ dependence is implicit through the argument $G$, $V$ and $C$, contrarily to the main text we drop them from the arguments of the functions $\wtI_2^{(0)}$ and  $\wtI_2^{(1)}$ to lighten the notations.}
\begin{align}
  \wtI_2^{(1)}(\Sigma_{(1)};G,V,C) &= \int_0^1 du \, \ln\left( G \, u^2 - 2 \, V \, u -C - i \, \lambda \right) \label{eqdefI20eps0} \\
  \wtI_2^{(2)}(\Sigma_{(1)};G,V,C) &= \int_0^1 du \, \ln^2\left( G \, u^2 - 2 \, V \, u -C - i \, \lambda \right)
  \label{eqdefI21eps0}
\end{align}
where $G$ is real and $V$ and $C$ are either real or complex. In the latter case, the vanishing quantity $\lambda$, inherited from the Feynman prescription for the propagators, can be neglected with respect to the non zero imaginary parts of $V$ and $C$. 
Eqs.~(\ref{eqdefI20eps0}) and (\ref{eqdefI21eps0}) become
\begin{align}
  \wtI_2^{(1)}(\Sigma_{(1)};G,V,C) &= \int_0^1 du \left[ \ln\left( G - i \, \lambda \right) + \ln\left( (u-u^{+}) \, (u-u^{-}) \right) \right] \label{eqdefI20eps1} \\
  \wtI_2^{(2)}(\Sigma_{(1)};G,V,C) &= \int_0^1 du \left[ \ln\left( G - i \, \lambda \right) + \ln\left( (u-u^{+}) \, (u-u^{-}) \right) \right]^2
  \label{eqdefI21eps1}
\end{align}
where $u^{+}$ and $u^{-}$ are the two roots of the polynomial in argument of the logarithm in eqs.~(\ref{eqdefI20eps0}) and (\ref{eqdefI21eps0})
\begin{align}
  u^{\pm} &= \frac{V \pm \sqrt{V^2 + G \, (C + i \, \lambda)}}{G}
  \label{eqdefupm1}
\end{align}
Expanding the terms in square bracket yields
\begin{align}
  \wtI_2^{(1)}(\Sigma_{(1)};G,V,C) &= \int_0^1 du \, \ln(u-u^{+}) + \int_0^1 du \, \ln(u-u^{-}) + \ln\left( G- i \, \lambda \right) + \eta\left( -u^{+},-u^{-} \right) \label{eqdefI20eps2} \\
  \wtI_2^{(2)}(\Sigma_{(1)};G,V,C) &= \int_0^1 du \, \ln^2(u-u^{+}) + \int_0^1 du \ln^2(u-u^{-}) + 2 \, \int_0^1 du \, \ln(u-u^{+}) \, \ln(u-u^{-}) \notag \\
  &\quad 2 \, \left[ \ln\left( G- i \, \lambda \right) + \eta\left( -u^{+},-u^{-} \right) \right] \, \left[ \int_0^1 du \, \ln(u-u^{+}) + \int_0^1 du  \, \ln(u-u^{-}) \right] \notag \\
  &\quad {} + \left[ \ln\left( G- i \, \lambda \right) + \eta\left( -u^{+},-u^{-} \right) \right]^2
  \label{eqdefI21eps2}
\end{align}
The integrals in the left-hand side of eq.~(\ref{eqdefI20eps2}) can be readily computed leading to
\begin{align}
  \wtI_2^{(1)}(\Sigma_{(1)};G,V,C) &= (1-u^{+}) \, \ln(1-u^{+}) + u^{+} \, \ln(- u^{+}) + (1-u^{-}) \, \ln(1-u^{-}) + u^{-} \, \ln(- u^{-}) \notag \\
  &\quad {} + \ln(G - i \, \lambda) + \eta(-u^{+}, -u^{-}) - 2 
  \label{eqdefI20eps3}
\end{align}
Given eq.~(\ref{eqdefI21eps2}), the new integrals, with respect to the one-loop two-point function, to be calculated are
\begin{align}
  \cali_1 &= \int^1_0 dx \, \ln^2(x-A)
  \label{eqdefcali10}
\end{align}
and
\begin{align}
  \cali_2 &= \int_0^1 dx \, \ln(x-A) \, \ln(x-B)
  \label{eqdefcali20}
\end{align}
where $A$ and $B$ are complex numbers with non vanishing imaginary parts. A primitive of $\ln^2(x-A)$ is 
\[
  (x-A) \, \ln^2(x-A) - 2 \, (x-A) \, \ln(x-A) + 2 \, x
\]
and thus the integral $\cali_1$ can be readily computed. For the integral $\cali_2$, after an integration by part, eq.~(\ref{eqdefcali20}) can be put under the following form
\begin{align}
  \cali_2 &= 1 + A \, \left[ \ln(1-A) - \ln(-A) \right] + (1-B) \, \ln(1-A) \, \ln(1-B) - \ln(1-A)  \notag \\
  &\quad  + B \, \ln(-A) \, \ln(-B) - \left[ (1-B) \, \ln(1-B) - 1 + B \, \ln(-B) \right] \notag \\
  &\quad {} - (A-B) \, \left[ \int_0^1 dx \, \frac{\ln(x-B) - \ln(A-B)}{x-A} + \ln (A-B) \, \int_0^1 \frac{dx}{x-A} \right]
  \label{eqdefcali21}
\end{align}
The first integral in the square bracket appears in the computation of the one-loop three-point function for example and can be expressed in terms of dilogarithms. Using this known result leads to
\begin{align}
  \cali_2 &= 2 - (1-A) \, \ln(1-A) - (1-B) \, \ln(1-B) - A \, \ln(-A) - B \,  \ln(-B) + B \, \ln(-A) \, \ln(-B) \notag \\
  &\quad {} + (1-B) \, \ln(1-A) \, \ln(1-B) - (A-B) \, \ln(A-B) \, \left[ \ln(1-A) - \ln(-A) \right] \notag \\
&\quad {} - (A-B) \, \left[ \dilog\left( \frac{A}{A-B} \right) - \dilog\left( \frac{A-1}{A-B} \right) + \eta\left( -B, \frac{1}{A-B} \right) \, \ln\left( \frac{A}{A-B} \right) \right. \notag \\
&\qquad \qquad \qquad \quad {} - \left. \eta\left(1-B,\frac{1}{A-B} \right) \, \ln\left( \frac{A-1}{A-B} \right) \right]
  \label{eqdefcali22}
\end{align}
Putting everything together, we end with
\begin{align}
  \hspace{2em}&\hspace{-2em}\wtI_2^{(2)}(\Sigma_{(1)};G,V,C) \notag \\
  &= (1-u^{+}) \, \ln^2(1-u^{+}) - 4 \, (1-u^{+}) \, \ln(1-u^{+}) + u^{+} \, \ln^2(-u^{+}) - 4 \, u^{+} \, \ln(- u^{+}) \notag \\
  &\quad {} + (1-u^{-}) \, \ln^2(1-u^{-}) - 4 \, (1-u^{-}) \, \ln(1-u^{-}) + u^{-} \, \ln^2(-u^{-}) - 4 \, u^{-} \, \ln(- u^{-}) \notag \\
  &\quad {} + 8 + 2 \, (1 - u^{-}) \, \ln(1-u^{+}) \, \ln(1-u^{-}) + 2 \, u^{-} \, \ln(-u^{+}) \, \ln(- u^{-}) \notag \\
&\quad {} - 2 \, (u^{+}-u^{-}) \, \left[ \dilog\left( \frac{u^{+}}{u^{+}-u^{-}} \right) - \dilog\left( \frac{u^{+}-1}{u^{+}-u^{-}} \right) + \eta\left( -u^{-}, \frac{1}{u^{+}-u^{-}} \right) \, \ln\left( \frac{u^{+}}{u^{+}-u^{-}} \right) \right. \notag \\
&\qquad \qquad \qquad \quad {} - \left. \eta\left( 1-u^{-}, \frac{1}{u^{+}-u^{-}} \right) \, \ln\left( \frac{u^{+}-1}{u^{+}-u^{-}} \right) \right] \notag \\
&\quad {} - 2 \, (u^{+}-u^{-}) \, \ln(u^{+}-u^{-}) \, \ln\left( \frac{u^{+}-1}{u^{+}} \right) + \left[ \ln\left( G- i \, \lambda \right) + \eta\left( -u^{+},-u^{-} \right) \right]^2 \notag \\
&\quad {} + 2 \, \left[ \ln\left( G- i \, \lambda \right) + \eta\left( -u^{+},-u^{-} \right) \right] \, \left[ (1-u^{+}) \, \ln(1-u^{+}) + u^{+} \, \ln(- u^{+}) \right. \notag \\
&\qquad \qquad \qquad \qquad \qquad \qquad \qquad \qquad \quad {} + \left. (1-u^{-}) \, \ln(1-u^{-}) + u^{-} \, \ln(- u^{-}) -2 \right]
  \label{eqdefcali23}
\end{align}
\\

\noindent
The formulae given by eqs.~(\ref{eqdefI20eps3}) and (\ref{eqdefcali23}) may not be suited for numerical applications in the case of special kinematics: $G \rightarrow 0$ or $u^{+} \rightarrow u^{-}$, etc. They have to be replaced by dedicated formula which will not be presented here for the sake of simplicity.

\section{One-loop three-point function integrated over a square \label{ol3ptsq}}

Despite the fact that the Feynman parameter integration domain is not the usual 2-simplex, the method developed in refs. \cite{paper1}, \cite{paper2} and \cite{paper3} can be applied to compute this function analytically. This will be the purpose of this appendix.\\

\noindent
To start with, we slightly change the notation to match the one of these articles\footnote{We also drop the dependence on $\rho$ and $\xi$ which plays no role here.}.
\begin{align}
   \wtI_{3}^{\,(0)}(K_{(2)};G,V,C) &= 2 \, \int_0^1 d x_1 \, \int_0^1 d x_2 \, \frac{1}{D(x_1,x_2) - i \, \lambda}
  \label{eq1I340}
\end{align}
where the short-hand notation $D(x_1,x_2)$ means
\begin{align}
  D(x_1,x_2) &= X^T \cdot \cG \cdot X - 2 \, \cV^T \cdot X - \cC
  \label{eqdefD}
\end{align}
with
\[
  X = \left[
  \begin{array}{c}
    x_1 \\
    x_2
  \end{array}
\right]
\]
and $\cG = 2 \, G$, $\cV = 2 \, V$ and $\cC = 2 \, C$.
Changing the power of the denominator in the integrand of eq. (\ref{eq1I340}) and applying once the ``Stokes-like'' identity, we get
\begin{align}
\wtI_{3}^{\,(0)}(K_{(2)};G,V,C) &= - \int_0^{\infty} \, \frac{d \chi}{\Delta_2 - \chi + i \, \lambda} \, \int_0^1 dx \, \left[ \left( 1 - \left( \cG^{-1} \cdot \cV \right)_1 \right) \, \frac{1}{D(1,x) + \chi - i \, \lambda} \right. \notag \\
&\qquad \qquad \qquad \qquad \qquad \qquad \qquad {} + \left( \cG^{-1} \cdot \cV \right)_1 \, \frac{1}{D(0,x) + \chi - i \, \lambda} \notag \\
&\qquad \qquad \qquad \qquad \qquad \qquad \qquad {} + \left( 1 - \left( \cG^{-1} \cdot \cV \right)_2 \right) \, \frac{1}{D(x,1) + \chi - i \, \lambda} \notag \\
&\qquad \qquad \qquad \qquad \qquad \qquad \qquad {} + \left. \left( \cG^{-1} \cdot \cV \right)_2 \, \frac{1}{D(x,0) + \chi - i \, \lambda} \right]
  \label{eq1T341}
\end{align}
with
\begin{equation}
\begin{aligned}
  D(1,x) &= x^2 \, \cG_{22} - 2 \, (\cV_2 - \cG_{12}) \, x - (\cC + 2 \, \cV_1 - \cG_{11}) \\
  D(0,x) &= x^2 \, \cG_{22} - 2 \, \cV_2 \, x - \cC \\
  D(x,1) &= x^2 \, \cG_{11} - 2 \, (\cV_1 - \cG_{12}) \, x - (\cC + 2 \, \cV_2 - \cG_{22}) \\
  D(x,0) &= x^2 \, \cG_{11} - 2 \, \cV_1 \, x - \cC
\end{aligned}
  \label{eqdefD1}
\end{equation}
\\

\noindent
At this stage, one can integrate over the variable $\chi$ using appendix D of ref.~\cite{paper1}. Then introducing the following quantities to get a more compact formula
\begin{equation}
\begin{aligned}
  \bbar_1 &= \det(\cG) \, \left( 1 - \left( \cG^{-1} \cdot \cV \right)_1 \right) = \cG_{12} \, \cV_2 - \cG_{22} \, \cV_1 + \cG_{11} \, \cG_{22} - \cG_{12}^2 \\
  \bbar_2 &= \det(\cG) \, \left( \cG^{-1} \cdot \cV \right)_1 = \cG_{22} \, \cV_1 - \cG_{12} \, \cV_2 \\
  \bbar_3 &= \det(\cG) \, \left( 1 - \left( \cG^{-1} \cdot \cV \right)_2 \right) = \cG_{12} \, \cV_1 - \cG_{11} \, \cV_2 + \cG_{11} \, \cG_{22} - \cG_{12}^2 \\
  \bbar_4 &= \det(\cG) \, \left( \cG^{-1} \cdot \cV \right)_2 = \cG_{11} \, \cV_2 - \cG_{12} \, \cV_1
  \label{eqdefbbari}
\end{aligned}
\end{equation}
\begin{align*}
  D^{[1]}(x) = D(1,x) &\qquad D^{[2]} = D(0,x) \\ 
  D^{[3]}(x) = D(x,1) &\qquad D^{[4]} = D(x,0)
\end{align*}
eq. (\ref{eq1T341}) becomes
\begin{align}
  \wtI_{3}^{\,(0)}(K_{(2)};G,V,C) &= \sum_{i=1}^4 \, \frac{\bbar_i}{\det(\cG)} \, \int_0^1 \, \frac{d x}{D^{[i]}(x) + \Delta_2} \, \left[ \ln \left( D^{[i]}(x) - i \, \lambda \right) - \ln\left( - \Delta_2 - i \, \lambda \right) \right]
  \label{eq1T343}
\end{align}
with
\begin{align}
  \Delta_2 &\equiv \cV^T \cdot \cG^{-1} \cdot \cV + \cC \notag \\
  &= \frac{\cG_{11} \, \cV_2^2 - 2 \, \cG_{12} \, \cV_1 \, \cV_2 + \cG_{22} \, \cV_1^2 + \cC ( \cG_{11} \, \cG_{22} - \cG_{12}^2)}{\cG_{11} \, \cG_{22} - \cG_{12}^2}
  \label{eqdefDelta2}
\end{align}

\noindent
Another way to evaluate eq.~(\ref{eq1T341}) is to apply once more the ``Stokes-like'' identity on the integration over $x$.
To do that, let us note that the four terms in the square brackets of this last equation have the form
\begin{align}
  K &= \int_0^{\infty} \frac{d \chi}{\chi - \Delta_2 - i \, \lambda} \, \int_0^1 \frac{d x}{\chi + E(x) - i \, \lambda}
  \label{eqdefK0}
\end{align}
where the polynomial $E(x)$ has the following structure
\begin{align}
  E(x) &= x^2 \, \hG - 2 \, \hV \, x - \hC
  \label{eqdefEx}
\end{align}
Then by changing the power of the denominator in eq. (\ref{eqdefK0}) and applying the identity [A.5] of ref.~\cite{paper1}, we end with
\begin{align}
  K &= \left( 1 - \frac{\hV}{\hG} \right) \, L_3^4\left(\Delta_2,\Delta_1,E(1)\right) + \frac{\hV}{\hG} \, L_3^4\left(\Delta_2,\Delta_1,E(0)\right)
  \label{eqdefK1}
\end{align}
where the function $L_3^4(\Delta_2,\Delta_1,D)$ is the same as the one which appears in the computation of the genuine one-loop three-point function (cf.~eq.~(2.35) of ref.~\cite{paper1}). Let us give it here for the sake of completeness
\begin{align}
  L_3^4\left(\Delta_2,\Delta_1,D\right) &= \int_0^1 \frac{d z}{(D+\Delta_1) \, z^2 + \Delta_2 - \Delta_1} \, \left[ \ln \left( (D+\Delta_1) \, z^2 - \Delta_1 - i \, \lambda \right) - \ln \left( -\Delta_2 - i \, \lambda \right) \right]
  \label{eqdefL34}
\end{align}
and
\begin{align}
  \Delta_1 &= \frac{\hV^2 + \hG \, \hC}{\hG}
  \label{eqdefDelta1}
\end{align}
The function $\wtI_{3}^{\,(0)}(K_{(2)};G,V,C)$ can then be put under the following form, introducing an upper index $[i]$ to label the sector
\begin{align}
  \wtI_{3}^{\,(0)}(K_{(2)};G,V,C) &= \sum_{i=1}^4 \frac{\bbar_i}{\det(\cG)} \, \sum_{j=1}^2 \frac{\bbsj{j}{i}}{\detgsj{i}} \, L_3^4\left(\Delta_2,\Delta_1^{[i]},\tD^{[i]}_j\right)
  \label{eq1I345}
\end{align}
The different quantities depending on the sector $i$ are given as a function of the elements of the matrix $\cG$, the elements of the vector $\cV$ and the scalar $\cC$.\\

\hspace{-0.5cm}\begin{minipage}[c]{0.7\textwidth}
  \textbf{For sector 1}\\
\begin{align*}
  \detgsj{1} &= \cG_{22} \notag \\
  \bbsj{1}{1} &= \cG_{22} - \cV_2 + \cG_{12} \notag \\
  \bbsj{2}{1} &= \cV_2 - \cG_{12} \notag \\
  \tD^{[1]}_1 &= \cG_{22} + 2 \, \cG_{12} + \cG_{11} - 2 \, ( \cV_1 + \cV_2) - \cC \notag \\
  \tD^{[1]}_2 &= \cG_{11} - 2 \, \cV_1 - \cC \notag \\
\Delta_1^{[1]} &= \frac{\cV_2^2 - 2 \, \cG_{12} \, \cV_2 + 2 \, \cG_{22} \, \cV_1 + \cC \, \cG_{22} - \cG_{11} \, \cG_{22} + \cG_{12}^2}{\cG_{22}}
\end{align*}
\end{minipage}
\begin{minipage}[c]{0.3\textwidth}
\textbf{For sector 2}\\
\begin{align*}
  \detgsj{2} &= \cG_{22} \notag \\
  \bbsj{1}{2} &= \cG_{22} - \cV_2 \notag \\
  \bbsj{2}{2} &= \cV_2 \notag \\
  \tD^{[2]}_1 &= \cG_{22} - 2 \, \cV_2 - \cC \notag \\
  \tD^{[2]}_2 &= - \cC \notag \\
\Delta_1^{[2]} &= \frac{\cV_2^2 + \cC \, \cG_{22}}{\cG_{22}}
\end{align*}
\end{minipage}

\vspace{1cm}

\hspace{-0.5cm}\begin{minipage}[c]{0.7\textwidth}
\textbf{For sector 3}\\
\begin{align*}
  \detgsj{3} &= \cG_{11} \notag \\
  \bbsj{1}{3} &= \cG_{11} - \cV_1+ \cG_{12} \notag \\
  \bbsj{2}{3} &= \cV_1 - \cG_{12} \notag \\
  \tD^{[3]}_1 &= \cG_{11} + 2 \, \cG_{12} + \cG_{22} - 2 \, ( \cV_1 + \cV_2) - \cC \notag \\
  \tD^{[3]}_2 &= \cG_{22} - 2 \, \cV_2 - \cC \notag \\
\Delta_1^{[3]} &= \frac{\cV_1^2 - 2 \, \cG_{12} \, \cV_1 + 2 \, \cG_{11} \, \cV_2 + \cC \, \cG_{11} - \cG_{11} \, \cG_{22} + \cG_{12}^2}{\cG_{11}}
\end{align*}
\end{minipage}
\begin{minipage}[c]{0.3\textwidth}
\textbf{For sector 4}\\
\begin{align*}
  \detgsj{4} &= \cG_{11} \notag \\
  \bbsj{1}{4} &= \cG_{11} - \cV_1 \notag \\
  \bbsj{2}{4} &= \cV_1 \notag \\
  \tD^{[4]}_1 &= \cG_{11} - 2 \, \cV_1 - \cC \notag \\
  \tD^{[4]}_2 &= - \cC \notag \\
\Delta_1^{[4]} &= \frac{\cV_1^2 + \cC \, \cG_{11}}{\cG_{11}}
\end{align*}
\end{minipage}

\vspace{1cm}

\noindent
Note that despite the fact that it is difficult to express the coefficients $\bbar_i$ in terms of elements of the inverse of the kinematical matrix $\cals$ as in the genuine one-loop case, the magic identities given in \cite{paper1} (cf.\ eqs. (2.36), (2.37)) related to the property of determinants still holds, namely
\begin{align}
  \Delta_2 - \Delta_1^{[i]} &= \frac{\bbar_i^2}{\det(\cG) \, \detgsj{i}} \notag \\
  \tD^{[i]}_j + \Delta_1^{[i]} &= \frac{\bbsjsq{j}{i}}{\detgsj{i}}
  \label{eqmagicid}
\end{align}
Thus the different roots are given again by
\begin{align}
  u_0^{[ij] \, 2} &= - \frac{\Delta_2 - \Delta_1^{[i]}}{\tD^{[i]}_j + \Delta_1^{[i]}} \notag \\
    &= - \frac{\bbar_i^2}{\bbsjsq{j}{i} \, \detg} \label{equ0sq} \\
    \baru^{[ij] \, 2} &= \frac{(\Delta_1^{[i]} + i \, \lambda) \, \detgj{i}}{\bbsjsq{j}{i}}
  \label{equbarsq}
\end{align}
\\

\noindent
Note that the results presented in this appendix are valid in the real mass case. In the case with complex masses, similar formulae can be derived along the same lines but using results of ref.~\cite{paper2} instead of ref.~\cite{paper1}.

\section{One-loop three-point function at $O(\varepsilon)$}\label{new_integral}

This appendix is dedicated to the computation of the one-loop three-point function at $O(\varepsilon)$ which appears for the topologies $\calt_{23\_3\_1\_1}$ and $\calt_{24\_2\_2\_1}$
\begin{align}
  \wtI_3^{\,(1)}(\Sigma_{(2)},G,V,C) &= \int_{\Sigma_{(2)}} d u_1 \, d u_2 \, \frac{\ln\left( D(u_1,u_2) - i \, \lambda \right)}{D(u_1,u_2) - i \, \lambda}
  \label{eqdefJp}
\end{align}
where
\begin{align}
  D(u_1,u_2) &= U^T \cdot G \cdot U - 2 \, V^T \cdot U - C
  \label{eqdefDu1u2}
\end{align}
with
\[
  U = \left[
  \begin{array}{c}
    u_1 \\
    u_2
  \end{array}
\right]
\]
To compute $\wtI_3^{\,(1)}$, a new integral, related to the former and on which it will be more handy to apply the method developed in ref.~\cite{paper1}, is defined
\begin{align}
  J_{\beta} &= \int_{\Sigma_{(2)}} d u_1 \, d u_2 \,  \left( D(u_1,u_2) - i \, \lambda \right)^{-1+\beta}
  \label{eqdefJbeta1}
\end{align}
The quantity $J_{\beta}$ is linked to $\wtI_3^{\,(1)}$ by the relation
\begin{align}
  \wtI_3^{\,(1)}(\Sigma_{(2)},G,V,C) &= \left. \frac{\partial}{\partial \, \beta} \, J_{\beta} \right|_{\beta=0}
  \label{eqrelJJbeta}
\end{align}
The computation of $J_{\beta}$ is similar to the infrared case treated in ref.~\cite{paper3}. Using eq. (2.45) of this reference, multiplying by $- 2^{-1-\varepsilon}/\Gamma(1+\varepsilon)$ to take into account the different normalisation and setting $\varepsilon = -\beta$, we get
\begin{align}
  J_{\beta} &= - \frac{1}{2} \, \frac{1}{1+\beta} \, \frac{1}{B(1-\beta,1+\beta)} \, \sum_{i \in S_3} \frac{\bbar_i}{\detg} \, \sum_{j \in S_3 \setminus\{i\}} \frac{\bbsj{j}{i}}{\detgj{i}} \, L_3^{\beta}(\Delta_2,\Delta_1^{\{i\}},\tD_{ij})
  \label{eqdefJbeta2}
\end{align}
where
\begin{align}
  L_3^{\beta}(\Delta_2,\Delta_1^{\{i\}},\tD_{ij}) &= \int_0^{\infty} \frac{d \chi}{\chi^{\nu} - \Delta_2 - i \, \lambda} \, \int_0^{\infty} \frac{d \rho}{(\chi^{\nu} + \rho^2 - \Delta_1^{\{i\}} - i \, \lambda) \, (\tD_{ij} + \chi^{\nu} + \rho^2 - i \, \lambda)^{\frac{1}{2}}}
  \label{eqdefJbeta21}
\end{align}
and $\nu = 1/(1+\beta)$.\\

\noindent
Using appendix D of ref.~\cite{paper1} to trade the $\rho$ integration for an integration between $0$ and $1$ and appendix A of ref.~\cite{paper3} to integrate over $\chi$, we get
\begin{align}
  J_{\beta} &= \frac{1}{2} \, \frac{1}{\beta} \, \sum_{i \in S_3} \frac{\bbar_i}{\detg}  \, \sum_{j \in S_3 \setminus\{i\}} \frac{\bbsj{j}{i}}{\detgj{i}} \, \int_0^1 \frac{d z}{z^2 \, (\tD_{ij} + \Delta_1^{\{i\}}) + \Delta_2 - \Delta_1^{\{i\}}}  \notag \\
  &\qquad {} \times \left[ \left( - \Delta_2 - i \, \lambda \right)^{\beta}  - \left( z^2 \, (\tD_{ij} + \Delta_1^{\{i\}}) - \Delta_1^{\{i\}} - i \, \lambda \right)^{\beta} \right]
  \label{eqdefJbeta3}
\end{align}

\noindent
But what is needed is the derivative of $J_{\beta}$ with respect to $\beta$ taken at $\beta = 0$ which leads to
\begin{align}
     \wtI_3^{\,(1)}(\Sigma_{(2)},G,V,C) &= \frac{1}{4} \, \sum_{i \in S_3} \, \frac{\bbar_i}{\detg}   \, \sum_{j \in S_3 \setminus\{i\}} \frac{\bbsj{j}{i}}{\detgj{i}} \, \int_0^1 \frac{d z}{z^2 \, (\tD_{ij} + \Delta_1^{\{i\}}) + \Delta_2 - \Delta_1^{\{i\}}}  \notag \\
      &\quad {} \times  \left[ \ln^2\left(  - \Delta_2 - i \, \lambda \right) - \ln^2\left( z^2 \, (\tD_{ij} + \Delta_1^{\{i\}}) - \Delta_1^{\{i\}} - i \, \lambda \right) \right]
  \label{eqdefJbeta4}
\end{align}
Thus we have to compute the integral between 0 and 1 of the ratio of the logarithm squared of a second order polynomial in the integration variable over another second order polynomial in the same variable.\\ 

\noindent
Let us describe the method used to compute analytically the kind of integrals appearing in eq.~(\ref{eqdefJbeta4}) in terms of Nielsen polylogarithms. For that, let us study the following integral along the same line as the quantity $K^{R}_{0,1}(A,B,u_0^2)$ appearing in the appendix E of ref.~\cite{paper1}
\begin{align}
^{(2)}K^{R}_{0,1}(A,B,u_0^2) &= \int_0^1 du \; \frac{\ln^2\left( A \, u^2 + B) \right) - \ln^2\left( A \, u_0^{2 \, R} + B) \right) }{u^2 - u_0^2} 
  \label{eqdefK2R01}
\end{align}
where $A$ is real, $B$ and $u_0^2$ are complex with a vanishing imaginary part. $u_0^{2 \, R}$ denotes the real part of $u_0^2$. Let us name $\baru$ a root of the polynomial $A \, u^2 + B$, $\baru = \sqrt{-B/A}$, eq.(\ref{eqdefK2R01}) can be re-written as
\begin{align}
\hspace{2em}&\hspace{-2em} ^{(2)}K^{R}_{0,1}(A,B,u_0^2) \notag \\
  &= \frac{1}{2 \, u_0} \, \int_0^1 du \, \left\{ \frac{1}{u - u_0} \, \left[ \left( \ln(A - i \, \lambda \, S_u) + \ln(u - \baru) + \ln(u + \baru) \right)^2 \right. \right. \notag \\
  &\qquad \qquad \qquad  \qquad \qquad \quad {} - \left. \left( \ln(A - i \, \lambda \, S_u) + \ln(\tuz - \baru) + \ln(\tuz + \baru) + \eta(\tuz-\baru,\tuz+\baru) \right)^2 \right] \notag \\
  &\qquad \qquad \qquad  \quad {} - \frac{1}{u + u_0} \, \left[ \left( \ln(A - i \, \lambda \, S_u) + \ln(u - \baru) + \ln(u + \baru) \right)^2 \right. \notag \\
  &\qquad \qquad \qquad  \qquad \qquad \qquad  {} - \left( \ln(A - i \, \lambda \, S_u) + \ln(-\tuz - \baru) + \ln(-\tuz + \baru) \right. \notag \\ 
  &\qquad \qquad \qquad  \qquad \qquad \qquad \qquad  {} + \left. \left. \left. \eta(-\tuz-\baru,-\tuz+\baru) \right)^2 \right] \right\} 
  \label{eqdefK2R011}
\end{align}
where
\[
  S_u = \sign(\Im(-\baru^2))
\]
and $\tuz$ is defined by
\[
  \tuz \equiv
  \left\{
  \begin{array}{ccc}
    i \, s \, \sqrt{-u^{2 \, R}_0} & \text{if} & u^{2 \, R}_0 < 0 \\
    & & \\
    \sqrt{u^{2 \, R}_0} & \text{if} & u^{2 \, R}_0 > 0 \\
  \end{array}
  \right.
\]
with $s = \pm 1$ is the sign of imaginary part of $u_0^2$, namely $\Im(u_0^2) = i \, s \lambda$.
Expanding the terms into the curly brackets of eq.~(\ref{eqdefK2R011}) leads to
\begin{align}
 \hspace{2em}&\hspace{-2em} ^{(2)}K^{R}_{0,1}(A,B,u_0^2) \notag \\
 &= \frac{1}{2 \, u_0} \, \left\{ \bRt(u_0,\baru,\tuz) + \bRt(u_0,-\baru,\tuz) - \bRt(-u_0,\baru,-\tuz) - \bRt(-u_0,-\baru,-\tuz)  \right. \notag \\ 
  &\quad {} + 2 \, \ln(A - i \, \lambda \, S_u) \, \left[  \bR(u_0,\baru,\tuz) + \bR(u_0,-\baru,\tuz) - \bR(-u_0,\baru,-\tuz) - \bR(-u_0,-\baru,-\tuz) \right] \notag \\
  &\quad {} + 2 \, \left( \bQ(u_0,\baru,\tuz) - \bQ(-u_0,\baru,-\tuz) \right) \notag \\
&\quad {} - \ln\left( \frac{u_0-1}{u_0} \right) \, \left[ 2 \, \eta(\tuz-\baru,\tuz+\baru) \, \left( \ln(A - i \, \lambda \, S_u) + \ln(\tuz - \baru) + \ln(\tuz+\baru) \right) \right. \notag \\
&\qquad {} + \left. \eta^2(\tuz-\baru,\tuz+\baru) \right] \notag \\
  &\quad {} + \ln\left( \frac{u_0+1}{u_0} \right) \, \left[ 2 \, \eta(-\tuz-\baru,-\tuz+\baru) \, \left( \ln(A - i \, \lambda \, S_u) + \ln(-\tuz - \baru) + \ln(-\tuz+\baru) \right) \right. \notag \\
  &\qquad {} + \left. \left. \eta^2(-\tuz-\baru,-\tuz+\baru) \right] \right\} 
  \label{eqdefK2R012}
\end{align}
where two new integrals have been defined
\begin{align}
  \bRt(u_0,\baru,\tuz) &= \int_0^1 du \; \frac{\ln^2(u-\baru) - \ln^2(\tuz-\baru)}{u-u_0} \label{eqdefR2} \\
  \bQ(u_0,\baru,\tuz) &= \int_0^1 du \; \frac{\ln(u-\baru) \, \ln(u+\baru) - \ln(\tuz-\baru) \, \ln(\tuz+\baru)}{u - u_0}
  \label{eqdefQ}
\end{align}
\\

\noindent
Using parity arguments, cf.~appendix E of ref.~\cite{paper1}, some integrals, for example $\bRt(u_0,\baru,\tuz)$ and  $-\bRt(-u_0,-\baru,-\tuz)$ can be glued together to give an integral between $-1$ and $1$. In addition, the $u_0$ appearing at the denominator can be traded safely for $\tuz$ (cf.~appendix E of \cite{paper1}). Let us introduce new notations, namely
\begin{align}
  \bRtp(\baru,\tuz) &= \int_{-1}^1 du \; \frac{\ln^2(u-\baru) - \ln^2(\tuz-\baru)}{u-\tuz} \label{eqdefR2p} \\
  \bQp(\baru,\tuz) &= \int_{-1}^1 du \; \frac{\ln(u-\baru) \, \ln(u+\baru) - \ln(\tuz-\baru) \, \ln(\tuz+\baru)}{u-\tuz}
  \label{eqdefbQ}
\end{align}
where the path of integration for the two integrals is along the real axis.
After some algebra, eq.~(\ref{eqdefK2R012}) becomes
\begin{align}
^{(2)}K^{R}_{0,1}(A,B,u_0^2) &= \frac{1}{2 \, \tuz} \left\{ \bRtp(\baru,\tuz) + \bRtp(-\baru,\tuz) + 2 \, \ln(A - i \, \lambda \, S_u) \, \calf(\tuz,\baru) + 2 \, \bQp(\baru,\tuz) \right. \notag \\
&\quad {} + \ln\left( \frac{u_0+1}{u_0-1} \right) \, \eta(\tuz-\baru,\tuz+\baru) \, \left[ 2 \, \left( \ln(A - i \, \lambda \, S_u) + \ln(\tuz-\baru) + \ln(\tuz+\baru) \right) \right. \notag \\
&\qquad \qquad \qquad \qquad \qquad \qquad \qquad \qquad {} + \left. \left. \eta(\tuz-\baru,\tuz+\baru) \right] \vphantom{\bRtp(\baru,\tuz)}\right\}
  \label{eqdefK2R013}
\end{align}
where the function $\calf(\tuz,\baru)$ is given by eq.~(E.17) of ref.~\cite{paper1}. At this point, it remains to compute $\bRtp(\baru,\tuz)$ and $\bQp(\baru,\tuz)$.

\subsection{Computation of $\bQp(\baru,\tuz)$}

For this calculation, $\baru$ and $\tuz$ are complex numbers, the imaginary part of $\baru$ never vanishes while the one of $\tuz$ may be zero. The product $\ln(u-\baru) \, \ln(u+\baru)$ can be written in the following way
\begin{align}
  \ln(u-\baru) \, \ln(u+\baru) &= \ln(-\baru) \, \ln(\baru) + \ln(\baru) \, \ln\left( 1- \frac{u}{\baru} \right) + \ln(- \baru) \, \ln\left( 1 + \frac{u}{\baru} \right) \notag \\
  &\quad {} + \ln\left( 1 - \frac{u}{\baru} \right) \, \ln\left( 1 + \frac{u}{\baru} \right)
  \label{eqbreakprod}
\end{align}
Furthermore, the last term of eq.~(\ref{eqbreakprod}) can be rewritten as
\begin{align}
  \ln\left( 1 - \frac{u}{\baru} \right) \, \ln\left( 1 + \frac{u}{\baru} \right) &= \frac{1}{2} \, \left[ \ln^2\left( 1 - \frac{u}{\baru} \right) + \ln^2\left( 1 + \frac{u}{\baru} \right) - \left( \ln\left( \frac{\baru - u}{\baru + u} \right)- \eta\left( 1 - \frac{u}{\baru},\frac{1}{1 + \frac{u}{\baru}} \right) \right)^2 \right]
  \label{eqbreakprod2}
\end{align}
It is easy to show that the $\eta(1-u/\baru,1/(1+u/\baru)) = 0$ because $u$ is real, thus the function $\bQp(\baru,\tuz)$ can be expressed in the following way
\begin{align}
\bQp(\baru,\tuz) &= \int_{-1}^1 \frac{du}{u - \tuz} \, \left[ \frac{1}{2} \left( \ln^2\left( 1 - \frac{u}{\baru} \right) + \ln^2\left( 1 + \frac{u}{\baru} \right) - \ln^2\left( \frac{\baru - u}{\baru + u} \right) \right) + \ln(-\baru) \, \ln(\baru) \right. \notag \\
&\quad {} + \left. \ln(\baru) \, \ln\left( 1- \frac{u}{\baru} \right) + \ln(- \baru) \, \ln\left( 1 + \frac{u}{\baru} \right)  - \ln(\tuz-\baru) \, \ln(\tuz+\baru) \right]
  \label{eqnewdefQb}
\end{align}
\\

\begin{figure}[h]
\begin{center}
  \includegraphics[scale=1.2]{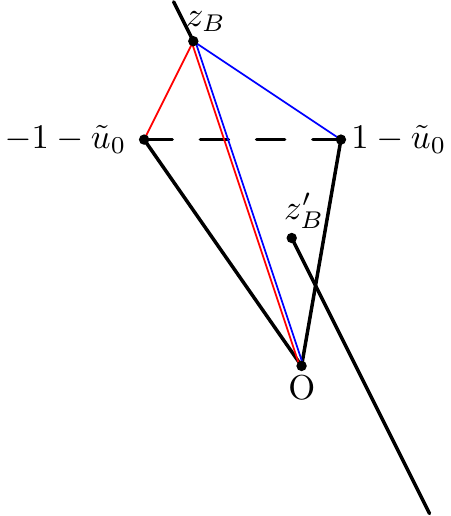}
\end{center}
\caption{\footnotesize Example of a contour deformation running through the branch point $z_B$. The blue (resp.~red) segments are the contour deformation for the integration between $0$ and $1-\tuz$ (resp.~$-1-\tuz$) }\label{cut_triangle}
\end{figure}

\noindent
Let us define and compute the following integral
\begin{align}
  \call_2(\baru,\tuz) &= \int_{-1}^1 \frac{du}{u - \tuz} \left\{ \ln^2\left( \frac{\baru-u}{\baru+u} \right) - \ln^2\left( \frac{\baru- \tuz}{\baru + \tuz} \right) \right\}
  \label{eqdefL2}
\end{align}
where the integration variable $u$ runs along the real axis. 
Setting $z=u-\tuz$, the integral $\call_2(\baru,\tuz)$ becomes
\begin{align}
  \call_2(\baru,\tuz) &= \int_{-1-\tuz}^{1-\tuz} \frac{dz}{z} \left[ \ln^2\left( \frac{\baru-z-\tuz}{\baru+z+\tuz} \right) - \ln^2\left( \frac{\baru-\tuz}{\baru+\tuz} \right) \right]
  \label{eqdefL21}
\end{align}
Then, the idea is to split the integral from $-1-\tuz$ to $1-\tuz$ in two parts
\begin{align}
  \int_{-1-\tuz}^{1-\tuz} dz &= \int_0^{1-\tuz} dz - \int_0^{-1-\tuz} dz
  \label{eqgnagna}
\end{align}
But because of the structure of the argument of the logarithm squared in eq.~(\ref{eqdefL21}), the $z$ variable cannot run along the straight line $[0,1-\tuz]$ (resp.~$[0,-1-\tuz]$) for the first (resp.~the second) integral of the right member of eq.~(\ref{eqgnagna}). Indeed, one of the cuts of the logarithm may cross these segments as shown in appendix~\ref{cut_log} and thus the contour must be deformed.
In order to avoid this crossing, the strategy to deform the contour is to go through one of the branch points. Instead of integrating on a segment between $0$ and $1-\tuz$ (resp. $-1-\tuz$), the integration contour is chosen to be a segment between $0$ and one of the branch points, then a segment from this branch point and $1-\tuz$ (resp. $-1-\tuz$), cf.~fig.~\ref{cut_triangle}.

\noindent
In this way, the integration contour never crosses the branch cuts of the logarithm. Of course, any of the two branch points can be chosen equivalently and even the branch point used for the contour deformation needs not to be the same for the two paths $0, 1-\tuz$ and $0, -1-\tuz$.\\

\noindent
To fix the idea, we will choose for the two contour deformations the branch point $z_B = \baru- \tuz$ (cf.~appendix~\ref{cut_log}). Let us introduce the integral $\calj_1$ given by
\begin{align}
  \calj_1 &= \int_{0}^{\beta-\tuz} \frac{dz}{z} \left[ \ln^2\left( \frac{\baru-z-\tuz}{\baru+z+\tuz} \right) - \ln^2\left( \frac{\baru-\tuz}{\baru+\tuz} \right) \right]
  \label{eqdefcalJb1}
\end{align}
which depends on a real parameter $\beta$. As discussed previously, the integration path is chosen to avoid any crossing with the branch cuts of the logarithm, namely
\begin{align}
  \calj_1 &= \int_{0}^{z_B} \frac{dz}{z} \left[ \ln^2\left( \frac{\baru-z-\tuz}{\baru+z+\tuz} \right) - \ln^2\left( \frac{\baru-\tuz}{\baru+\tuz} \right) \right] \notag \\
  &\quad {} + \int_{z_B}^{\beta-\tuz} \frac{dz}{z} \left[ \ln^2\left( \frac{\baru-z-\tuz}{\baru+z+\tuz} \right) - \ln^2\left( \frac{\baru-\tuz}{\baru+\tuz} \right) \right]
  \label{eqdefcalJb2}
\end{align}
the integration paths for the two integrals are the segments $[0,z_B]$ and $[z_B,\beta-\tuz]$. The following change of variable 
\begin{equation}
  v = \frac{\baru - z - \tuz}{\baru + z + \tuz}
  \label{eqchgtvarJ1}
\end{equation}
leading to
\[
  \frac{dz}{z} = - \left[ \frac{1}{v+1} - \frac{1}{v - \frac{\baru-\tuz}{\baru+\tuz}} \right] \, dv
\]
is performed in both integrals of eq.~(\ref{eqdefcalJb2}).
Thus this equation becomes
\begin{align}
  \calj_1 &= \int^{\frac{\baru-\tuz}{\baru+\tuz}}_{0} dv \, \left[ \frac{1}{v+1} - \frac{1}{v-\frac{\baru-\tuz}{\baru+\tuz}} \right] \, \left( \ln^2(v) - \ln^2\left( \frac{\baru-\tuz}{\baru+\tuz} \right) \right) \notag \\
  &\quad {} - \int_{0}^{\frac{\baru-\beta}{\baru+\beta}} dv \, \left[ \frac{1}{v+1} - \frac{1}{v-\frac{\baru-\tuz}{\baru+\tuz}} \right] \, \left( \ln^2(v) - \ln^2\left( \frac{\baru-\tuz}{\baru+\tuz} \right) \right)
  \label{eqdefcalJb3}
\end{align}
\\

\noindent
If the other branch point $z^{\prime}_B=-\baru-\tuz$ (cf.~appendix~\ref{cut_log}) has been chosen instead of $z_B$ for the contour deformation, the correct change of variable would have been
\begin{equation}
  v = \frac{\baru + z + \tuz}{\baru - z - \tuz}
  \label{eqchgtvarJ2}
\end{equation}
yielding
\begin{align}
  \calj_1 &= \int^{\frac{\baru+\tuz}{\baru-\tuz}}_{0} dv \, \left[ \frac{1}{v+1} - \frac{1}{v-\frac{\baru+\tuz}{\baru-\tuz}} \right] \, \left( \ln^2(v) - \ln^2\left( \frac{\baru-\tuz}{\baru+\tuz} \right) \right) \notag \\
  &\quad {} - \int_{0}^{\frac{\baru+\beta}{\baru-\beta}} dv \, \left[ \frac{1}{v+1} - \frac{1}{v-\frac{\baru+\tuz}{\baru-\tuz}} \right] \, \left( \ln^2(v) - \ln^2\left( \frac{\baru-\tuz}{\baru+\tuz} \right) \right)
  \label{eqdefcalJb4}
\end{align}
\\

\noindent
Changing $v = (\baru-\tuz)/(\baru+\tuz) \, t$ in the first integral of eq.~(\ref{eqdefcalJb3}) and $v = (\baru-\beta)/(\baru+\beta) \, t$ in the second one leads to the following result
\begin{align}
  \calj_1 &= \int^{1}_{0} dt \, \left[ \frac{1}{t+\frac{\baru+\tuz}{\baru-\tuz}} - \frac{1}{t-1} \right] \, \left( \ln^2(t) + 2\, \ln\left( \frac{\baru-\tuz}{\baru+\tuz} \right) \, \ln(t) \right) \notag \\
  &\quad {} - \int_{0}^{1} dt \, \left[ \frac{1}{t+\frac{\baru+\beta}{\baru-\beta}} - \frac{1}{t-\frac{\baru+\beta}{\baru-\beta} \, \frac{\baru-\tuz}{\baru+\tuz}} \right] \, \left( \ln^2(t) + 2 \, \ln\left( \frac{\baru-\beta}{\baru+\beta} \right) \, \ln(t) \right. \notag \\
  &\qquad \qquad \qquad \qquad \qquad \qquad \qquad \qquad \qquad {} + \left. \ln^2\left( \frac{\baru-\beta}{\baru+\beta} \right) - \ln^2\left( \frac{\baru-\tuz}{\baru+\tuz} \right) \right)
  \label{eqdefcalJb5}
\end{align}
The logarithms having $t$ in their argument have been split because $t$ runs on the segment $[0,1]$.\\

\noindent
Coming back to function $\call_2(\baru,\tuz)$ and using eq.~(\ref{eqdefcalJb5}), we get
\begin{align}
  \call_2(\baru,\tuz)  &= \left. \calj_1 \right|_{\beta=1} - \left. \calj_1 \right|_{\beta=-1} \notag \\
&= \int_{0}^{1} dt \, \left[ \frac{1}{t+\frac{\baru-1}{\baru+1}} - \frac{1}{t-\frac{\baru-1}{\baru+1} \, \frac{\baru-\tuz}{\baru+\tuz}} \right] \,  \left( \ln^2(t) + 2 \, \ln\left( \frac{\baru+1}{\baru-1} \right) \, \ln(t) \right. \notag \\
  &\qquad \qquad \qquad \qquad \qquad \qquad \qquad \qquad \qquad {} + \left. \ln^2\left( \frac{\baru+1}{\baru-1} \right) - \ln^2\left( \frac{\baru-\tuz}{\baru+\tuz} \right) \right) \notag \\
  &\quad {} - \int_{0}^{1} dt \, \left[ \frac{1}{t+\frac{\baru+1}{\baru-1}} - \frac{1}{t-\frac{\baru+1}{\baru-1} \, \frac{\baru-\tuz}{\baru+\tuz}} \right] \, \left( \ln^2(t) + 2 \, \ln\left( \frac{\baru-1}{\baru+1} \right) \, \ln(t) \right. \notag \\
  &\qquad \qquad  \qquad \qquad \qquad \qquad \qquad \qquad \qquad {} + \left. \ln^2\left( \frac{\baru-1}{\baru+1} \right) - \ln^2\left( \frac{\baru-\tuz}{\baru+\tuz} \right) \right)
  \label{eqdefL22}
\end{align}

\noindent
Using the fact that, for a complex number $z \in \mathds{C} \setminus [1,\infty[$, we have the following results\footnote{These results can be shown easily by an integration by part from the integral representations of the functions $\dilog(z)$ and $S_{2,1}(z)$.} 
\begin{align}
  \int_0^1 dt \, \frac{\ln^2(t)}{t - \frac{1}{z}} &= -2 \, S_{2,1}(z) \label{eqrelS1} \\
  \int_0^1 dt \, \frac{\ln(t)}{t - \frac{1}{z}} &= \dilog(z) \label{eqrelS2}
\end{align}
where the $S_{n,p}(z)$ are the Nielsen polylogarithms, cf.~\cite{Kolbig:1969zza}, defined as
\begin{align}
  S_{n,p}(z) &= \frac{(-1)^{n+p-1}}{(n-1)! \, p!} \, \int_0^1 dt \, \frac{\ln^{n-1}(t) \, \ln^p(1-z \, t)}{t}
  \label{eqdefNpolylog}
\end{align}
Thus, the $t$ integration can be easily performed, yielding
\begin{align}
  \call_2(\baru,\tuz)
  &= 2 \, \left[ S_{2,1}\left( \frac{1-\baru}{1+\baru} \right) - S_{2,1}\left( \frac{1+\baru}{1-\baru} \right) - S_{2,1}\left(  \frac{\tuz+\baru}{\tuz-\baru} \, \frac{1-\baru}{1+\baru} \right) + S_{2,1}\left(  \frac{\tuz+\baru}{\tuz-\baru} \, \frac{1+\baru}{1-\baru} \right) \right] \notag \\
&\quad {} + 2 \, \ln\left( \frac{\baru+1}{\baru-1} \right) \, \left[ \dilog\left( \frac{1+\baru}{1-\baru} \right) - \dilog\left( \frac{\tuz+\baru}{\tuz-\baru} \, \frac{1-\baru}{1+\baru} \right) \right. \notag \\
&\qquad \qquad \qquad \qquad \qquad {} + \left. \dilog\left( \frac{1-\baru}{1+\baru} \right) - \dilog\left( \frac{\tuz+\baru}{\tuz-\baru} \, \frac{1+\baru}{1-\baru} \right) \right] \notag \\
&\quad {} + \left[ \ln^2\left( \frac{\baru-1}{\baru+1} \right) - \ln^2\left( \frac{\baru-\tuz}{\baru+\tuz} \right) \right] \, \left\{ \ln\left( \frac{2 \, \baru}{\baru-1} \right) - \ln\left( \frac{2 \, \baru}{\baru-1} \, \frac{\tuz+1}{\tuz-\baru} \right) \right. \notag \\
&\qquad \qquad \qquad \qquad \qquad \qquad \qquad \qquad \qquad {} - \left. \ln\left( \frac{2 \, \baru}{\baru+1} \right) + \ln\left( \frac{2 \, \baru}{\baru+1} \, \frac{\tuz-1}{\tuz-\baru} \right) \right\}
  \label{eqdefL23}
\end{align}
In addition, from ref.~\cite{Kolbig:1969zza}, we get that $S_{2,1}(z) \equiv \trilog(z)$ and that
\begin{equation}
  \trilog(z) = \trilog\left( \frac{1}{z} \right) - \frac{\pi^2}{6} \, \ln(-z) - \frac{1}{6} \, \ln^3(-z)
  \label{eqinvtrilog}
\end{equation}
therefore eq.~(\ref{eqdefL23}) becomes
\begin{align}
  \call_2(\baru,\tuz) &= 2 \, \left[ \trilog\left(  \frac{\baru+1}{\baru-1} \, \frac{\baru+\tuz}{\baru-\tuz} \right) - \trilog\left( \frac{\baru-1}{\baru+1}  \, \frac{\baru+\tuz}{\baru-\tuz} \right)  \right] \notag \\
  &\quad {} + 2 \, \ln\left( \frac{\baru-1}{\baru+1} \right) \, \left[ \dilog\left( \frac{\baru+1}{\baru-1} \, \frac{\baru+\tuz}{\baru-\tuz} \right) + \dilog\left( \frac{\baru-1}{\baru+1}  \, \frac{\baru+\tuz}{\baru-\tuz} \right) \right] \notag \\
  &\quad {} +  \frac{2}{3} \, \ln^3\left( \frac{\baru-1}{\baru+1} \right) + \left[ \ln^2\left( \frac{\baru-1}{\baru+1} \right) - \ln^2\left( \frac{\baru-\tuz}{\baru+\tuz} \right)\right] \notag \\
&\quad {} \times \left\{  \ln\left( \frac{\tuz-1}{\tuz-\baru} \right) - \ln\left( \frac{\tuz+1}{\tuz-\baru} \right)
 + \eta\left( \frac{2 \, \baru}{\baru+1}, \frac{\tuz-1}{\tuz-\baru} \right) - \eta\left( \frac{2 \, \baru}{\baru-1}, \frac{\tuz+1}{\tuz-\baru} \right)\right\} \notag \\
  \label{eqdefL24}
\end{align}
If the branch point $z^{\prime}_B$ had been chosen instead of $z_B$ for the two contour deformations, we would have got the right-hand side of eq.~(\ref{eqdefL24}) with $\baru$ changed in $-\baru$, but since $\call_2(\baru,\tuz) = \call_2(-\baru,\tuz)$ as it is obvious from its definition, cf.~eq.~(\ref{eqdefbQ}), the results are the same as it should be.\\

\noindent
To finish the computation of $\bQp(\baru,\tuz)$, let us introduce the quantity 
\begin{align}
  \call^{(n)}_1(\baru,\tuz) &= \int_{-1}^1 \frac{du}{u-\tuz} \, \left[ \ln^n\left( 1 - \frac{u}{\baru} \right) - \ln^n\left( 1 - \frac{\tuz}{\baru} \right) \right]
  \label{eqdefcall11}
\end{align}
Setting $z = u - \tuz$ and taking $n=2$ leads to
\begin{align}
  \call^{(2)}_1(\baru,\tuz) &= \int_{-1-\tuz}^{1-\tuz} \frac{dz}{z} \, \left[ \ln^2\left( 1 - \frac{z+\tuz}{\baru} \right) - \ln^2\left( 1 - \frac{\tuz}{\baru} \right) \right]
  \label{eqdefcall12}
\end{align}
As for the computation of the integral $\call_2(\baru,\tuz)$, the idea will be to split the integral on the left-hand side of eq.~(\ref{eqdefcall12}) as follows
\begin{align}
  \int_{-1-\tuz}^{1-\tuz} \frac{dz}{z} \, F^{\prime}(z) &= \int_0^{1-\tuz} \frac{dz}{z} \, F^{\prime}(z) - \int_0^{-1-\tuz} \frac{dz}{z} \, F^{\prime}(z)
  \label{eqfracintL1}
\end{align}
with $F^{\prime}(z) = \ln^2(1 - (z+\tuz)/\baru) - \ln^2(1 - \tuz/\baru)$. But the integration path for the two integrals on the left-hand side of eq.~(\ref{eqfracintL1}) cannot be a straight line between the end points of the integral because, in some kinematical configurations, there could be a crossing between the cut of the function $F^{\prime}(z)$ and the segments $[0,1-\tuz]$ or $[0,-1-\tuz]$. The choice for the contour deformation will be the same as for the computation of $\call_2$, namely a segment from the lower end point to the branch point and another segment from the branch point to the higher end point of the integral.\\

\noindent
Let us consider now the integral $\calj_2$ such that
\begin{align}
  \calj_2 &= \int^{\beta-\tuz}_0 \frac{dz}{z} \; \left[ \ln^2\left( 1- \frac{z+\tuz}{\baru} \right) - \ln^2\left( 1 - \frac{\tuz}{\baru} \right) \right]
  \label{eqdefcalJ31}
\end{align}
with $\beta = \pm 1$. The change of variable $v = 1 - (z+\tuz)/\baru$ is performed which leads to
\begin{align}
  \calj_2 &= \int_0^{\frac{\baru-\beta}{\baru}} \frac{dv}{v - \frac{\baru-\tuz}{\baru}} \, \left[ \ln^2(v) - \ln^2\left( \frac{\baru-\tuz}{\baru} \right) \right] - \int_0^{\frac{\baru-\tuz}{\baru}} \frac{dv}{v - \frac{\baru-\tuz}{\baru}} \, \left[ \ln^2(v) - \ln^2\left( \frac{\baru-\tuz}{\baru} \right) \right]
  \label{eqdefcalJ34}
\end{align}
Setting $v = (\baru-\beta)/\baru \, t$ in the first integral and $v = (\baru-\tuz)/\baru \, t$ in the second one to make the new integration variable running on the segment $[0,1]$ leads to
\begin{align}
  \calj_2 &= \int_0^1 \frac{dt}{t - \frac{\baru-\tuz}{\baru-\beta}} \, \left[ \ln^2(t) + 2 \, \ln(t) \, \ln\left( \frac{\baru-\beta}{\baru} \right) + \ln^2\left( \frac{\baru-\beta}{\baru} \right) - \ln^2\left( \frac{\baru-\tuz}{\baru} \right) \right] \notag \\
  &\quad {} - \int_0^1 \frac{dt}{t-1} \, \left[ \ln^2(t) + 2 \, \ln(t) \, \ln\left( \frac{\baru-\tuz}{\baru} \right) \right]
  \label{eqdefcalJ35}
\end{align}
Thus, the integral $\call^{(2)}_1(\baru,\tuz)$ becomes
\begin{align}
  \call^{(2)}_1(\baru,\tuz) &= \left. \calj_2 \right|_{\beta=1} - \left. \calj_2 \right|_{\beta=-1} \notag \\
      &= 2 \, \left[ \trilog\left( \frac{\baru+1}{\baru-\tuz} \right) - \trilog\left( \frac{\baru-1}{\baru-\tuz} \right) \right] \notag \\
      &\quad {} + 2 \, \left[ \ln\left( \frac{\baru-1}{\baru} \right) \, \dilog\left( \frac{\baru-1}{\baru-\tuz} \right) - \ln\left( \frac{\baru+1}{\baru} \right) \, \dilog\left( \frac{\baru+1}{\baru-\tuz} \right) \right] \notag \\
      &\quad {} + \left[ \ln^2\left( \frac{\baru-1}{\baru} \right) - \ln^2\left( \frac{\baru-\tuz}{\baru} \right) \right] \, \ln\left( \frac{\tuz-1}{\tuz-\baru} \right) \notag \\
      &\quad {} - \left[ \ln^2\left( \frac{\baru+1}{\baru} \right) - \ln^2\left( \frac{\baru-\tuz}{\baru} \right) \right] \, \ln\left( \frac{\tuz+1}{\tuz-\baru} \right) 
  \label{eqdefcall13}
\end{align}
\\

\noindent
On the same lines, the quantity $\call^{(1)}_1(\baru,\tuz)$ can be readily computed and the result is
\begin{align}
  \call^{(1)}_1 &= \dilog\left( \frac{\tuz+1}{\tuz-\baru} \right) - \dilog\left( \frac{\tuz-1}{\tuz-\baru} \right) + \ln\left( \frac{\tuz+1}{\tuz-\baru} \right) \, \left[ \eta\left( -1-\baru, \frac{1}{\tuz-\baru} \right) - \eta\left( -\baru, \frac{1}{\tuz-\baru} \right) \right]\notag \\
  &\quad {} - \ln\left( \frac{\tuz-1}{\tuz-\baru} \right) \, \left[ \eta\left( 1-\baru, \frac{1}{\tuz-\baru} \right) - \eta\left( -\baru, \frac{1}{\tuz-\baru} \right) \right] 
  \label{eqdefcalle11}
\end{align}
\\

\noindent
All the ingredients for the computation of the function $\bQp(\baru,\tuz)$ are gathered and the final result is given by
\begin{align}
  \bQp(\baru,\tuz) &= \frac{1}{2} \, \left[ \call^{(2)}_1(\baru,\tuz) + \call^{(2)}_1(-\baru,\tuz) - \call_2(\baru,\tuz) \right] + \ln\left( \baru \right) \, \call^{(1)}_1(\baru,\tuz) + \ln\left( - \baru \right) \, \call^{(1)}_1(- \baru,\tuz) \notag \\
  &\quad {} + \left\{ \frac{1}{2} \, \eta^2\left( \baru-\tuz, \frac{1}{\baru} \right) + \frac{1}{2} \, \eta^2\left( \baru+\tuz, \frac{1}{\baru} \right) - \frac{1}{2} \,  \eta^2\left( \baru-\tuz, \frac{1}{\baru+\tuz} \right)\right. \notag \\
  &\quad {}  + \ln\left( \baru - \tuz \right) \, \left[ \eta\left( \baru-\tuz,\frac{1}{\baru} \right) - \eta\left( \baru-\tuz, \frac{1}{\baru+\tuz} \right) \right] \notag \\
  &\quad {}  + \ln\left( \baru + \tuz \right) \, \left[ \eta\left( \baru+\tuz,\frac{1}{\baru} \right) + \eta\left( \baru-\tuz, \frac{1}{\baru+\tuz} \right) \right] \notag \\
  &\quad {} - \left. i \, \pi \, \left[ \left( S(\baru)+S(\tuz-\baru) \right) \, \ln\left( \baru + \tuz \right) + S(\baru) \, \eta\left( \baru+\tuz,\frac{1}{\baru} \right) \right] \right\} \, \ln\left( \frac{\tuz-1}{\tuz+1} \right)
  \label{eqdefbQp1}
\end{align}
Note that if the imaginary part of $\tuz$ is zero, all the $\eta$ functions vanish.

\subsection{Computation of $\bRtp(\baru,\tuz)$}

In this case, there is nothing new to compute, we just have to rearrange the integrand of eq.~(\ref{eqdefR2p}) which can get the following form
\begin{align}
  \bRtp(\baru,\tuz) &= \int_{-1}^{1} \frac{du}{u - \tuz} \, \left\{ \left[ \ln(-\baru) + \ln\left( 1 - \frac{u}{\baru} \right) \right]^2 \right. \notag \\
  &\qquad \qquad \qquad \quad {} - \left. \left[ \ln(-\baru) + \ln\left( 1-\frac{\tuz}{\baru} \right) + \eta\left( -\baru, 1-\frac{\tuz}{\baru} \right) \right]^2 \right\}
  \label{eqdefR2p1}
\end{align}
then expanding the squared terms and expressing in terms of the function $\call^{(n)}_1(\baru,\tuz)$, we get
\begin{align}
  \bRtp(\baru,\tuz) &= \call^{(2)}_1(\baru,\tuz) + 2 \, \ln(- \baru) \, \call^{(1)}_1(\baru,\tuz) \notag \\
  &\quad {} + \eta\left( \tuz-\baru, -\frac{1}{\baru} \right) \, \left[ \eta\left(  \tuz-\baru, -\frac{1}{\baru} \right) + 2 \, \ln\left( \tuz -\baru \right)  \right] \, \ln\left( \frac{\tuz-1}{\tuz+1} \right)
  \label{eqdefR2p2}
\end{align}
\\

\noindent
Putting eqs.~(\ref{eqdefbQp1}) and  (\ref{eqdefR2p2}) into eq.~(\ref{eqdefK2R013}) leads to the following result
\begin{align}
\hspace{2em}&\hspace{-2em}^{(2)}K^{R}_{0,1}(A,B,u_0^2) \notag \\   
&= \frac{1}{2 \, \tuz} \left\{ \vphantom{\frac{u_0-1}{u_0}} 2 \, \call^{(2)}_1(\baru,\tuz) + 2 \, \call^{(2)}_1(-\baru,\tuz) - \call_2(\baru,\tuz) + 2 \, \ln(- \baru^2) \, \left[ \call^{(1)}_1(\baru,\tuz) + \call^{(1)}_1(-\baru,\tuz) \right] \right. \notag \\
  &\qquad \qquad {} + \left\{ 2 \, \eta^2\left( \baru-\tuz, \frac{1}{\baru} \right) + 2 \, \eta^2\left( \baru+\tuz, \frac{1}{\baru} \right) - \eta^2\left( \baru-\tuz, \frac{1}{\baru+\tuz} \right)\right. \notag \\
  &\qquad \qquad \quad {}  + 2 \, \ln\left( \baru - \tuz \right) \, \left[ 2 \, \eta\left( \baru-\tuz,\frac{1}{\baru} \right) - \eta\left( \baru-\tuz, \frac{1}{\baru+\tuz} \right) + i \, \pi \left( S(\baru-\tuz) - S(\baru) \right) \right] \notag \\
  &\qquad \qquad \quad {}  + 2 \, \ln\left( \baru + \tuz \right) \, \left[ 2 \, \eta\left( \baru+\tuz,\frac{1}{\baru} \right) + \eta\left( \baru-\tuz, \frac{1}{\baru+\tuz} \right) + i \, \pi \left( S(\baru-\tuz) - S(\baru) \right) \right] \notag \\
  &\qquad \qquad \quad {} - \left. 2 \, i \, \pi \, S(\baru) \, \left[ \eta\left( \baru-\tuz,\frac{1}{\baru} \right) + \eta\left( \baru+\tuz,\frac{1}{\baru} \right) \right] \right\} \, \ln\left( \frac{\tuz-1}{\tuz+1} \right) \notag \\
  &\qquad \qquad {} + \left[ 2 \, \left( \ln(A - i \, \lambda \, S_u) + \ln(\tuz-\baru) + \ln(\tuz+\baru) \right) + \eta(\tuz-\baru,\tuz+\baru) \right] \notag \\
  &\qquad \qquad \quad {} \times \left. \ln\left( \frac{u_0+1}{u_0-1} \right) \, \eta(\tuz-\baru,\tuz+\baru) + 2 \, \ln(A - i \, \lambda \, S_u) \, \calf(\tuz,\baru)  \vphantom{\frac{u_0-1}{u_0}} \right\}
  \label{eqdefK2R014}
\end{align}
This last result enables the computation of the integral $\wtI_3^{\,(1)}$ given by eq.~(\ref{eqdefJbeta4}) which will be a sum of functions $^{(2)}K^{R}_{0,1}(A,B,u_0^2)$ which was the goal of this appendix.
Note that eq.~(\ref{eqdefK2R014}) is valid for the real mass case. An extension to the complex mass case does not present new difficulties and can be carried out following the lines of section $2$ in ref.~\cite{paper2}.

\section{Study of the cut of the logarithm squared}\label{cut_log}

\subsection{Case of $\call_2(\baru,\tuz)$}\label{cut_log1}

In this appendix, we will study the following integral
\begin{align}
  \call &= \int_{-1-\tuz}^{1-\tuz} \frac{dz}{z} \, \ln^2\left( \frac{1-w}{1+w} \right)
  \label{eqdefL0}
\end{align}
with $w = (z+\tuz)/\baru$ where $\tuz$ and $\baru$ are genuine complex numbers. For a generic complex number $w$, let us note $w_R = \Re(w)$ and $w_I = \Im(w)$. The real and imaginary parts of $w$ are given by
\begin{align}
  w_R &= \frac{(z_R+\tuzr) \, \baru_R + (z_I + \tuzi) \, \baru_I}{|\baru|^2} \label{eqdefWr} \\
  w_I &= \frac{(z_I+\tuzi) \, \baru_R - (z_R + \tuzr) \, \baru_I}{|\baru|^2}
  \label{eqdefWi}
\end{align}
and the argument of the logarithm in eq.~(\ref{eqdefL0}) is given in terms of $w_R$ and $w_I$ by
\begin{align}
  \frac{1-w}{1+w} &= \frac{1 - |w|^2 - 2 \, i \, w_I}{|1+w|^2}
  \label{eqdefarglog0}
\end{align}
The cut of the logarithm is given by the two conditions
\[
  \begin{array}{ccl}
  \Im\left( \frac{1-w}{1+w} \right) = 0 & \rightarrow & w_I = 0 \\
  & \text{and} & \\
  \Re\left( \frac{1-w}{1+w} \right) \leq 0 & \rightarrow & 1 - |w|^2 \le 0
  \end{array}
\]
In terms of the variable $z$ these two conditions translate into
\[
  \left.
  \begin{array}{ccl}
    \Im\left( \frac{1-w}{1+w} \right) = 0 & \rightarrow & z_I = \frac{\baru_I}{\baru_R} \, (z_R+\tuzr) - \tuzi \\
   & & \\
  \Re\left( \frac{1-w}{1+w} \right) \leq 0 & \rightarrow & 
  \left\{
  \begin{array}{lcl}
    z_R > \baru_R - \tuzr & \text{if} & \baru_R > 0 \quad  \text{and} \quad w^0_R \ge 1 \\
    z_R < -\baru_R - \tuzr & \text{if} & \baru_R > 0 \quad \text{and} \quad w^0_R \le -1 \\
    z_R < \baru_R - \tuzr & \text{if} & \baru_R < 0 \quad \text{and} \quad w^0_R \ge 1 \\
    z_R > -\baru_R - \tuzr & \text{if} & \baru_R < 0 \quad \text{and} \quad w^0_R \le -1
  \end{array}
\right.
  \end{array}
\right.
\]
with $w^0_R = (z_R+\tuzr)/\baru_R$. Thus the cuts are some parts of a straight line $z_I = A_2 \, z_R + B_2$ with $A_2=\baru_I/\baru_R$ and $B_2 = (\baru_I \, \tuzr - \baru_R \, \tuzi)/\baru_R$ in the $z$ plane and the two branch points which correspond to $w^0_R = \pm 1$, are given by
\begin{align}
  z_B &= \baru_R - \tuzr + i \, (\baru_I - \tuzi) = \baru - \tuz \label{eqdefzB} \\ 
  z_B^{\prime} &= -\baru_R - \tuzr - i \, (\baru_I + \tuzi) = -\baru - \tuz \label{eqdefzBp}
\end{align}

\subsection{Case of $\call_1^{(2)}(\baru,\tuz)$}\label{cut_log2}

In this case , let us study the cuts of the following function
\begin{align}
  \call^{\prime}(z) &= \ln^2(1-w)
  \label{eqdeffuncLp1}
\end{align}
with $w=(z+\tuz)/\baru$. The real and imaginary parts of $w$ is given by eqs.~(\ref{eqdefWr}) and (\ref{eqdefWi}) and the branch cut of the function $\call^{\prime}(z)$ is given by the conditions
\[
  \left.
  \begin{array}{ccl}
    \Im\left( 1-w \right) = 0 & \rightarrow & z_I = \frac{\baru_I}{\baru_R} \, (z_R+\tuzr) - \tuzi \\
    & \text{and} & \\
  \Re\left( 1-w \right) \leq 0 & \rightarrow & 
  \left\{
  \begin{array}{lcl}
    z_R > \baru_R - \tuzr & \text{if} & \baru_R > 0 \quad  \text{and} \quad w^0_R \ge 1 \\
    z_R < \baru_R - \tuzr & \text{if} & \baru_R < 0 \quad \text{and} \quad w^0_R \ge 1
  \end{array}
\right.
  \end{array}
\right.
\]
In this case, there is only one branch cut and so only one branch point which correspond to $z_B$. 

\section{Further checks of the two-loop scalar amplitudes}\label{fcheck}

In this appendix, we provide further checks of formulae found for the different scalar two-loop amplitudes. The idea behind these checks is the following. A two-loop $N$-point scalar amplitude in momentum space is of the type
\begin{align}
  ^{(2)}I^{n}_{N} &= \int \frac{d^n k_1}{(2 \, \pi)^n} \, \int \frac{d^n k_2}{(2 \, \pi)^n} \, \prod_{i=1}^{I} \, \frac{1}{q_i^2 - m_i^2 + i \, \lambda}
  \label{eqgenetwoloop}
\end{align}
Taking the derivative of eq.~(\ref{eqgenetwoloop}) with respect to the mass squared $m_k^2$, we get
\begin{align}
  \frac{\partial \, ^{(2)}I^{n}_{N}}{\partial \, m_k^2} &= \int \frac{d^n k_1}{(2 \, \pi)^n} \, \int \frac{d^n k_2}{(2 \, \pi)^n} \, \left( \prod_{i=1 \, i \neq k}^{I} \, \frac{1}{q_i^2 - m_i^2 + i \, \lambda} \right) \, \frac{1}{(q_k^2 - m_k^2 + i \, \lambda)^2}
  \label{eqgenetwoloop1}
\end{align}
The last term in eq.~(\ref{eqgenetwoloop1}) can be viewed as two propagators of a scalar particle having a mass $m_k$ connected through a three-leg vertex whose external leg has a zero momentum, in other words, it represents a zero momentum insertion in the internal line propagating the scalar particle of mass $m_k$. We will use this idea to relate a $N$-point two-loop scalar amplitude to a $N+1$-point two-loop scalar amplitude whose one of the external momenta is taken to be zero. This idea is far from being new, cf.~\cite{Kotikov:1991pm,Kotikov:1991hm} for example. Several examples of such relations are given. Notice that, in this appendix, the different functions $\calft$ will be labelled for each topology with the same letter used to label the kinematics.\\

\subsection{Relations between topology $\calt_{22\_0\_2\_1}$ and $\calt_{23\_1\_2\_1}$}\label{relt22021_t23121}

A quick proof of such a relation can be provided by taking the derivative with respect to $m_3^2$ of eq,~(\ref{eqdefT22n2}). Using eq.~(\ref{eqdefFbar}), we readily get
\begin{align}
  \frac{\partial \, ^{(2)}I_{2}^{n}\left( \kappa_a \, ; \calt_{22\_0\_2\_1} \right)}{\partial \, m_3^2} &= - (4 \, \pi)^{-4 + 2 \, \varepsilon} \, \frac{\Gamma(1 + 2 \, \varepsilon)}{2 \, \varepsilon} \, \int_0^1 d \rho \int_0^1 d \xi \, \rho^{-1+\varepsilon} \, \left( 1 - \rho + \rho \, \xi \, (1-\xi) \right)^{-2 + 3 \, \varepsilon} \notag \\
  &\quad {} \times  (1-\rho) \, \left( \calft_a(\rho,\xi) - i \, \lambda \right)^{- 2 \, \varepsilon}
  \label{eqreltypeI1}
\end{align}
whose right-hand side has the same structure as the right-hand side of eq.~(\ref{eqUV3t6}).
Now, concerning the kinematics of the topology $\calt_{23\_1\_2\_1}$, setting $m_2=m_3$ and $p_2=0$ in eqs.~(\ref{eqdefGt3t}), (\ref{eqdefVt3t}) and (\ref{eqdefCt3t}), they become
\begin{align}
  \tG_c &=  0 \label{eqdefGt3tp0} \\
  \tV_c &=  0 \label{eqdefVt3tp0} \\
  \tC_c &=   \rho \, (1-\rho) \, \xi \, (1-\xi) \, p_1^2 - (1 - \rho + \rho \, \xi \, (1-\xi)) \, \left( \rho \, \xi \, m_1^2 + (1-\rho) \, m_3^2 + \rho \, (1-\xi) \, m_4^2 \right) \label{eqdefCt3tp0}
\end{align}
which implies that the polynomial $\calft_c(u,\rho,\xi)$ (cf.~eq.~(\ref{eqUV3t7})) does not depend on $u$ and thus
\begin{align}
  \calft_c(u,\rho,\xi) &= -\rho \, (1-\rho) \, \xi \, (1-\xi) \, p_1^2 + (1 - \rho + \rho \, \xi \, (1-\xi)) \notag \\
  &\quad {} \times \left( \rho \, \xi \, m_1^2 + (1-\rho) \, m_3^2 + \rho \, (1-\xi) \, m_4^2 \right)
  \label{eqUV3t7pp}
\end{align}
The right-hand side of eq.~(\ref{eqUV3t7pp}) is the same as the right-hand side of eq.~(\ref{eqdefFbar}) provided the substitution $p_1 \rightarrow p$ and $m_4 \rightarrow m_2$. Thus we get the following relation
\begin{align}
  \frac{\partial \, ^{(2)}I_{2}^{n}\left( \{p^2,m_1^2,m_2^2,m_3^2\} \, ; \calt_{22\_0\_2\_1} \right)}{\partial \, m_3^2} &= {} ^{(2)}I_{3}^{n}\left( \{ p^2,0,p^2,m_1^2, m_3^2,m_3^2,m_2^2\} \, ; \calt_{23\_1\_2\_1} \right)
  \label{eqrelTypeI2}
\end{align}
\\

\noindent
We can go one step further and verify that the eq.~(\ref{eqrelTypeI2}) is also verified using the final formulae for these two topologies. This will provide a test of eqs.~(\ref{eqdeftot1}) and (\ref{eqUV3tresfin2}). To do that, we take the derivative with respect to $m_3^2$ of eq.~(\ref{eqdeftot1}), this yields
\begin{align}
  \hspace{2em}&\hspace{-2em}\frac{\partial \, ^{(2)}I_{2}^{n}\left( \kappa_a \, ; \calt_{22\_0\_2\_1} \right)}{\partial \, m_3^2} = -(4 \, \pi)^{-4 + 2 \, \varepsilon} \, \frac{\Gamma(1 + 2 \, \varepsilon)}{2 \, \varepsilon} \, (1 + 2 \, \varepsilon + 4 \, \varepsilon^2) \notag \\
  &\quad {} \times \left\{ \frac{1}{\varepsilon} -1 - 2 \, \ln(m_3^2 - i \, \lambda) + \varepsilon \left[ \vphantom{\frac{(1-\rho)^2 \, \xi \, (1-\xi)}{(1-\rho + \rho \, \xi \, (1-\xi))^2}} 2 \, \ln^2(m_3^2 - i \, \lambda) + 4 \, \ln(m_3^2 - i \, \lambda) + 1 + 3 \, I_1 + I_3 + I_4   \right. \right. \notag \\
&\qquad \quad {}  -2 \, \int_0^1 d \xi \int_0^1 \frac{d \rho}{\rho} \, \left[ (1-\rho) \, \ln(\calft_a(\rho,\xi) - i \, \lambda) - \ln(m_3^2 - i \, \lambda) \right] \notag \\
&\qquad \quad {}  -2 \, \int_0^1 d \xi \int_0^1 d \rho \, \left( \frac{1-\rho}{\rho \, (1-\rho+\rho \, \xi \, (1-\xi))^2} \,  \ln(\calft_a(\rho,\xi) - i \, \lambda) - \frac{1-\rho}{\rho} \,  \ln(\calft_a(\rho,\xi) - i \, \lambda) \right. \notag \\
&\qquad \qquad \qquad \qquad \qquad \qquad {} + \frac{1-\rho}{\rho} \, \ln(1-\rho+\rho \, \xi \, (1-\xi)) \, \left( 1- \frac{1}{(1-\rho+\rho \, \xi \, (1-\xi))^2} \right) \notag \\
&\qquad \qquad \qquad \qquad \qquad \qquad {} + \left. \left. \left. \frac{1}{\rho} \, \left( \frac{1-\rho}{(1-\rho+\rho \, \xi \, (1-\xi))^2} - 1 \right) \right) \right] \right\}
  \label{eqdeftot1p0}
\end{align}
Furthermore, using the definition of $I_3$ (cf.~eq.~(\ref{eqUVdefI3p1})), the derivative of eq.~(\ref{eqdeftot1}) can be written as
\begin{align}
  \hspace{2em}&\hspace{-2em}\frac{\partial \, ^{(2)}\!I_{2}^{n}\left( \kappa_a \, ; \calt_{22\_0\_2\_1} \right)}{\partial \, m_3^2} = -(4 \, \pi)^{-4 + 2 \, \varepsilon} \, \frac{\Gamma(1 + 2 \, \varepsilon)}{2 \, \varepsilon}\notag \\
  &\quad {} \times \left\{ \frac{1}{\varepsilon} +1 - 2 \, \ln(m_3^2 - i \, \lambda) + \varepsilon \left[ \vphantom{\frac{1-\rho}{(1-\rho + \rho \, \xi \, (1-\xi))^2}} 2 \, \ln^2(m_3^2 - i \, \lambda) + 3 + 3 \, I_1 + 3 \, I_3 + I_4  \right. \right. \notag \\
  &\qquad \quad {}  -2 \, \int_0^1 d \xi \int_0^1 \frac{d \rho}{\rho} \, \left[ \frac{(1-\rho)}{(1 - \rho + \rho \,  \xi \, (1-\xi))^2} \, \ln(\calft_a(\rho,\xi) - i \, \lambda) - \ln(m_3^2 - i \, \lambda) \right] \notag \\
  &\qquad \quad {} - \left. \left.  2 \, \int_0^1 d \xi \int_0^1 \frac{d \rho}{\rho} \, \left( \frac{1-\rho}{(1-\rho + \rho \, \xi \, (1-\xi))^2} - 1 \right) \right] \right\}
  \label{eqdeftot1p1}
\end{align}
Computing the last integral using the method of appendix~\ref{intsecpar}, we end with
\begin{align}
  \hspace{2em}&\hspace{-2em}\frac{\partial \, ^{(2)}\!I_{2}^{n}\left( \kappa_a \, ; \calt_{22\_0\_2\_1} \right)}{\partial \, m_3^2} = -(4 \, \pi)^{-4 + 2 \, \varepsilon} \, \frac{\Gamma(1 + 2 \, \varepsilon)}{2 \, \varepsilon}\notag \\
  &\quad {} \times \left\{ \frac{1}{\varepsilon} +1 - 2 \, \ln(m_3^2 - i \, \lambda) + \varepsilon \left[ \vphantom{\frac{1-\rho}{(1-\rho + \rho \, \xi \, (1-\xi))^2}} 2 \, \ln^2(m_3^2 - i \, \lambda) + 1 + 3 \, I_1 + 3 \, I_3 + I_4  \right. \right. \notag \\
  &\qquad \quad {}  - \left. \left. 2 \, \int_0^1 d \xi \int_0^1 \frac{d \rho}{\rho} \, \left[ \frac{1-\rho}{(1 - \rho + \rho \,  \xi \, (1-\xi))^2} \, \ln(\calft_a(\rho,\xi) - i \, \lambda) - \ln(m_3^2 - i \, \lambda) \right] \right] \right\}
  \label{eqdeftot1p2}
\end{align}
The right-hand side of eq.~(\ref{eqdeftot1p2}) has the same structure as the right-hand side of eq.~(\ref{eqUV3tresfin2}). 
But it easy to see that
\begin{align}
  \left. \wtI_{2}^{\,(1)}(\Sigma_{(1)};\tG_c,\tV_c,\tC_c,\rho,\xi) \vphantom{\frac{\rho}{\rho}} \right|_{\substack{p_2 \rightarrow 0\\ p_1 \rightarrow p\\ p_3 \rightarrow -p\\
    m_2 \rightarrow m_3 \\ m_4 \rightarrow m_2}} &= \ln\left( \calft_a(\rho,\xi) - i \, \lambda \right)
  \label{eqlimitw1}
\end{align}
and 
\begin{align}
  \left. \wtI_{2}^{\,(i)}(\Sigma_{(1)};\tG_c,\tV_c,\tC_c,0,\xi) \vphantom{\frac{\rho}{\rho}} \right|_{\substack{p_2 \rightarrow 0\\ p_1 \rightarrow p\\ p_3 \rightarrow -p\\
    m_2 \rightarrow m_3 \\ m_4 \rightarrow m_2}} &= \ln^{i} \left( m_3^2 - i \, \lambda \right)
  \label{eqlimitw2}
\end{align}
Using eqs~(\ref{eqlimitw1}) and (\ref{eqlimitw2}) in eq.~(\ref{eqdeftot1p2}), we are back to eq.~(\ref{eqrelTypeI2}) thus providing a cross-check of eqs~(\ref{eqdeftot1}) and (\ref{eqUV3tresfin2}).

\subsection{Relations between topology $\calt_{22\_0\_2\_1}$ and $\calt_{24\_2\_2\_2}$}\label{relt22021_t24222}

Let us differentiate two times eq.~(\ref{eqdeftot1}) with respect to $m_1^2$ and to $m_2^2$. A straightforward computation shows that
\begin{align}
  \hspace{2em}&\hspace{-2em}\frac{\partial \, ^{(2)}\!I_{2}^{n}\left( \kappa_a \, ; \calt_{22\_0\_2\_1} \right)}{\partial \, m_1^2} \notag \\ 
  &= -(4 \, \pi)^{-4 + 2 \, \varepsilon} \, \frac{\Gamma(1 + 2 \, \varepsilon)}{2 \, \varepsilon} \,  \left\{ \frac{1}{\varepsilon} +1 - 2 \, \ln(m_1^2 - i \, \lambda) + \varepsilon \left[ \vphantom{\frac{1-\rho}{(1-\rho + \rho \, \xi \, (1-\xi))^2}} 2 \, \ln^2(m_1^2 - i \, \lambda) \right. \right. \notag \\
  &\quad {} + 1 -2 \, \ln(m_1^2 - i \, \lambda) + \frac{2}{\xi^+ - \xi^-} \, \left( \dilog\left( \xi^- \right) - \dilog\left( \xi^+ \right) \right) - \frac{\pi^2}{6} + 2 \, \dilog\left(  \frac{m_1^2 - m_2^2}{m_1^2 - i \, \lambda} \right) \notag \\
  &\quad {}  - \left. \left. 2 \, \int_0^1 d \xi \int_0^1 d \rho \, \frac{\xi}{(1 - \rho + \rho \,  \xi \, (1-\xi))^2} \, \left[ \ln(\calftp_a(\rho,\xi) - i \, \lambda) - \ln(\xi \, m_1^2 + (1-\xi) \, m_2^2 - i \, \lambda) \right] \right] \right\}
  \label{eqdeftot1p3}
\end{align}
Then taking the derivative of eq.~(\ref{eqdeftot1p3}) with respect to $m_2^2$ leads to
\begin{align}
  \hspace{2em}&\hspace{-2em}\frac{\partial^2 \, ^{(2)}\!I_{2}^{n}\left( \kappa_a \, ; \calt_{22\_0\_2\_1} \right)}{\partial \, m_1^2 \, \partial \, m_2^2} \notag \\ 
  &= (4 \, \pi)^{-4 + 2 \, \varepsilon} \, \Gamma(1 + 2 \, \varepsilon) \,  \int_0^1 d \xi \int_0^1 d \rho \, \frac{\rho \, \xi \, (1-\xi)}{1 - \rho + \rho \,  \xi \, (1-\xi)} \, \left( \calft_a(\rho,\xi) - i \, \lambda \right)^{-1} 
  \label{eqdeftot1p4}
\end{align}
The right-hand side of eq.~(\ref{eqdeftot1p4}) has the same structure as the right-hand side of eq.~(\ref{eqUV8t8}) for $\varepsilon=0$. To go further, let us take the limit $p_2,p_4 \rightarrow 0$ with $m_1=m_2$ and $m_3=m_4$ for the kinematics $\kappa_i$, this leads for the matrix $\tG_i$, the vector $\tV_i$ and the scalar $\tC_i$ to the following result
\begin{align}
  \tG_i &=  \left[
  \begin{array}{cc}
    0 & 0 \\
    0 & 0
  \end{array}
\right] \label{eqdefaGt8tw} \\
\tV_i &= \left[
\begin{array}{c}
  0 \\
  0
\end{array}
\right] \label{eqdefVt8tw} \\
\tC_i &= \rho \, (1-\rho) \, \xi \, (1-\xi) \, p_3^2 - (1 - \rho + \rho \, \xi \, (1-\xi)) \, \left( \rho  \, \xi \, m_1^2 + \rho \, (1-\xi) \, m_4^2 + (1-\rho) \, m_5^2 \right)
  \label{eqdefCt8tw}
\end{align}
The polynomial given by eq.~(\ref{eqUV8t7}) does not depend on $u_1$ and $u_2$ in this limit, thus the integration over the square $K_{(2)}$ can be readily done bringing
\begin{align}
  \left. \wtI_{3}^{\,(0)}(K_{(2)};\tG_i,\tV_i,\tC_i,\rho,\xi) \vphantom{\frac{\rho}{\rho}} \right|_{\substack{p_2,p_4 \rightarrow 0\\ p_1 \rightarrow p\\ p_3 \rightarrow -p\\
    m_2 \rightarrow m_1 \\ m_3 \rightarrow m_2 \\ m_4 \rightarrow m_2 \\ m_5 \rightarrow m_3}} &= \left( \calft_a(\rho,\xi) - i \lambda \right)^{-1}
  \label{eqlimitw3}
\end{align}
Putting this last result in eq.~(\ref{eqUV8t8}), we get that
\begin{align}
  \frac{\partial^2 \, ^{(2)}\!I_{2}^{n}\left( \{p^2,m_1^2,m_2^2,m_3^2\} \, ; \calt_{22\_0\_2\_1} \right)}{\partial \, m_1^2 \, \partial \, m_2^2} &= ^{(2)}\!I_{4}^{4}\left( \{p^2,0,p^2,0,p^2,p^2,m_1^2,m_1^2,m_2^2,m_2^2,m_3^2\} \, ; \calt_{24\_2\_2\_2} \right)
  \label{eqrelTypeI2p}
\end{align}

\subsection{Relations between topology $\calt_{23\_1\_2\_1}$ and $\calt_{24\_2\_2\_1}$}\label{relt23121_t24221}

Let us start by taking the derivative of eq.~(\ref{eqUV3tresfin2}) with respect to $m_3^2$. To do that, we have to compute the derivative with respect to $m_3^2$ of $\wtI_{2}^{\,(k)}(\Sigma_{(1)};\tG_c,\tV_c,\tC_c,\rho,\xi)$
\begin{align}
  \frac{\partial \, \wtI_{2}^{\,(k)}(\Sigma_{(1)};\tG_c,\tV_c,\tC_c,\rho,\xi)}{\partial \, m_3^2} &\equiv \frac{\partial}{\partial \, m_3^2} \, \int_0^1 du \, \ln^k(\tG_c \, u^2 - 2 \, \tV_c \, u - \tC_c - i \, \lambda) \notag \\
  &= (1-\rho) \, (1-\rho+\rho \, \xi \, (1-\xi)) \notag \\
  &\quad {} \times k \, \int_0^1 du \, (1-u) \, \frac{\ln^{k-1}(\tG_c \, u^2 - 2 \, \tV_c \, u - \tC_c - i \, \lambda)}{\tG_c \, u^2 - 2 \, \tV_c \, u - \tC_c - i \, \lambda}
  \label{eqderwrtm320}
\end{align}
But taking the limit $p_3 \rightarrow 0$, with $m_3=m_4$ in the kinematics $\kappa_f$, eqs.~(\ref{eqdefaGt7t}), (\ref{eqdefVt7t}) and (\ref{eqdefCt7t}) become
\begin{align}
  \tG_f &= (1-\rho)^2 \, \left[
  \begin{array}{cc}
    p_2^2 & 0 \\
    0 & 0
  \end{array}
\right] \label{eqdefaGt7tw} \\
\tV_f &= \frac{1}{2} \, (1 - \rho) \, \left[
\begin{array}{c}
  \rho \, \xi \, (1-\xi) \, 2 \, p_2 \cdot p_4 + (1 - \rho + \rho \, \xi \, (1-\xi)) \, (p_2^2 - m_2^2 + m_4^2) \\
  0
\end{array}
\right] \label{eqdefVt7tw} \\
\tC_f &= \rho \, (1-\rho) \, \xi \, (1-\xi) \, p_4^2 - (1 - \rho + \rho \, \xi \, (1-\xi)) \, \left( \rho  \, \xi \, m_1^2 + (1-\rho) \, m_4^2 + \rho \, (1-\xi) \, m_5^2 \right)
  \label{eqdefCt7tw}
\end{align}
From the definition of $\wtI_{3}^{\,(0)}(\Sigma_{(2)};\tG_f,\tV_f,\tC_f,\rho,\xi)$ (cf.~eq.~(\ref{eqdeftI3})) and the structure of eqs.~(\ref{eqdefaGt7tw}), (\ref{eqdefVt7tw}) and (\ref{eqdefCt7tw}), the following relation holds
\begin{align}
  \hspace{2em}&\hspace{-2em}\frac{\partial \, \wtI_{2}^{\,(k)}(\Sigma_{(1)};\tG_c,\tV_c,\tC_c,\rho,\xi)}{\partial \, m_3^2} \notag \\
  &= \left. k \, (1-\rho) \, (1-\rho+\rho \, \xi \, (1-\xi)) \, \wtI_{3}^{\,(k-1)}(\Sigma_{(2)};\tG_f,\tV_f,\tC_f,\rho,\xi) \vphantom{\frac{A^2}{B^2}} \right|_{\substack{p_1 \rightarrow p_3\\p_3 \rightarrow 0\\p_4 \rightarrow p_1\\m_4 \rightarrow m_3\\m_5 \rightarrow m_4}}
  \label{eqderwrtm31}
\end{align}
Having established eq.~(\ref{eqderwrtm31}), it is easy to show that
\begin{align}
\hspace{2em}&\hspace{-2em}\frac{\partial \, ^{(2)}I_{3}^{n}\left( \{p_1^2,p_2^2,p_3^2,m_1^2,m_2^2,m_3^2,m_4^2\} \, ; \calt_{23\_1\_2\_1} \right))}{\partial \, m_3^2} \notag \\
&= ^{(2)}\!I_{4}^{4}\left( \{p_3^2,p_2^2,0,p_1^2,p_1^2,p_2^2,m_1^2,m_2^2,m_3^2,m_3^2,m_4^2\} \, ; \calt_{24\_2\_2\_1} \right)
  \label{eqrelTypeI30}
\end{align}
This gives a cross-check of eqs.~(\ref{eqUV3tresfin2}) and (\ref{eqUV7t8}).

\vspace{1cm}

\bibliographystyle{unsrt}
\bibliography{../../biblio,../../publi}

\end{fmffile}

\end{document}